\documentclass[twocolumn]{aastex631}

\usepackage{xcolor}
\usepackage{gensymb}

\newcommand{\ignore}[1]{}

\def\Ha{\mbox{H$\alpha$}}
\def\Msun{\mbox{${\rm M}_{\odot}$}}

\def\Rsun{\mbox{${\rm R}_{\odot}$}}

\def\kms{\mbox{km\,s$^{-1}$}}

\usepackage{xspace}
\usepackage{amsmath}

\newcommand{\spec}[3]{\mbox{#1\,{\sc #2}\,$\lambda$#3}\xspace}

\shorttitle{Masses and orbits of Be binaries}
\shortauthors{Klement et al.}

\graphicspath{{./}{figures/}}

\begin{document}

\title{The CHARA Array interferometric program on the multiplicity of classical Be stars: new detections and orbits of stripped subdwarf companions}

\author[0000-0002-4313-0169]{Robert Klement}
\affiliation{The CHARA Array of Georgia State University, Mount Wilson Observatory, Mount Wilson, CA 91023, USA}
\affiliation{European Organisation for Astronomical Research in the Southern Hemisphere (ESO), Casilla 19001, Santiago 19, Chile}

\author[0000-0003-1013-5243]{Thomas Rivinius}
\affiliation{European Organisation for Astronomical Research in the Southern Hemisphere (ESO), Casilla 19001, Santiago 19, Chile}

\author[0000-0001-8537-3583]{Douglas R. Gies}
\affiliation{Center for High Angular Resolution Astronomy, Department of Physics and Astronomy,\\ Georgia State University, P.O. Box 5060, Atlanta, GA 30302-5060, USA}

\author[0000-0003-1637-9679]{Dietrich Baade}
\affiliation{European Organisation for Astronomical Research in the Southern Hemisphere (ESO), \\ Karl-Schwarzschild-Str.\ 2, 85748 Garching bei M\"unchen, Germany}

\author[0000-0003-2125-0183]{Antoine Mérand}
\affiliation{European Organisation for Astronomical Research in the Southern Hemisphere (ESO), \\ Karl-Schwarzschild-Str.\ 2, 85748 Garching bei M\"unchen, Germany}

\author[0000-0002-3380-3307]{John D. Monnier}
\affiliation{Department of Astronomy, University of Michigan, 1085 S. University Ave, Ann Arbor, MI 48109, USA}

\author[0000-0001-5415-9189]{Gail H. Schaefer}
\affiliation{The CHARA Array of Georgia State University, Mount Wilson Observatory, Mount Wilson, CA 91023, USA}

\author[0000-0001-9745-5834]{Cyprien Lanthermann}
\affiliation{The CHARA Array of Georgia State University, Mount Wilson Observatory, Mount Wilson, CA 91023, USA}

\author[0000-0002-2208-6541]{Narsireddy Anugu}
\affiliation{The CHARA Array of Georgia State University, Mount Wilson Observatory, Mount Wilson, CA 91023, USA}

\author[0000-0001-6017-8773]{Stefan Kraus}
\affiliation{Astrophysics Group, School of Physics and Astronomy, University of Exeter, Stocker Road, Exeter, EX4 4QL, UK}

\author[0000-0002-3003-3183]{Tyler Gardner}
\affiliation{Astrophysics Group, School of Physics and Astronomy, University of Exeter, Stocker Road, Exeter, EX4 4QL, UK}

\begin{abstract}

Rapid rotation and nonradial pulsations enable Be stars to build decretion disks, where the characteristic line emission forms. A major but unconstrained fraction of Be stars owe their rapid rotation to mass and angular-momentum transfer in a binary. The faint, stripped companions can be helium-burning subdwarf OB-type stars (sdOBs), white dwarfs (WDs), or neutron stars. We present optical/near-IR CHARA interferometry of 37 Be stars selected for spectroscopic indications of low-mass companions. From multi-epoch $H$- and/or $K$-band interferometry plus radial velocities and parallaxes collected elsewhere, we constructed 3D orbits and derived flux ratios and absolute dynamical masses of both components for six objects, quadrupling the number of anchor points for evolutionary models. In addition, a new wider companion was identified for the known Be + sdO binary 59~Cyg, while auxiliary VLTI/GRAVITY spectrointerferometry confirmed circumstellar matter around the sdO companion to HR\,2142. On the other hand, we failed to detect any companion to the six Be stars with $\gamma$\,Cas-like X-ray emission, with sdOB and main-sequence companions of the expected spectroscopic mass being ruled out for the X-ray-prototypical stars $\gamma$\, Cas and $\pi$\, Aqr, leaving the elusive WD companions as the most likely companions, as well as a likely explanation of the X-rays. No low-mass main-sequence close companions were identified in the other stars.

\end{abstract}


\section{Introduction} \label{sec:intro}

Be stars, i.e., B-type stars with Balmer line emission, are common objects with a long observational history. Classical Be stars, which represent the vast majority of observed Be stars, are rapidly rotating main sequence (MS) B-type stars in which the line emission originates from self-ejected, ionized circumstellar disks in Keplerian rotation \citep{2013A&ARv..21...69R}. The origin of the near-critical rotation, which is an essential property of classical Be stars, is highly debated. It might be acquired during MS evolution of a single star as the stellar core contracts if the star is a moderately rapid rotator already at zero age MS \citep{2000A&A...361..101M,2008A&A...478..467E,2013A&A...553A..25G}, or it can otherwise be gained during a phase of mass and angular momentum transfer during evolution in an interacting binary system \citep{1982IAUS...98..327R, 1991A&A...241..419P,2005ApJS..161..118M,2010ApJ...722..605H}. In fact, given that B-type stars are commonly found in binary systems \citep{2015A&A...580A..93D,2021MNRAS.tmp.1986V}, a large fraction is expected to go through binary mass-transferring phase which spins up the mass gainer,  or may result in stellar mergers \citep{2013ApJ...764..166D,2014ApJ...782....7D}. In Be stars, the formation of the circumstellar disk can then be understood as the most efficient way for the rapid rotator to shed its excess angular momentum, with gas being launched into orbit from the stellar equator with the likely help of non-radial pulsations \citep{2003A&A...411..229R,2011A&A...527A..84K,2022AJ....163..226L}. The so-called viscous decretion disk (VDD) itself grows and transfers the angular momentum outwards by means of astrophysical viscosity \citep{1991MNRAS.250..432L,2011IAUS..272..325C,2019NewA...70....7M}. 

There is a growing number of Be stars that are found to reside in binary systems with stripped, evolved companions, a fact which unambiguously points towards past binary transfer leading to their formation \citep[e.g.,][]{2018ApJ...853..156W, 2021AJ....161..248W,2023AJ....165..203W}. The progenitors of such Be + stripped star systems are close binaries (orbital periods of the order of days to weeks), where the more massive component is the first to evolve off the main-sequence, which is followed up by its Roche lobe overflow, mass transfer to the less massive component, and expansion of the orbit following a mass ratio reversal. After this case~B mass transfer, the mass gainer is spun-up and rejuvenated, and the mass donor goes through a short-lived and unstable `pre-subdwarf' phase. At this point, the donor is still bloated and bright \citep[Rivinius et al., submitted;][]{2023arXiv230707766V}, before becoming a faint subdwarf OB-type star (sdOB), sometimes also referred to as a helium star, i.e., a stripped stellar core which continues the nuclear burning of helium \citep{1981PASP...93..297P,2016PASP..128h2001H}. 

Following the mass transfer, the spun-up mass gainer may proceed to form a decretion disk and become observable as a classical Be star on a circularized orbit \citep[e.g.,][]{2010ApJ...724..546S} with the faint sdOB companion. The less massive sdOB companions would eventually enter the white dwarf (WD) cooling phase after exhausting their helium supplies, but this generally should not happen, as the corresponding timescales are longer than the MS lifetime of the newly formed Be star. Stripped companions that are more massive than $\sim1.6$\,{\Msun} should explode in core-collapse supernovae \citep{2019ApJ...878...49W} to become neutron stars (NS) or black holes (BH), although the latter have yet to be confirmed observationally \citep[][see below]{2023arXiv230808642J}. Some sdOBs may also fill their Roche lobe again during a late shell helium-burning phase, leading to a second phase of mass transfer to the Be star, after which the companion becomes a WD \citep{1981A&A....96..142D,2020A&A...637A...6L}.

Binary population synthesis simulations generally predict that a large fraction of presently observed Be stars are formed by binary mass transfer, with the majority expected to have sdOB or WD companions, and with NS or BH companions expected to be rare \citep{1991A&A...241..419P, 2001A&A...367..848R, 2014ApJ...796...37S, 2021ApJ...908...67S}. Ancillary pieces of observational evidence supporting the binary origin of Be stars are a lack of detected MS close companions among Be stars \citep[][with the latter study concerning specifically field Be stars earlier than B1.5]{2000ASPC..214..668G,2020A&A...641A..42B}, a prevalence of spectral energy distribution (SED) turndown in the radio among Be stars that is indicative of tidally truncated circumstellar disks \citep{2017A&A...601A..74K,2019ApJ...885..147K}, and a non-negligible fraction of runaway stars among Be stars \citep{2001ApJ...555..364B,2018MNRAS.477.5261B}.

Several potential mass-transferring progenitors of classical Be stars have been identified, although the exact range of parameters required to form a Be star remains poorly constrained \citep{2015A&A...573A.107H,2022MNRAS.516.3602E}. The first case of the unstable Be + pre-subdwarf stage has been confirmed only recently \citep{2022A&A...659L...3F}, and there are now six examples of this short-lived stage (Rivinius et al., submitted). These objects are identifiable thanks to the brightness of the low-mass pre-subdwarf star, whose prominent narrow absorption lines exhibit large radial velocity (RV) variations, while quasi-static spectral features are observed from the newly formed Be star. However, it should be noted that the Be + pre-sdOB binaries can be confused with hierarchical triple systems, in which the Be star is the outer companion to an inner binary consisting of two main sequence stars \citep{2020A&A...637L...3R}, an example of which is the Be star $\nu$~Gem \citep{2021ApJ...916...24K}. The unstable phase is expected to last only on the order of $10^4$\,yr, after which the pre-sdOB contracts, gets hotter, and settles to the stable helium-burning sdOB state.

The sdOB companions are difficult to find because they can be completely outshined by the bright Be stars \citep{2000ASPC..214..668G}. However, they can be directly detected in far-UV (FUV) spectroscopy, where the flux contribution of the hot companion can reach up to $\sim15$\%, as was found for $\varphi$~Per \citep[][]{1998ApJ...493..440G}, or by near-IR interferometry, if the companion is bright enough to contribute at least $\sim0.3$\% of the flux of the primary Be star \citep{2015A&A...577A..51M,2022ApJ...926..213K}. Targeted explorations in the FUV have led to the confirmation of $15$ Be + sdO binaries \citep{1998ApJ...493..440G,2008ApJ...686.1280P,2013ApJ...765....2P,2016ApJ...828...47P,2017ApJ...843...60W,2018ApJ...853..156W,2021AJ....161..248W} and one Be + sdB binary \citep{2022ApJ...940...86K}. Five of these were subsequently also detected by near-IR interferometry from the CHARA Array \citep{2015A&A...577A..51M, 2022ApJ...926..213K,2022ApJ...940...86K}. 

A direct sign of an O-type companion is also the presence of the \spec{He}{ii}{4686} spectral line in absorption \citep{2018ApJ...865...76C,2023AJ....165..203W} or in emission, and in the latter case it indicates the presence of circumcompanion gas, possibly in the form of a disk as the companion accretes gas from the outer disk of the Be star \citep{1981PASP...93..297P,2000ASPC..214..581R}. The presence of a hot companion can also be deduced indirectly from spectral effects caused by heating up of the part of the disk facing the companion, such as additional phase-locked emission peaks in \ion{He}{1} lines moving with larger velocity amplitudes than the companion \citep{2000A&A...358..208S,2001A&A...368..471H}. Furthermore, orbital phase-locked transient sharp absorption line features originating from disk self-absorption (i.e.\ shell lines, see below) can originate from enhanced ionization in the disk region facing the hot companion \citep{1981PASP...93..297P,2004A&A...427..307R}, or even from gas streams during an ongoing mass transfer from the Be-star disk to the companion \citep{2016ApJ...828...47P}.

WD companions to Be stars are not detectable in the FUV or without enormous efforts in the optical \citep{2020ApJ...902...25G}. But, they should be detectable in X-rays if there is an ongoing accretion of the Be disk material onto the WD. Indeed, several candidates for Be + WD systems were recently found in the Magellanic Clouds (MC) as luminous super-soft X-ray sources, in which the X-rays originate from nuclear burning of the accreted material on the surface of the WD \citep[][and references therein]{2023RAA....23b5021Z}. 

It has also been suggested that accretion from Be disks onto WD companions can explain the formation of peculiar, hard X-rays detected for $\gamma$~Cas and about two dozen other stars \citep{2016AdSpR..58..782S,2020MNRAS.493.2511N}, which otherwise appear as normal early-type classical Be stars \citep{2022MNRAS.510.2286N}. In this case, an accretion disk would be formed around the WD and the X-rays would originate directly from gravitational energy conversion similar to the case of cataclysmic variables \citep[e.g.,][]{2018PASJ...70..109T}. While a large fraction of Be stars with $\gamma$~Cas-like X-rays are confirmed spectroscopic binaries \citep{2022MNRAS.510.2286N}, no signature of an sdOB (or any other) companion has been detected in the FUV, optical, or near-IR, which favors the Be + WD interpretation \citep{2023ApJ...942L...6G}. NS companions in propeller regime were also suggested to explain the observed X-rays \citep{2017MNRAS.465L.119P}, but this scenario appears unlikely due to the spectroscopic estimates of the companion masses being generally too low \citep{2022MNRAS.510.2286N} as well as other reasons \citep{2017MNRAS.469.1502S}. More recent results from ongoing all-sky X-ray surveys indicate an incidence of up to $\sim12$\% of bright and hard X-ray sources associated with Be stars at distances between 100 and 1000\,pc, while on the other hand, there is still no known example of a luminous super-soft X-ray source in the Milky Way (MW) asscociated with a Be star \citep{2023arXiv230713308N}.

The existence of high-mass X-ray binaries with a Be component (i.e., the Be X-ray binaries) implies that in some cases, the stripped companions remain massive enough to explode in core-collapse supernovae. The accretion of Be disk material onto the compact companion, which in all confirmed cases is a NS \citep{2011Ap&SS.332....1R}, then generates conspicuous luminous and hard X-rays, observable from extra-galactic regions \citep{2005A&AT...24..151R}. There are currently about 70 identified Be X-ray binaries in the MW\footnote{\url{http://xray.sai.msu.ru/~raguzova/BeXcat/}}, a number which confirms that this evolutionary outcome is rather rare \citep{2022MNRAS.516.3366N}. The only reported case of a Be star with a BH companion, which would however be in a quiescent state, i.e., not actively accreting \citep{2014Natur.505..378C}, has recently been ruled out and proposed as another probable case of a Be + sdO binary, even though the emission in \spec{He}{ii}{4686} is much larger than observed for the confirmed Be + sdO binaries $\varphi$~Per or 59~Cyg \citep{2022arXiv220812315R,2023arXiv230808642J}. 

Given that the binary evolution channel is important and possibly dominant, Be stars are extremely useful laboratories for testing and calibrating models of binary evolution including different phases of mass transfer between the components. The sdOB companions found around Be stars specifically are thought to be at the low-mass end of stars stripped in binaries, which are expected to form a continuous population with stripped Wolf-Rayet stars at the high-mass end \citep{2018A&A...615A..78G}. The widespread presence of stripped companions to Be stars is also of broad astrophysical importance as constituting a significant source of ionizing radiation \citep{2019A&A...629A.134G}, and a pool of possible progenitors of hydrogen-poor supernovae and $\gamma$-ray bursts. However, in order to calibrate the binary evolutionary sequence leading to and following the formation of post-mass-transfer Be binaries, it is necessary to obtain the fundamental parameters of the two presently observed binary components. Without relying on model assumptions, this can be done only by obtaining 3D orbital solutions from the combination of spectroscopic and astrometric time-series of observations, aided by the knowledge of the distance from parallax measurements.

While spectrocopic orbital solutions are available for a number of Be stars, RV measurements of the primary Be star are hindered by broadened spectroscopic features and low RV amplitudes due to the much less massive companion. The situation is improved in the case of Be-shell stars seen close to edge-on orientation, which present sharp spectral features due to self-absorption in the circumstellar disk. As for the sdOB secondaries, their faintness prevents reliable RV measurements except when it is hot enough to contribute significantly in the FUV (and if FUV spectra exist in sufficient quality). Thus, even though it has been possible to measure double-lined spectroscopic orbits for a few Be stars \citep{1981PASP...93..297P,2013ApJ...765....2P,2023AJ....165..203W}, single-lined spectroscopic orbits are more commonly available in the literature \citep[e.g.,][]{2009A&A...506.1319R,1988A&AS...75..311D,1997A&A...328..551K}. In the latter case, knowledge of the distance is crucial for the derivation of all orbital parameters and subsequently the dynamical masses.

Optical/near-IR interferometry is the only astronomical technique capable of spatially resolving the orbits of nearby and bright Be binaries with periods of the order of months, which is typical for both confirmed Be + sdOB binaries, and close spectroscopic Be binaries with unknown types of companions. The first Be + sdOB binary with an interferometrically mapped orbit was $\varphi$~Per, which together with the spectroscopic orbit led to the first precise dynamical masses for a Be star and its stripped companion \citep{2015A&A...577A..51M}. Recently, similar orbital analyses resulted in the first orbits and dynamical masses for the Be + sdOB binaries V2119~Cyg, 60~Cyg \citep[both][]{2022ApJ...926..213K}, and $\kappa$~Dra \citep{2022ApJ...940...86K}. An interferometric multiplicity survey of Be stars with a limiting magnitude difference for companion detection of 3.0 to 3.5 mag in the visible was performed by the Navy Precision Optical Interferometer (NPOI) and the Mark III Stellar Interferometer, with several targets in common with the present study \citep{2021ApJS..257...69H}. The results (all compatible with ours) will be mentioned in later subsections dedicated to the individual targets.

In this work, we present the results of a large interferometric program with the CHARA Array aiming at the detection and orbital mapping of confirmed and candidate Be binaries. The CHARA data are complemented by a small amount of mostly archival VLTI data. The ultimate goal is to obtain a representative set of dynamical masses of Be stars and their stripped companions in order to anchor the binary evolution models among B-type stars. The full sample consists of more than 40 Be stars (37 presented here) which include Be + sdOB binaries, single-lined spectroscopic (SB1) binaries with undetected companions, Be stars emitting $\gamma$~Cas-like X-rays (i.e., Be + WD binary candidates), and binary candidates based on the presence of the radio SED turndown \citep{2017A&A...601A..74K, 2019ApJ...885..147K} or the morphology of the H$\alpha$ emission line suggestive of tidal distortions \citep{2019IAUS..346..105R}. The observational facilities and the observing campaign are described in Sects.~\ref{sec:observations} and \ref{sec:campaign}. We present updated orbits for the three Be + sdOB binaries 28~Cyg, V2119~Cyg, and 60~Cyg in Sect.~\ref{sec:orbits_updated} and orbits for three new detections in Sect.~\ref{sec:orbits_new}. In Sect.~\ref{sec:non-detections}, marginal detections, detections of wide companions, and non-detections are discussed, which is followed by a general discussion and conclusions section.

\section{Observations} \label{sec:observations}


\subsection{Interferometry}

\subsubsection{The CHARA Array}

The Center for High Angular Resolution Astronomy (CHARA) Array located at Mt. Wilson, California is an optical/near-IR interferometer with the longest baselines in the world ($B_{\rm max} \sim 330$\,m), translating to sub-milliarcsec angular resolution \citep[$\lambda / 2 B_{\rm max} \sim0.5$\,mas in the near-IR $H$ band;][]{2005ApJ...628..453T,2020SPIE11446E..05S}. The six 1-meter telescopes of CHARA are arranged in a fixed Y-shaped configuration and each of the six light beam paths is equipped with two adaptive optics (AO) systems, one at the telescope, and the other in the optical laboratory just before entering the beam-combining instrument \citep{2013JAI.....240007C,2016AAS...22742702T,2020SPIE11446E..22A}.  

The $H$-band six-beam combiner MIRC-X \citep[upgraded from MIRC and commissioned in 2018 September;][]{2020AJ....160..158A} and the $K$-band six-beam combiner MYSTIC \citep[installed in 2021 July;][]{2023JATIS...9b5006S} operate simultaneously, enabling dense $(u,v)$ coverage consisting of 15 independent baselines and 10 independent closed triangles in each band. The main observables are squared visibilities (VIS2) for each baseline and closure phases (CP) for each baseline triangle; while VIS2 is sensitive to the angular extent of the observed target, and decreases with increasing projected baseline if the target is spatially resolved, CP reveals deviations from point symmetry in the resolved target \citep[for an introduction to optical interferometry and its observables, we refer to][]{2003RPPh...66..789M}.

MIRC-X and MYSTIC observations typically use a dispersion prism leading to a spectral resolution of $R\sim50$, which achieves sensitivity to targets with magnitudes below 6.5\,mag in average conditions (7.5\,mag in best conditions), and a bandwidth-smearing interferometric field of view (FoV) of $\sim55$ and $\sim65$\,mas, respectively, for the two instruments. For brighter targets, higher resolutions of $R \sim100$ and $\sim190$ enabling a larger FoV of $\sim100$\,mas and $\sim200$\,mas, respectively, are also available at the expense of the overall sensitivity.

In terms of detection of binary companions, CHARA/MIRC(-X) routinely reaches a contrast ratio of 1\% or lower \citep[corresponding to a magnitude difference of more than 5\,mag; e.g.,][]{2022ApJ...926..213K,2022ApJ...941..118D,2023A&A...672A...6L}, with the best-performance limit being $\sim 0.3$\% \citep[$\Delta {\rm mag} \sim 6.4$;][]{2015A&A...579A..68G}. Thanks to the six telescopes and the resulting dense coverage of the $(u,v)$ plane, CHARA is capable of detecting faint companions with snapshot observations, i.e., 10--20 minutes integration of fringes on the science target. With the necessary overheads and an observation of a calibrator, the total observing time is around one hour when using a calibrator-science (CAL-SCI) sequence. It should be noted, however, that the data quality is highly dependent on the atmospheric conditions, most importantly the seeing. 

The CHARA data used in this study were taken mostly with $R=50$ and were reduced with the dedicated MIRC-X reduction pipeline\footnote{\url{https://gitlab.chara.gsu.edu/lebouquj/mircx_pipeline}} \citep[version 1.3.5;][]{2020AJ....160..158A}. The calibrator stars used in this work were selected with the Searchcal software developed and maintained by the Jean-Marie Mariotti Center\footnote{\url{https://www.jmmc.fr/english/tools/proposal-preparation/search-cal/}} \citep[JMMC;][]{2016A&A...589A.112C}, while the calibrator angular diameters were adopted from the JMMC catalog of stellar diameters\footnote{\url{https://vizier.u-strasbg.fr/viz-bin/VizieR-3?-source=II/346}} \citep{2014ASPC..485..223B, 2017yCat.2346....0B} and are listed in Table~\ref{tab:interf_calibrators}. Following the data reduction, each calibrator dataset was checked for signs of binarity or other issues, and several bad calibrators were indeed identified and reported. The calibrated OIFITS files corresponding to all interferometric observations analyzed in this work will be available in the Optical Interferometry Database\footnote{\url{http://oidb.jmmc.fr/index.html}} \citep{2014SPIE.9146E..0OH} and the CHARA Data Archive\footnote{\url{https://www.chara.gsu.edu/observers/database}}. The interferometric program is further described in Sect.~\ref{sec:campaign}.

\subsubsection{VLTI}

The Very Large Telescope Interferometer (VLTI) is a near-IR interferometric facility located at Cerro Paranal, Chile, which offers similar capabilities as CHARA when it comes to detection of faint binary companions, with a sparser $(u,v)$ coverage per snapshot 4-telescope observation, but with larger telescopes and generally much better observing conditions \citep{2018SPIE10701E..03W,2018SPIE10701E..14M}. The observations used here were taken with the four movable 1.8-m Auxiliary Telescopes (ATs), each equipped with an AO system, in the `large' configuration (maximum baseline of $\sim130$\,m), translating to $\lambda / 2 B_{\rm max} \sim1.75$\,mas in the near-IR $K$ band. 

The $K$-band 4-telescope beam combiner GRAVITY \citep{2017A&A...602A..94G} offers high spectral resolution ($R\sim4000$) spectro-interferometry covering the entire $K$ band as one of the three available spectral modes. Each single-field GRAVITY observation, in which light from the science target is also used for fringe tracking, results in the $K$-band spectrum, four CPs, six absolute visibilities $|$V$|$, and six differential phases (DPHI), the latter tracking the photocenter displacement in spectral lines with respect to the surrounding continuum \citep{2003RPPh...66..789M}. The high spectral resolution capabilities are particularly useful for acquiring spatial and kinematic information about regions associated with spectral (emission) lines, such as Br$\gamma$. The precision of the wavelength calibration for GRAVITY in the high spectral resolution mode is $0.1$\,nm, which translates to an RV uncertainty of $\sim15$\,{\kms} \citep{2017ApJ...845...57S,2023A&A...672A.119G}. 

GRAVITY data used in this study were all obtained in the high spectral resolution mode and were reduced using the official ESO pipeline for GRAVITY in the EsoReflex environment \citep{2013A&A...559A..96F}. The calibrator stars were selected in the same way as for CHARA. The data include two new snapshots for HR~2142, and a few archival\footnote{\url{http://archive.eso.org/eso/eso_archive_main.html}} observations for $\epsilon$~Cap, $\pi$~Aqr, $o$~Aqr, and $\beta$~Psc. For the latter, we make use only of the broad-band data, and reserve the analysis of the spectro-interferometry in the Br$\gamma$ emission line to an upcoming paper. For the purposes of companion detection, the GRAVITY data were rebinned to a spectral resolution of $R\sim50$, and the Br$\gamma$ region was excluded. 

Lastly, we also inspected archival data from the VLTI/PIONIER instrument \citep{2011A&A...535A..67L} for the sample stars $\beta$~CMi, $\epsilon$~Cap, $\pi$~Aqr, $o$~Aqr, and $\beta$~Psc, which were downloaded already reduced and calibrated from the Optical interferometry DataBase\footnote{\url{http://oidb.jmmc.fr}} (OiDB). However, the constraining power of these data was generally surpassed by the more recent CHARA and GRAVITY data.

\subsection{Spectroscopy}

Given the large amount of interferometric data presented in this work, the spectroscopic analysis is restricted to visual inspection of publically available spectra in the BeSS database\footnote{\url{http://basebe.obspm.fr/basebe/}} \citep{BeSS,2018sf2a.conf..459N}. In addition, we also made use of archival UV spectra from the International Ultraviolet Explorer (IUE, taken with the large aperture), available from the INES database\footnote{\url{http://ines.ts.astro.it/cgi-ines/IUEdbsMY}}. Further analysis of both amateur and professional spectra is reserved for an upcoming study.

{\catcode`\&=11
\gdef\2014AandA...567A..57K{\cite{2014A&A...567A..57K}}
}
{\catcode`\&=11
\gdef\harmanec2020{\cite{2020A&A...639A..32H}}
}
\begin{deluxetable*}{llccCchchhhccccC} 
\tablecaption{Be + sdOB binaries \label{tab:sdOB_binaries}}
\tablewidth{700pt}
\tabletypesize{\scriptsize}
\tablehead{
\colhead{HD} & \colhead{Name} & \colhead{Type} & \colhead{$H$} & \colhead{$d$} & \colhead{$P$} & \nocolhead{$\Delta H$} & \colhead{a''} & \nocolhead{a} & \nocolhead{M Be} & \nocolhead{M sdO} & \colhead{sdOB evidence} & \colhead{Ref.} & \colhead{$N_H$} & \colhead{$N_K$} & \colhead{$\Delta  H$} \\
\nocolhead{HD} & \nocolhead{Name} & \nocolhead{} & \colhead{[mag]} & \colhead{[pc]} & \colhead{[d]} & \nocolhead{[mag]} & \colhead{[mas]} & \nocolhead{[AU]} & \nocolhead{[Msun]} & \nocolhead{[Msun]} & \nocolhead{} & \nocolhead{} & \nocolhead{} & \nocolhead{} & \nocolhead{sdOB det.}
}
 \startdata
  12302  & V780~Cas  & B1:\,V:pe   & 7.09 & 971\pm20   & \nodata & & \nodata & & & & visible spec. & 0 & 5 & 5 & 4.8{\rm:}  \\ 
  29441  & V1150~Tau & B2.5\,Vne   & 7.55 & 622\pm17   & \nodata & & \nodata & & & & FUV & Wa21 & 2 & 2 & >3.7 \\ 
  41335  & HR 2142   & B1\,Ve       & 5.01 & 507\pm53   & 80.9    & 3.8 & 1.9 & 1.0 & 18.0\pm5.1 & 0.68\pm0.13 & FUV & Pe16 & 4 & 6 & 3.8 \\  
  51354  & QY~Gem    & B3\,ne      & 7.18 & 540\pm16   & \nodata & & \nodata & & & & FUV & Wa21 & 1 & 1 & >4.1 \\ 
  161306 & \nodata   & B3/5\,Vnne  & 6.58 & 451\pm5     & 99.3    & 3.9 & 1.8 & 0.8 & 6.316\pm0.250 & 0.806\pm0.086 & visible spec. & Ko14 & 3 & 5 & 3.9 \\ %
  183537 & 7 Vul     & B5\,Vn      & 6.51 & 277\pm4     & 69.5    & 4.6 & 2.0 & 0.6 & 4.35\pm0.20 & 0.485\pm0.018  & visible spec. & Ha20 & 5 & 6 & 4.6\\ %
  191610 & 28 Cyg    & B3\,IV\,e    & 5.16 & 255\pm7     & 359.1   & 5.2 & 7.4 & 1.9 & 5.20\pm1.16 & 1.98\pm1.17 & FUV cand. & Wa18 & 8 & 4 & 5.3 \\ %
  194335 & V2119 Cyg & B2\,IIIe    & 6.23 & 364\pm7     & 63.1    & 4.1 & 1.8 & 0.7 & 8.42\pm0.36 & 1.51\pm0.30 & FUV & Wa21 & 5 & 2 & 4.2 \\ %
  200120 & 59 Cyg    & B1\,Ve  & 4.46 & 871\pm300 & 28.2\tablenotemark{a}    &  & 0.4\tablenotemark{a} & 0.35-0.40 & 6.3-9.4 & 0.62-0.91 & FUV & Pe13 & 1 & 1 & >4.9\\ %
  200310 & 60 Cyg    & B1\,Ve      & 5.95 & 415\pm5   & 147.6   & 4.8 & 3.0 & 1.1 & 7.93\pm1.19 & 1.10\pm0.24 & FUV & Wa17 & 7 & 2 & 4.7  \\ %
  204722 & V2162~Cyg & B1.5\,IV:np & 7.64 & 763\pm13   & \nodata & & \nodata & & &  & visible spec. & 0 & 1 & 1 & >2.9 \\ %
  214168 & 8 Lac A   & B1\,IVe     & 5.85 & 521\pm26   & \nodata & & \nodata & & &  & FUV cand. & Wa18 & 1 & 1 & >3.5 \\ %
\enddata
\tablecomments{Type is the spectral type, $H$ is the $H$-band magnitude, $d$ is the distance, and $P$ and $a''$ are the orbital period and the angular semimajor axis, both taken from this work unless noted otherwise. The sdOB evidence column lists whether the sdOB companion was detected in FUV spectra (FUV), marginally detected in FUV spectra (FUV cand.), or whether there is only indirect evidence for a hot companion in visible spectra (visible spec.), with the corresponding reference in the Ref.\ column. $N_H$ and $N_K$ list the total number of interferometric snapshots taken in the two near-IR bands $H$ (MIRC-X) and $K$ (MYSTIC and GRAVITY). Finally, $\Delta H$ gives the the magnitude difference between the components in case of detections (colon indicates a marginal detection), or the lower limit on $\Delta H$ ($\rho<25$\,mas) in case of non-detections.}
\tablenotetext{a}{Taken from \citet{2013ApJ...765....2P}; $a''$ calculated assuming orbital inclination of 80$^{\circ}$.}
\tablerefs{0 - this work; Wa21 - \citet{2021AJ....161..248W}; Pe16 - \citet{2016ApJ...828...47P}; Ko14 - \2014AandA...567A..57K; Ha20 - \harmanec2020; Wa18 - \citet{2018ApJ...853..156W}; Pe13 - \citet{2013ApJ...765....2P}; Wa17 - \citet{2017ApJ...843...60W}}
\end{deluxetable*}

\section{Interferometric program on the close binarity of Be stars}
\label{sec:campaign}

{\catcode`\&=11
\gdef\Ne10{\cite{2010A&A...516A..80N}}}
{\catcode`\&=11
\gdef\Ru09{\cite{2009A&A...506.1319R}}}
{\catcode`\&=11
\gdef\Ko97{\cite{1997A&A...328..551K}}}
{\catcode`\&=11
\gdef\Du88{\cite{1988A&AS...75..311D}}}
{\catcode`\&=11
\gdef\Hr22{\cite{2022A&A...666A.136H}}}
{\catcode`\&=11
\gdef\Ri06{\cite{2006A&A...459..137R}}}
\begin{deluxetable*}{llccClhchhhhcccC}
\tablecaption{SB1 binaries\label{tab:SB1_binaries}}
\tablewidth{700pt}
\tabletypesize{\scriptsize}
\tablehead{
\colhead{HD} & \colhead{Name} & \colhead{Type} & \colhead{$H$} & \colhead{$d$} & \colhead{$P$} & \nocolhead{$\Delta H$} & \colhead{$a''$} & \nocolhead{$a$} & \nocolhead{$M_{\rm Be}$} & \nocolhead{$M_{\rm comp}$} & \nocolhead{SB1 sol.} & \colhead{Ref.} & \colhead{$N_H$} & \colhead{$N_K$} & \colhead{$\Delta H$}\\
\nocolhead{HD} & \nocolhead{Name} & \nocolhead{Type} & \colhead{[mag]} & \colhead{[pc]} & \colhead{[d]} & \nocolhead{[mag]} & \colhead{[mas]} & \nocolhead{[AU]} & \nocolhead{\Msun} & \nocolhead{\Msun} & \nocolhead{} & \nocolhead{} & \nocolhead{} & \nocolhead{} & \nocolhead{Comp.\ det.}
}
\startdata
      23862 & 28 Tau & B8\,Vpe & 5.07 & 138\pm3 & 218.0 &  & 7.6 & 1.04-1.05 & 2.9 & 0.254-0.338 & H$\alpha$ & Ne10 & 4 & 3 & >5.7 \\ 
      37202 & $\zeta$~Tau & B1\,IVe-shell & 3.07 & 136\pm16 & 133.0 &  & 8.5 & 1.16-1.17 & 11 & 0.87-1.02 & & Ru09 & 1 & 0 & >5.0 \\ 
      58715 & $\beta$~CMi & B8\,Ve & 3.07  & 49.6\pm0.5 & 170.4 &  & 19.2 & 0.95 & 3.5 & 0.21 & & Du17 & 2 & 1 & >6.0 \\ 
      109387 & $\kappa$ Dra\tablenotemark{a} & B6\,IIIe & 3.96 & 142\pm6  & 61.5 & 4.56 & 3.4 & 0.487 & 3.65\pm0.48 & 0.426\pm0.043 & & Kl22 & 4 & 1 & 4.6 \\
      142926 & 4 Her & B7\,IVe-shell & 5.89 & 165.1\pm1.3 & 46.2 &  & 2.5 & 0.378-0.382 & 3.2 & 0.18-0.29 & & Ko97 & 2 & 0 & >5.1 \\ 
      162732 & 88 Her & B6\,IIInp-shell & 6.91 & 315\pm3 & 86.7 &  & 1.9 & 0.61-0.62 & 3.65 & 0.47-0.50 & & Du88 & 1 & 0 & >4.6 \\ 
      184279 & V1294~Aql & B0.5\,IVe & 6.68 & 1432\pm169 & 192.9 & & 1.2 & 1.71-1.72 & 16.9 & 1.17-1.36 & & Ha22 & 1 & 1 & >4.0 \\ 
      205637 & $\epsilon$ ~Cap & B3\,IIIe & 4.91 & 271\pm14  & 128.5 &  & 3.4 & 0.91 & 5.4 & 0.69-0.74 & & Ri06 & 1 & 4 & >4.3 \\  
\enddata
\tablecomments{Type is the spectral type, $H$ is the $H$ band magnitude, $d$ is the distance, and $P$ is the orbital period taken from the given reference (Ref.). The semimajor axis $a''$ was estimated using the published orbital solution and the distance, with the mass of the primary Be star assumed according to its spectral type, except for $\kappa$~Dra. $N_H$ and $N_K$ list the total number of interferometric snapshots taken in the two near-IR bands $H$ (MIRC-X) and $K$ (MYSTIC). Finally, $\Delta H$ gives the the magnitude difference between the components in case of detections (colon indicates a marginal detection), or the lower limit on $\Delta H$ ($\rho<25$\,mas) in case of non-detections.}
\tablenotetext{a}{Interferometric detection of sdB companion along with orbital solution is published \citep{2022ApJ...940...86K}. }
\tablerefs{Ne10 - \Ne10; Ru09 - \Ru09; Du17 - \citet{2017ApJ...836..112D}; Kl22 - \citet{2022ApJ...940...86K}; Ko97 - \Ko97; Du88 - \Du88, Ha22 - \Hr22, Ri06 - \Ri06}
\end{deluxetable*}

\subsection{Target sample}

CHARA with its 6-telescope beam combiners MIRC-X ($H$ band) and MYSTIC ($K$ band) recently concluded a 2-year-long observing campaign (2020~Dec -- 2022~Oct, PI: Klement) aimed at the detection of faint and close binary companions to more than $40$ classical Be stars. Since MYSTIC went online only in 2021 July, targets observed early in the campaign only have MIRC-X $H$-band data available. The observations and the associated results (including the auxiliary GRAVITY data) are presented in Tables~\ref{tab:detections} and \ref{tab:detlims}.

The targets were selected from among confirmed close binaries as well as close binary candidates among nearby and bright classical Be stars with a declination above $-20^{\circ}$. All targets had to be bright enough to be observable by CHARA/MIRC-X and MYSTIC, for which the best-performance limit is $H \approx 7.5$ and $K \approx7.5$, respectively. Two targets slightly fainter than 7.5\,mag in $H$ and $K$ were successfully observed as well (V1150~Tau and V2162~Cyg), although the quality of the data turned out to be lower than for most of the brighter targets. The targets were also selected to be nearby enough so that the expected binary orbit has a semimajor axis of at least 0.5\,mas ($d < 1$\,kpc for a 0.5\,AU orbit), with the one exception being the binary candidate V1294~Aql at a distance of $\sim1.4$\,kpc. We used the spectrophotometric distances based on Gaia EDR3 \citep{2016A&A...595A...1G, 2021A&A...649A...1G} according to \citet{2021AJ....161..147B}\footnote{\url{https://vizier.cds.unistra.fr/viz-bin/VizieR?-source=I/352&-to=3}} when available, or from the revised reduction of Hipparcos data\footnote{\url{https://vizier.cds.unistra.fr/viz-bin/VizieR?-source=I/311&-to=3}} \citep{2007A&A...474..653V}. Gaia EDR3 parallaxes are not available for four of the brightest targets in the sample: $\zeta$~Tau, $\beta$~CMi, $\gamma$~Cas, and $\eta$~Tau ($G<4$). We also note that caution should be exercised when dealing directly with Gaia parallaxes for bright targets \citep[$G\lesssim12$;][]{2021A&A...649A...4L}, which in our case constitutes the entire sample, but the generally good quality of the astrometric solutions (given by the RUWE parameter, see below), and the adoption of the distances calculated with spectrophotometric priors according to \citet{2021AJ....161..147B} should limit the possible bright-star bias. For convenience, the targets were divided into several groups which are introduced below, and all of the targets are discussed individually in later sections. The adopted distances for all targets are listed in Tables \ref{tab:sdOB_binaries} to \ref{tab:candidate_binaries}.

The first target group henceforth referred to as \textit{Be + sdOB binaries} totals 12 targets and contains both confirmed and candidate Be + sdOB systems, where the evidence for the sdOB nature of the companion comes from directly detected spectral features in the FUV or from signatures of the radiative influence of a hot companion on the Be star disk in the visible. The observed targets are listed in Table~\ref{tab:sdOB_binaries} along with some basic characteristics relevant for this work, as well as the number of interferometric snapshots taken for this study, and (upper limits on) the $H$-band magnitude differences between the Be stars and their (unseen) companions. The second category dubbed \textit{SB1 binaries} includes eight targets and consists of confirmed or strong candidate SB1 binaries with close companions of unknown nature, which could be faint sdOBs, WDs or the elusive late-type MS companions (Table~\ref{tab:SB1_binaries}). The third group - \textit{$\gamma$~Cas-like binaries} - is composed of six Be stars with $\gamma$~Cas-like X-rays, i.e., candidate Be binaries with accreting WD companions (Table~\ref{tab:gamCas_binaries}). Five of the \textit{$\gamma$~Cas-like binaries} in our sample are confirmed SB1 binaries, while FR~CMa is a candidate SB1 binary \citep{2022MNRAS.510.2286N}.

{\catcode`\&=11
\gdef\Ne12{\cite{2012A&A...537A..59N}}}
\begin{deluxetable*}{llccClhchhhhcccC}
\tablecaption{$\gamma$~Cas-like binaries\label{tab:gamCas_binaries}}
\tablewidth{700pt}
\tabletypesize{\scriptsize}
\tablehead{
\colhead{HD} & \colhead{Name} & \colhead{Type} & \colhead{$H$} & \colhead{$d$} & \colhead{$P$} & \nocolhead{$\Delta H$} & \colhead{$a''$} & \nocolhead{$a$} & \nocolhead{$M_{\rm Be}$} & \nocolhead{$M_{\rm comp}$} & \nocolhead{SB1 sol.} & \colhead{Ref.} & \colhead{$N_H$} & \colhead{$N_K$} & \colhead{$\Delta H$}\\
\nocolhead{HD} & \nocolhead{Name} & \nocolhead{Type} & \colhead{[mag]} & \colhead{[pc]} & \colhead{[d]} & \nocolhead{[mag]} & \colhead{[mas]} & \nocolhead{[AU]} & \nocolhead{\Msun} & \nocolhead{\Msun} & \nocolhead{} & \nocolhead{} & \nocolhead{} & \nocolhead{} & \nocolhead{Comp.\ det.}
}
\startdata
5394   & $\gamma$~Cas & B0.5\,IVe       & 2.22 & 168\pm4 & 203.5 & & 9.7 & \sim{1.63} & 13 & \sim0.98 & & Ne12 & 2 & 1 & >6.7 \\ 
44458  & FR~CMa       & B1.5\,IVe       & 5.64 & 510\pm12  & \nodata & & \nodata & & & & & Na22 & 1 & 1 & >3.7 \\ 
45995  & HR~2370      & B2\,IVe       & 5.97 & 655\pm22  & 103.1 & & 1.5 & & & & & Na22 & 1 & 1 & >4.5 \\
183362 & V558~Lyr     & B3\,Ve          & 6.18 & 551\pm12  & 83.3 & & 1.4 & & & & & Na22 & 1 & 1 & >4.4 \\
12882  & V782~Cas   & B2.5\,III & 6.45 & 827\pm23  & 122 & & 1.2 & & & & & Na22 & 3 & 3 & >4.1\tablenotemark{a} \\
212571 & $\pi$ Aqr    & B1\,III-IVe     & 5.36 & 333\pm11  & 84.1 & 4.29 & 2.9 & $\sim0.96$ & 14.0 & $\sim2.31$ & & Bj02 & 3 & 2 & >6.1  \\ 
\enddata
\tablecomments{Type is the spectral type, $H$ is the $H$ band magnitude, $d$ is the distance, and $P$ is the orbital period taken from orbital solution in the corresponding reference (Ref.). The semimajor axis $a''$ was estimated from the orbital solution and the distance, with the mass of the primary Be star assumed according to its spectral type. $N_H$ and $N_K$ list the total number of interferometric snapshots taken in $H$ (MIRC-X) and $K$ (MYSTIC), and finally, $\Delta H$ gives the lower limit of the magnitude difference between the components ($\rho<25$\,mas).}
\tablenotetext{a}{Lower limit for the inner binary V782~Cas~A.}
\tablerefs{Ne12 - \Ne12; Na22 - \citet{2022MNRAS.510.2286N}; Bj02 - \citet{2002ApJ...573..812B}}
\end{deluxetable*}


{\catcode`\&=11
\gdef\Co21{\cite{2021A&A...647A.164C}}}
\begin{deluxetable*}{llclCccccC}
\tablecaption{Candidate binaries\label{tab:candidate_binaries}}
\tablewidth{700pt}
\tabletypesize{\scriptsize}
\tablehead{
\colhead{HD} & \colhead{Name} & \colhead{Type} & \colhead{$H$} & \colhead{$d$} & \colhead{Evidence} & \colhead{Ref.} & \colhead{$N_H$} & \colhead{$N_K$} & \colhead{$\Delta H$}\\
\colhead{} & \colhead{} & \colhead{} & \colhead{[mag]} & \colhead{[pc]} & \nocolhead{(d)} &  \nocolhead{}  & \nocolhead{$N_H$} & \nocolhead{$N_K$} & \nocolhead{Comp.\ det.}
} 
\startdata
22192 & $\psi$ Per & B5\,IIIe-shell & 4.12 & 167\pm3 & SED & Kl19 & 2 & 0 & >5.6 \\ 
23630 & $\eta$ Tau & B7\,IIIe & 2.84 & 124\pm7 & SED & Kl19 & 1 & 0 & >6.0 \\ 
25940 & 48 Per     & B4\,Ve & 3.91 & 156\pm8 & SED & Kl19 & 1 & 0 & >5.8 \\ 
32343 & 11 Cam     & B3\,Ve & 5.02 & 207\pm6 & SED & Kl19 & 2 & 2 & 4.2{\rm:} \\ 
32991 & 105 Tau    & B3\,Ve & 4.96 & 301\pm4 & SED & Kl19 & 2 & 1 & >5.2 \\ 
35439 & 25 Ori     & B1\,V & 5.42 & 339\pm13 & H$\alpha$ & 0 & 1 & 1 & >4.4 \\ 
138749 & $\theta$~CrB & B6\,III & 4.45 & 122\pm3 & \nodata & \nodata & 2 & 2 & >5.7 \\ 
166014 & $o$~Her  & B9.5\,III & 3.96 & 106.7\pm1.7 & \nodata & \nodata & 1 & 0 & >5.0 \\ 
187811 & 12~Vul    & B3\,Ve & 5.05 & 192\pm5 & $^{12}$CO & Co21 & 3 & 0 & >5.8 \\ 
209409 & $o$ Aqr   & B7\,IIIe-shell & 4.70 & 143\pm3 & SED & Kl19 & 1 & 4 & >4.3 \\ 
217050 & EW Lac    & B3:\,IV:e-shell & 5.26 & 286\pm5 & SED & Kl19 & 1 & 0 & >5.5 \\ 
217891 & $\beta$ Psc & B4\,V & 4.54 & 124\pm2 & SED & Kl19 & 1 & 3 & >4.5 \\ 
\enddata
\tablecomments{Type is the spectral type, $H$ is the $H$ band magnitude, $d$ is the distance, Evidence lists the type of evidence for possible binarity from the corresponding reference (Ref.), $N_H$ and $N_K$ list the total number of interferometric snapshots taken in $H$ (MIRC-X) and $K$ (MYSTIC). Finally, $\Delta H$ gives the the magnitude difference between the components in case of detections (colon indicates a marginal detection), or the lower limit on $\Delta H$ ($\rho<25$\,mas) in case of non-detections.}
\tablerefs{0 - this work; Kl19 - \citet{2019ApJ...885..147K}; Co21 - \Co21}
\end{deluxetable*}

The next subsample (\textit{Candidate binaries}) (12 stars) mostly contains binary candidates categorized as such based on the apparent tidal influence of the otherwise unseen companion on the circumstellar disk of the Be star (Table~\ref{tab:candidate_binaries}). One possible effect is the occurrence of a turndown in the spectral slope at long wavelengths (eight stars in the sample), which could be caused by disk truncation \citep{2017A&A...601A..74K, 2019ApJ...885..147K}. In addition, the presence of an orbiting companion can be deduced from the morphology and variability of strong emission lines such as H$\alpha$, which sometimes show complex effects from the tidal influence of the companion on the line-emitting regions of the Be star disk (25~Ori). The binary candidate status of 12~Vul is based on the detection of $^{12}$CO bandhead molecular emission, suggested to originate from a cool evolved companion \citep{2021A&A...647A.164C}. Finally, we observed two stars for which there is little evidence suggesting close binarity ($\theta$~CrB and $o$~Her), but for simplicity, these were still included in the \textit{Candidate binaries} subsample.

Next, we have a group of four objects, which were included in the sample because they are known as hierarchical triple or higher-order multiple systems containing a Be star, whose detailed orbital analysis is reserved for a dedicated study (in prep.). Lastly, three of the observed binary candidates (V742~Cas, V447~Sct, and V1362~Cyg) were discovered to be in the Be + (bloated) pre-subdwarf phase. The analysis of these three systems is published in a separate study (Rivinius et al., submitted).

Thus, as a part of the CHARA campaign, data were collected for a total of 45 classical Be stars. Excluding the four hierarchical triples and the three Be stars with bloated pre-subdwarf companions, both studied in separate papers, we obtained a total of seven interferometric detections (plus two marginal detections) of close, faint companions out of 38 Be stars, specifically six from among the \textit{Be + sdOB binaries}, and one from among the \textit{SB1 binaries}. Preliminary results following the initial companion detections were already published for 28~Cyg, V2119~Cyg, and 60~Cyg \citep{2022ApJ...926..213K}, and here we present the updated orbital solutions based on the complete set of interferometric measurements. The results for the first confirmed Be + sdB system $\kappa$~Dra that was observed as part of this campaign were published in a separate study \citep{2022ApJ...940...86K} and this object is therefore not analyzed further in this work. First orbital solutions for the three newly constrained systems are presented in Sect.~\ref{sec:orbits_new}. For the rest of the sample Be stars, the interferometric measurements enabled only a derivation of the limiting magnitude differences between the Be stars and the (suspected) companions. In several cases, the circumstellar disks were (partially) resolved and we were able to contrain their sizes and shapes.

\subsection{Data analysis}

For the detection of the companions and the derivation of limiting magnitude differences in cases of non-detection, we used a grid search algorithm implemented in the dedicated Python code CANDID \citep{2015A&A...579A..68G}. CANDID corrects for the effects of bandwidth smearing for large angular separations between the components, but with only one exception (wide companion for 48~Per), all new companions were detected well within the interferometric FoV, which is determined by the observed wavelength $\lambda$, the spectral resolution $R$, and the maximum baseline $B_{\rm max}$ from
\begin{equation}
    {\rm FoV} = R \frac{\lambda}{B_{\rm max}}.
\end{equation}
For the successfully detected companions, their relative positions - given as the angular separation $\rho$ and position angle (PA) from North to East - and flux ratios - defined as the ratio of the companion flux to the Be star flux including the contribution from its disk $f = f_{\rm comp} / f_{\rm Be}$ - were refined with a bootstrapping algorithm with 500 iterations (Table~\ref{tab:detections}). 

In the case of non-detections, the lower limits on the magnitude differences were computed using the robust fake-companion injection procedure included in CANDID \citep{2015A&A...579A..68G}. The resulting values of $\Delta {\rm mag}$ are given for a $3\sigma$ detection and 90\% confidence interval, which means that 90\% of the fake injected companions brighter than $\Delta {\rm mag}$ were detected by the grid search algorithm at a 3$\sigma$ confidence level. The limits were derived for angular separations corresponding to the interferometric FoV. Specifically, for data taken with $R=50$, we searched up to a separation of 50 and 60\,mas in the $H$ and $K$ band, respectively, for $R=100$ up to 100\,mas, and for $R=190$ up to 200\,mas ($H$ band data only, Table~\ref{tab:detlims}).

The grid search algorithm as well as the computation of the detection limits were applied only to the CP part of the datasets in order to limit the possible biases introduced by (1) imperfect calibration of VIS2, which is more susceptible to atmospheric effects than CP, and (2) the ellipsoidal shape of the circumstellar disk reflected in VIS2 for partially resolved targets, as this can bias the companion detection, and also because CANDID assumes a simple uniform disk (UD) geometry for the primary star.

Importantly, CP itself can also be biased by a resolved circumstellar disk if it deviates from point-source symmetry. This is indeed the case for some very nearby Be stars with large disks, for which an interferometric signal from a density wave in the disk masquerades as a signal from a faint binary companion. In this case, we have to distinguish between two main types of global oscillations in Be star disks. The first type of long-term one-armed spiral-shaped over-densities propagating through the disk is thought to arise from gravitational effects from the rotationally distorted central stars \citep{1991PASJ...43...75O,1992A&A...265L..45P,1997A&A...318..548O}. The (qua\-si-)\-periods of these oscillations are of the order of several years to decades, and are observable in the strong variations of the violet-to-red peak ratios ($V/R$) of emission lines \citep[e.g.,][]{1991A&A...241..159M}. The second type of disk oscillation is a two-armed structure that is excited by an orbiting companion, and the associated $V/R$ variability is phase-locked with the binary orbit \citep{2016MNRAS.461.2616P,2018MNRAS.473.3039P}. However, in this case the amplitude of the variability is expected to be much smaller than in the case of the one-armed oscillations, which is confirmed by the several known cases \citep{2023Galax..11...83M}, and in the broad-band interferometric data, we would generally not expect to detect these. Thus, the disk asymmetry that can masquerade as a binary signal would likely come from the long-term one-armed density waves. 

For the general case of a deformed disk introducing a signal in the CP, the position of the asymmetry would also come from within the circumstellar disk that should be (partially) resolved in VIS2 along the corresponding PA. The resulting angular separation for a disk asymmetry would therefore be too small, and the orbital movement derived from repeated observations too slow for a physical orbital solution to be attained (as opposed to the case of an actual orbiting companion). There is also the theoretical possibility that a non-zero CP may originate from the irradiated outer parts of the Be disk (if it has a hot companion), but the situation would probably be very similar to the above-mentioned case of the two-armed phase-locked density waves.

For the analysis of the spectro-interferometric GRAVITY data ($R=4000$) for HR~2142, and for constraining the shapes of resolved circumstellar disks in CHARA data and spectrally rebinned GRAVITY data, we used the PMOIRED code\footnote{\url{https://github.com/amerand/PMOIRED}}, which enables the fitting of a suite of geometrical models \citep{2022SPIE12183E..1NM}. Of particular use was the model of a Keplerian disk\footnote{\url{https://github.com/amerand/PMOIRED/blob/master/examples/Be model comparison with AMHRA.ipynb}} (a brief description is given in Sect.~\ref{sec:pmoired_keplerian}), which was included in the interpretation of the geometric distribution of line-emitting components (Sect.~\ref{sec:orbits_new}). 

To represent the Be-star disk when fitting the broad-band VIS2 data, we used an 2D Gaussian, in which the disk inclination $i_{\rm disk}$ is determined from $r = \cos{i_{\rm disk}}$, where $r$ is the ratio between the minor and major axes. As this assumes a geometrically thin disk, the resulting $i_{\rm disk}$ values might be biased by the presence of emitting gas outside of the plane of the disk, e.g., from disk flaring, which generally occurs in Be stars \citep{2013A&ARv..21...69R}. This would be the case specifically for disks seen close to edge-on, where $i_{\rm disk}$ could be systematically biased to lower values \citep{2015A&A...577A..51M}. The disk PA corresponds to the position angle of the disk major axis measured from North to East. Finally, we stress that the error margins given for the parameters resulting from interferometric fitting do not include systematic errors that could affect the data, such as miscalibration due to varying atmospheric conditions or an inaccurate estimate of angular diameters of the calibrators.

Targets that were identified as possible high-contrast binaries based on initial observations were observed repeatedly until the astrometric orbit could be constrained, or the presence of the companion ruled out (excluding the two marginal cases, which need further observations). Previously published RV measurements were recovered from the literature and included in the combined orbital solutions. Typically, RVs of only one of the two components were available, which meant that we had to rely on the measured distances in order to obtain the physical sizes of the relative orbits from the measured angular sizes. This introduces an additional source of error, as the parallax measurements might be biased by the photocenter displacement due to the binary orbital motions. Nevertheless, due to the high contrast and low mass ratios (here defined as $q=M_{\rm comp.} / M_{\rm Be}$), the photocenter displacement of the Be star is small and should not have a significant influence on the parallax measurement (unless another star contaminates the field of view). In the Gaia DR3 catalog, the RUWE parameter\footnote{\url{https://gea.esac.esa.int/archive/documentation/GDR3/Gaia_archive/chap_datamodel/sec_dm_main_source_catalogue/ssec_dm_gaia_source.html\#gaia_source-ruwe}}  reflects the quality of the astrometric solutions, and it should be sensitive to binarity-induced motions that can deteriorate the solution \citep[if it is at elevated values, typically $>1.4$;][]{2021A&A...649A...2L,2021A&A...649A..13M}. 

For the simultaneous fitting of the astrometric positions and the RVs for the detected binaries, we used the IDL package orbfit-lib\footnote{\url{https://www.chara.gsu.edu/analysis-software/orbfit-lib}} \citep{2016AJ....152..213S} to obtain the full 3D orbital solutions including the inclinations and the physical sizes of the orbit, and finally the dynamical masses for the binary components. The uncertainties of the RVs and the positions were scaled so that the RV and interferometric datasets contribute approximately equally to the total residuals. All orbital parameters obtained from the fit to the observational data, as well as additional parameters computed using the distance measurements, are listed in Table~\ref{tab:orbital_params}.

\section{Updated orbits for targets with previously detected close companions}
\label{sec:orbits_updated}

\begin{deluxetable*}{lCCCCCC}
\tablecaption{New orbital parameters from the combined SB1 and astrometric solutions} 
\tablewidth{0pt}
\tablehead{
\nocolhead{} & \colhead{28~Cyg} & \colhead{V2119~Cyg} & \colhead{60~Cyg} & \colhead{HR~2142} & \colhead{HD~161306} & \colhead{7~Vul}
}
\label{tab:orbital_params}
\startdata
$P$ [d]                        & 359.260\pm0.041        & 63.1475\pm0.0029  & 147.617\pm0.038  & 80.8733\pm0.0044 & 99.315\pm0.055   & 69.5258\pm0.0051 \\
$T$ [RJD\tablenotemark{a}]     & 59524.8\pm0.3          & 59526.0\pm0.2     & 59598.5\pm0.9    & 59541.3\pm0.2    & 59759.0\pm1.0    & 59847.6\pm0.2  \\
$e$                            & 0                      & 0                 & 0.20\pm0.01    & 0                & 0                & 0                \\
$a''$ [mas]                    & 7.426 \pm 0.032        & 1.809\pm0.012     & 3.032\pm0.014    & 1.914\pm0.010    & 1.764\pm0.012    & 2.005\pm0.019    \\
$i$ [{\degree}]                & 118.7\pm0.2            & 49.5\pm0.5        & 84.6\pm0.3       & 77.7\pm0.7       & 34.0\pm1.3       & 98.7\pm1.2   \\
$\Omega$ [{\degree}]           & 146.0\pm0.3            & 116.2\pm0.9       & 196.8\pm0.3      & 210.4\pm0.5      & 48.1\pm3.6       & 151.0\pm0.6    \\
$\omega_{\rm Be}$ [{\degree}]           & 90                     & 90                & 332.4\pm2.5      & 90               & 90               & 90               \\
$K_\mathrm{Be}$ [km/s]         & 5.4\pm1.7\tablenotemark{b} & 12.2\pm2.0\tablenotemark{b} & 11.2\pm1.7 & 7.0\pm0.4 & 5.6\pm0.5 & 8.7\pm0.2  \\ 
$K_\mathrm{sdO}$ [km/s]        & 44.9\pm1.0 & 74.0\pm0.6 & 83.0\pm2.1\tablenotemark{b} & 120.5\pm13.4\tablenotemark{b} & 43.1\pm2.6\tablenotemark{b} & 77.2\pm1.5\tablenotemark{b} \\
$\gamma$ [km/s]                & -27.1\tablenotemark{c} & -18.3\pm0.5   & -13.1\pm0.9    & 45.7\pm0.3     & -10.5\pm0.4    & -16.5\pm0.2    \\
\hline
$a$ [AU]                       & 1.89\pm0.06            & 0.658\pm0.014   & 1.258\pm0.016    & 0.97\pm0.10    & 0.796\pm0.011  & 0.5553\pm0.0096  \\
$q$                            & 0.12\tablenotemark{c}  & 0.165\pm0.027   & 0.135\pm0.024    & 0.0584\pm0.0072  & 0.130\pm0.014    & 0.1123\pm0.0036  \\
$M_\mathrm{total}$ [\Msun]     & 7.02\pm0.59            & 9.55\pm0.58     & 12.19\pm0.47     & 18.6\pm5.9       & 6.81\pm0.27      & 4.73\pm0.24 \\
$M_\mathrm{Be}$ [\Msun]        & 6.26\pm0.38            & 8.20\pm0.35     & 10.75\pm0.49     & 17.6\pm5.7       & 6.02\pm0.26      & 4.25\pm0.23      \\
$M_\mathrm{sdO}$ [\Msun]       & 0.76\pm0.28            & 1.35\pm0.27     & 1.45\pm0.23      & 1.03\pm0.22      & 0.784\pm0.074    & 0.477\pm0.020    \\
\enddata
\tablecomments{$P$ is the orbital period, $T$ is the epoch of periastron or the epoch of the quadrature when the primary Be is moving towards us, $e$ is the eccentricity, $a''$ is the angular semimajor axis of the orbit, $i$ is the orbital inclination, $\Omega$ is the longitude of the ascending node, $\omega$ is the longitude of the periastron (fixed at 90{\degree} for circular orbits), $K$ are the velocity semiamplitudes for the two components, $\gamma$ is the systemic velocity, $a$ is the orbital semimajor axis, $q=M_{\rm sdOB} / M_{\rm Be}$ is the mass ratio, $M_\mathrm{total}$ is the total mass, and $M_{\rm Be}$ and $M_{\rm sdOB}$ are the dynamical masses for the two components.}
\tablenotetext{a}{RJD defined here as {\rm RJD} = {\rm HJD} - 2,400,000}
\tablenotetext{b}{Calculated from the other parameters and the distance.}
\tablenotetext{c}{Fixed in the current solution due to poor phase coverage of the RV measurements.}
\end{deluxetable*}

\textbf{28 Cyg} (HR\,7708, HD\,191610, HIP\,99303) is an early-type Be star classified as B3\,IVe \citep{1982ApJS...50...55S} whose hot, stripped sdO companion was initially detected in 25 out of 46 IUE archival spectra by cross-correlation with a 45\,kK synthetic spectrum, resulting in $f_{\rm FUV} = 4.2\pm2.6$\% \citep{2018ApJ...853..156W}. However, using the same technique, the companion remained undetected in three recent high-quality HST/STIS spectra \citep[$f_{\rm FUV}<2.4$\%,][]{2021AJ....161..248W}. The difference was suggested to be due to a variable obscuration by nearby circumstellar gas \citep{2021AJ....161..248W}. The companion was later detected on three occasions with CHARA/MIRC-X interferometry ($f_{\rm H}=0.80\pm0.07$\%), leading to the confirmation of its sdO nature, and to a possible estimate of the combined astrometric and SB1 orbital solution \citep{2022ApJ...926..213K}. We also refer to \citet{2022ApJ...926..213K} for a description of the observational history regarding the disk variability as seen in Balmer emission lines. 

\begin{figure*}
\epsscale{1.1}
\plottwo{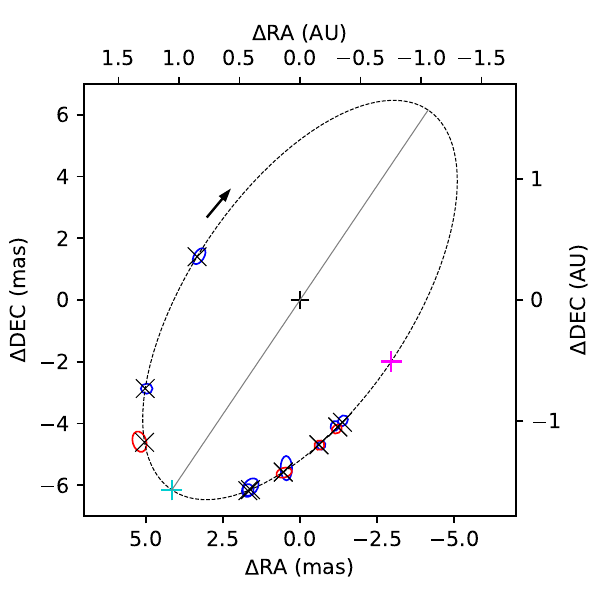}{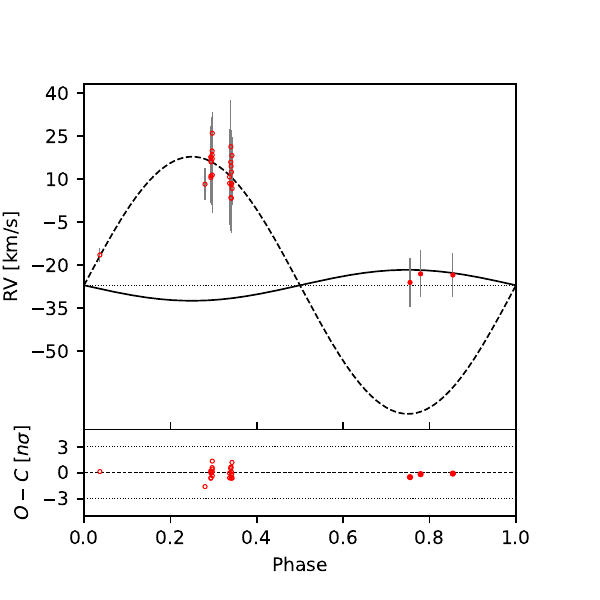}
\caption{Left: Relative astrometric orbit for the sdOB companion of 28~Cyg. The black plus sign corresponds to the location of the primary Be star, while the best-fit companion orbit is shown as a black dashed line. The error ellipses correspond to the interferometric measurements with 5-$\sigma$ uncertainties (MIRC-X in blue and MYSTIC in red). The associated calculated positions are plotted as black crosses. The gray solid line indicates the line of nodes, the cyan plus sign shows the ascending node, the magenta plus sign shows the inferior conjunction (periastron in case of an eccentric orbit), and the arrow shows the direction of the orbital motion. Right: RV curves of the primary Be star (solid line) and the secondary sdOB (dashed line) with measurements for the primary (filled red circles) and/or the secondary sdO component (empty red circles), when available. The lower panel shows the $O-C$ residuals in units of $\sigma$. \label{fig:orb_28Cyg}}
\end{figure*}

Interferometric observations obtained following the published preliminary results have enabled a combined orbital solution (Table~\ref{tab:orbital_params}, Fig.~\ref{fig:orb_28Cyg}), now based on a total of 12 interferometric measurements of the companion positions (Table~\ref{tab:detections}). The well-constrained orbital period $P$ is longer compared to the preliminary solution, and in fact 28~Cyg has thus far the longest $P$ ($\sim$1\,yr) and the widest orbit ($a \sim 2$\,AU) among close Be star binaries with constrained orbits. The RV curve of the companion is based on the 25 published measurements from cross-correlation of IUE spectra \citep{2018ApJ...853..156W}. While there are also three RV measurements available for the primary Be star based on the HST spectra \citep{{2021AJ....161..248W}}, these are not precise or numerous enough to enable a double-lined spectroscopic solution. 

Unfortunately, the low precision and very poor phase coverage of the RV measurements for the sdO make the resulting systemic velocity $\gamma$ and the sdO velocity semiamplitude $K_{\rm sdO}$ very uncertain. While the total mass of the system ($7.02$\,{\Msun}) should be reliable, as it is based only on $P$ and the distance, the mass ratio $q$ and therefore the individual masses remain poorly constrained due to the above mentioned uncertainty in $K_{\rm sdO}$. We tried fixing $\gamma$ at literature values available from Simbad\footnote{\url{https://simbad.u-strasbg.fr/simbad/}}, but this resulted in an unrealistically high $q$. Therefore, we proceeded to fix $q$ at the typical value for Be + sdO binaries of $0.12$ \citep{2021AJ....161..248W,2023ApJ...942L...6G}, which meant fixing $\gamma$ at $-27.1$\,{\kms}. This resulted in $K_{\rm sdO} = 44.9\pm1.0$\,{\kms}, and using the distance we obtained $K_{\rm Be} = 5.4\pm1.7$\,{\kms}. The latter value is in good agreement with the three Be star RVs mentioned above (right panel of Fig.~\ref{fig:orb_28Cyg}), providing further credibility to the orbital solution and the assumed $q$ of 0.12.

With the help of the Gaia distance measurement, the resulting masses are $M_{\rm Be} = 6.26\pm0.38$\,{\Msun} and $M_{\rm sdO} = 0.76\pm0.28$\,{\Msun}. $M_{\rm Be}$ appears to be in a reasonable agreement with predictions according to its spectral type, as for instance the updated online table from \citet[][version 2022.04.16]{2013ApJS..208....9P}, which contains mean parameters of MS dwarf stars based on literature and catalog survey\footnote{\url{https://www.pas.rochester.edu/~emamajek/EEM_dwarf_UBVIJHK_colors_Teff.txt}} assigns a mass of 6.1\,{\Msun} to a B2.5\,V star. To obtain dynamical masses independent of the assumption on $q$, new RVs, ideally for both components, are needed.

The variable detectability of the companion in the FUV does not show any dependence on the orbital phase according to our orbital solution, and the suggested obscuration of the companion by nearby circumstellar material is thus independent of the orbital motion. This was actually apparent already from the initial set of FUV detections, in which many spectra that resulted in non-detections were taken around the same time as spectra showing detections \citep{2018ApJ...853..156W}. If real, the variability of the strength of the spectral features of the sdO would have to occur on very fast time scales of the order of tens of minutes, given the cadence of the IUE spectra. On the other hand, the $f_{\rm FUV}$ of $4.2\pm2.6$\% from the IUE spectra is actually compatible with the upper limit derived from the HST spectra of $f_{\rm FUV} < 2.4$\%, which means that any obscuration of the sdO FUV spectrum might not in fact be needed to explain the IUE results. Instead, this could be due to the sdO being at the limit of detectability in the FUV, and the FUV spectra having variable signal-to-noise ratio (SNR). Another argument against the obscuration scenario is the wide orbit, which makes it less likely that there is a significant amount of gas from the Be disk in the vicinity of the companion (cf., the case of HR~2142 below). 

In the near-IR multi-epoch interferometry, there is some evidence for variability (Table~\ref{tab:detections}) in the eight $H$ band measurements, with an average $f_H = 0.78\pm0.02$\%, and the full range being $0.64\pm0.05$\% to $1.13\pm0.09$\% (Table~\ref{tab:detections}). The average value however remains consistent with the published preliminary result \citep{2022ApJ...926..213K}. In the $K$ band, on the other hand, the four new measurements agree within the uncertainties, with an average $f_K = 0.72\pm0.02$\%, which suggests that the $H$-band variability could be spurious and resulting from the fact that we are close to the typical CHARA companion detectability limit. The slightly lower contribution of the companion in the $K$ band is expected due to its high temperature and IR excess from the Be-star disk, which increases with wavelength. 

Finally, while the primary Be star and its disk are only marginally resolved, the VIS2 data indicate a low $i_{\rm disk}$ compatible with the orbital $i$, and fitting a Gaussian to all data in the respective bands simultaneously results in an $H$-band FWHM of $0.15\pm0.01$\,mas, and $K$-band FWHM of $0.17\pm0.01$\,mas. The increasing size of the Be disk with wavelength is according to the expectations \citep[e.g.,][]{2015MNRAS.454.2107V}.

\textbf{V2119 Cyg} (HR\,7807, HD\,194335, HIP\,100574) is a B2\,IIIe star \citep{1982ApJS...50...55S} that was initially reported as a Be + sdO binary candidate based on cross-correlation analysis of four IUE spectra using a 45\,kK model spectrum \citep{2018ApJ...853..156W}. Three more recent HST/STIS exposures \citep{2021AJ....161..248W} enabled refining the sdO stellar parameters, with $f_{\rm FUV} = 4.7\pm0.7$\%, $T_{\rm eff} = 38.2$\,kK, and a radius of $0.51\pm0.08$\,{\Rsun}. Both the IUE and HST detections enabled the measurement of the companion's RVs. The sdO companion was also subsequently detected interferometrically (four times), and its preliminary combined orbital solution was derived \citep{2022ApJ...926..213K}. A brief observational history regarding Balmer line emission was presented in the same work. Recently, V2119~Cyg was found to present weak orbital phase-locked $V/R$ variability in H$\alpha$ (with a semi-amplitude of $\sim0.02$), indicating a disk asymmetry caused by the tidal influence of the companion \citep{2023Galax..11...83M}. 

\begin{figure*}
\epsscale{1.1}
\plottwo{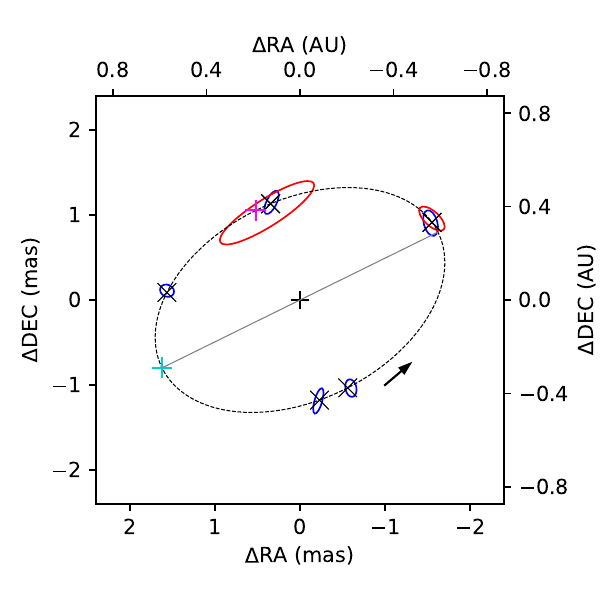}{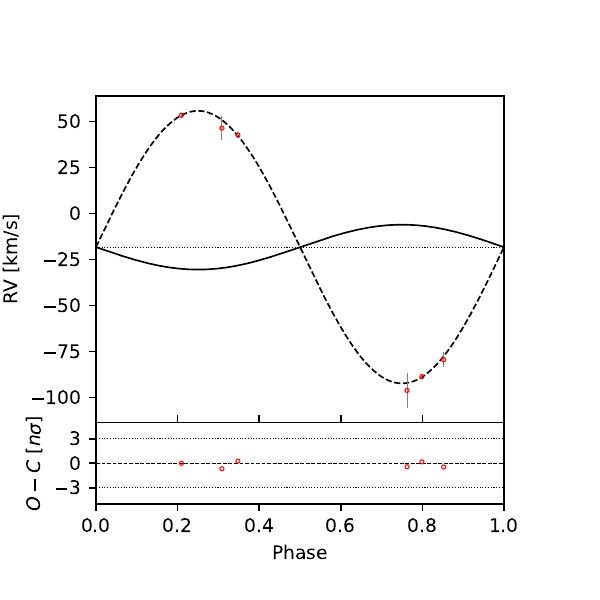}
\caption{Same as Fig.~\ref{fig:orb_28Cyg} but for V2119~Cyg. \label{fig:orb_V2119Cyg}}
\end{figure*}

Three additional interferometric measurements that were since obtained have confirmed the previously published orbital solution. The final solution (Table~\ref{tab:orbital_params}, Fig.~\ref{fig:orb_V2119Cyg}) is based on seven interferometric detections (Table~\ref{tab:detections}) and six published RVs of the sdO companion \citep{2018ApJ...853..156W,2021AJ....161..248W}. With the precision of the RV measurements and their reasonably good phase coverage, $K_\mathrm{sdO}$ is well constrained. Unfortunately, we are lacking RV measurements for the primary Be star.

Using the Gaia distance results in dynamical masses $M_{\rm Be} = 8.20\pm0.35$\,{\Msun} and $M_{\rm sdO} = 1.35\pm0.27$\,\Msun. As for the Be star, its mass falls between the masses given for B1.5V and B2V star by \citet{2013ApJS..208....9P}, which is in good agreement with the spectral classification taken from the literature. Additional RV measurements ideally for the Be star would be needed to improve the precision of the results.

According to the interferometric measurements, the flux of the companion appears relatively stable, with $f_H = 2.13\pm0.05$\%, and $f_K = 2.21\pm0.09$\%, although the latter value is based only on two measurements. The flux ratios in the two bands are thus compatible within the error bars and we do not clearly observe a decrease of the flux ratio with increasing wavelength, expected for a very hot companion. The $H$-band flux ratio remains consistent with the preliminary result \citep{2022ApJ...926..213K}.

The Be star and its disk are only marginally resolved in the VIS2 data, and fitting all epochs simultaneously results in ${\rm FWHM} = 0.16\pm0.01$\,mas in the $H$ band, and ${\rm FWHM} = 0.22\pm0.01$\,mas in the $K$ band (see an example plot with $H$-band data in Fig.~\ref{fig:V2119Cyg_mircx}). The inclination and PA of the disk cannot be constrained well with these data.

\textbf{60~Cyg} (HR\,8053, HD\,200310, HIP\,103732) is a B1Ve star \citep{1982ApJS...50...55S} that was reported as an SB1 based on RV variations of H$\alpha$ emission wings corresponding to the motions of the primary Be star \citep{2000A&A...356..913K}. The companion was later found to be a hot sdO from cross-correlation analysis of 23 FUV spectra from the IUE. The spectra were initially shifted and averaged using the sdO RVs estimated from a `trial' mass ratio $q$ of 0.131, which was based on the SB1 orbit and an assumed Be star mass and system inclination \citep{2017ApJ...843...60W}. Further optimization of the initial model $T_{\rm eff}$ and $q$ resulted in maximization of the cross-correlation signal for $T_{\rm eff} = 42\pm4$\,kK, $q = 0.146\pm0.023$, and $f_{\rm FUV} = 3.39\pm0.15$\%. Subsequently, the companion was detected interferometrically (five times) and a combined astrometric and SB1 orbital solution \citep[based on the Be star RVs from][]{2000A&A...356..913K} was derived, leading to preliminary estimates of the dynamical masses \citep{2022ApJ...926..213K}. A brief history of Balmer line emission is also given by \citet{2022ApJ...926..213K}. Similar to V2119~Cyg, 60~Cyg was recently found to show weak phase-locked $V/R$ variations \citep{2023Galax..11...83M}.

\begin{figure*}
\epsscale{1.1}
\plottwo{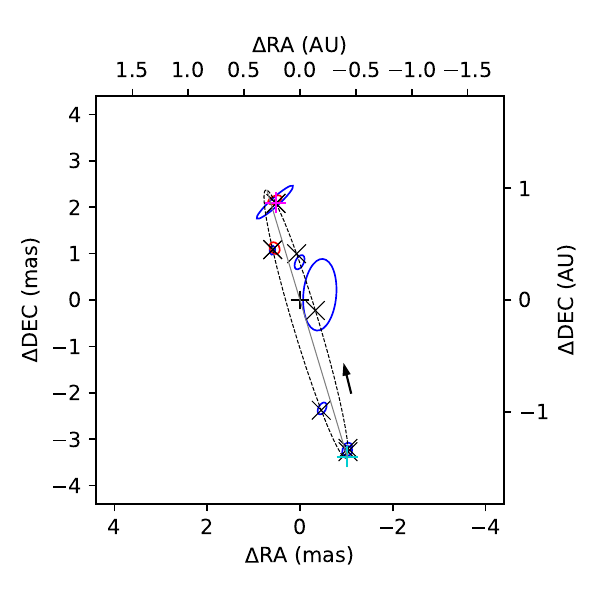}{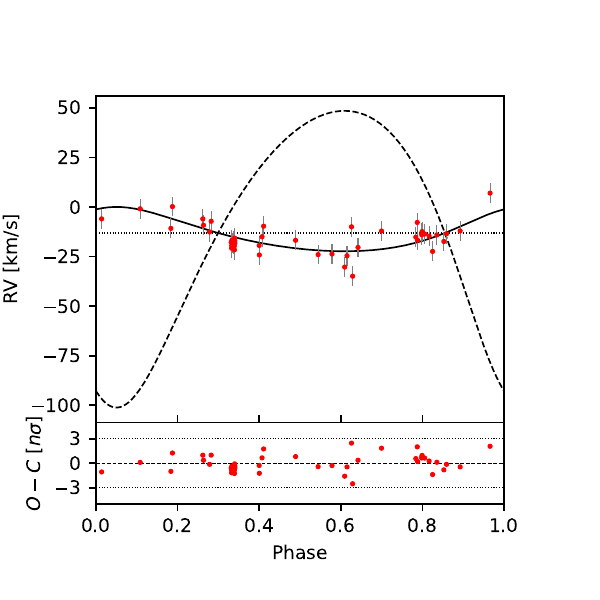}
\caption{Same as Fig.~\ref{fig:orb_28Cyg} but for 60~Cyg. \label{fig:orb_60Cyg}}
\end{figure*}

New interferometric data confirm the previously derived orbit, including the non-zero eccentricity ($e = 0.20\pm0.01$), and close to edge-on inclination ($i = 84.6\pm0.3$\degree), with the final solution (Table~\ref{tab:orbital_params}, Fig.~\ref{fig:orb_60Cyg}) relying on nine interferometric detections (Table~\ref{tab:detections}) and 47 published RVs of the primary Be star \citep{2000A&A...356..913K}. 60~Cyg is thus the second confirmed eccentric Be + sdO binary after 59~Cyg \citep{2013ApJ...765....2P}, although tidal interactions during mass transfer are expected to circularize the orbit. While for 59~Cyg we were able to detect a third outer component on an eccentric orbit, which probably affects tidally the inner Be + sdO binary (see Sect.~\ref{sec:non-detections}), this is not the case for 60~Cyg, and the origin of the non-zero $e$ thus remains unclear. As for the very high $i$ of $84.6\pm0.3${\degree}, it is noteworthy that no shell lines are observed in the spectra, even though the disk and orbital inclinations should be equal given the binary mass-transfer origin of the Be-star rapid rotation. This  therefore implies a very thin disk, with a half-opening angle in line-emitting regions of $\lesssim5${\degree}, while the typically quoted value is at least $10${\degree} \citep[see Sect.~5.1.1 in][]{2013A&ARv..21...69R}.

The mass ratio derived from the FUV spectra is in good agreement with $q=0.135\pm0.024$ resulting from our orbital solution. Similar to 28~Cyg and V2119~Cyg, the precision of the dynamical masses - $M_{\rm Be} = 10.75\pm0.49$\,{\Msun} and $M_{\rm sdO} = 1.45\pm0.23$\,{\Msun} - suffers from the scarcity and/or precision of the RV measurements. The Be star mass is in reasonable agreement with the spectral type, where for a B1V star, a mass of 11.8\,{\Msun} is expected according to \citet{2013ApJS..208....9P}. However, it should be noted that for a Be star seen close to edge-on such as 60~Cyg, the gravity darkening should cause the star to appear cooler than a non-rotating counterpart of the same mass, which does not seem to be the case here. We also note that the higher dynamical masses compared to those given by \citet{2022ApJ...926..213K} are mostly due to the adoption of the Gaia distance from \citet{2021AJ....161..147B}, which is larger than the one calculated directly from the parallax.

The component flux ratio derived from interferometry in the $H$ band show a somewhat complicated picture. Neglecting the measurement with large uncertainty corresponding to a very small angular separation of $<0.5$\,mas (${\rm RJD}= 59725.977$), we are left with four precise measurements in the range of 1.22\% to 1.69\%, including the $1\sigma$ error margins, and two lower-quality measurements with larger errors indicating a higher flux ratio between 1.84 and 2.85\%. Inspection of the orbital phase for these two measurements reveals that one of them was taken at periastron and the other one almost exactly at apastron. While it is conceivable that there is an orbital phase dependence of the sdO brightness due to the slightly eccentric orbit, this would need to be confirmed with additional measurements. Overall, excluding the lower quality flux-ratio measurements with errors above 0.5\% results in an average $f_H=1.36\pm0.03$\%, and $f_K=1.42\pm0.04$\% from the two measurements in the $K$ band. The $H$-band value is larger by about $3\sigma$ than the preliminary one ($1.16\pm0.07$\%), however, this actually improves the agreement with the $f_H$ expected from the observed FUV flux ratio \citep[Table~5 in][]{2022ApJ...926..213K}.

Finally, the Be star disk was found to be only marginally resolved without the possibility to constrain well the disk inclination and PA. As for the angular size, the data indicate ${\rm FWHM} = 0.15\pm0.01$\,mas in the $H$ band, and ${\rm FWHM} = 0.19\pm0.01$\,mas in the $K$ band.

\section{First orbits for targets with newly detected close companions}
\label{sec:orbits_new}

\textbf{HR\,2142} (HD~41335, HIP\,28744) is a B1Ve star \citep{2018MNRAS.474.5287A}, although historically it was also classified as B2IVne \citep{1982ApJS...50...55S}. It has had mostly stable emission in the Balmer lines, although the H$\alpha$ line often shows complex features such as transitions from double-peaked to triple-peaked profiles. HR\,2142 was found to be an interacting binary with the detection of periodic occurrences of shell profiles in Balmer lines \citep{1972PASP...84..334P}. An SB1 orbital solution was derived from the broad wings of Balmer and helium lines, resulting in a circular orbit, and $K_\mathrm{Be}$ of $9.4\pm0.9$\,km\,s$^{-1}$ \citep{1983PASP...95..311P}, which was later revised to $7.1\pm0.5$\,km\,s$^{-1}$ \citep{2016ApJ...828...47P}. Infrared space photometry revealed an excess compatible with a classical Be star disk, and an Be + sdOB post-mass transfer nature of HR~2142 was suggested \citep{1991A&A...250..437W}. The hot sdO companion was confirmed with cross-correlation analysis of averaged IUE spectra and found to contribute $f_{\rm FUV} \geq 0.9$\% and to have $T_{\mathrm{eff}}=43\pm5$\,kK \citep{2016ApJ...828...47P}. However, the companion was convincingly detected only at orbital phases spanning about a quarter of the orbit. A weak SED turndown indicative of the Be star disk truncation was detected as well \citep{2019ApJ...885..147K}.  HR~2142 also has a wider speckle companion separated by 0.56 to 0.59~arcsec with a PA of $\sim271^{\circ}$ according to measurements taken between the years 1993 and 2015 available from the Washington Visual Double Star Catalog\footnote{\url{http://www.astro.gsu.edu/wds/}} \citep[WDS,][]{2023yCat....102026M}.

The phase-locked shell phases were interpreted as originating from gas streams crossing the gap carved by the orbiting companion in the Be disk, with both the orbital and the disk planes seen close to edge-on orientation. The Be disk is then divided into an inner part and a circumbinary part extending beyond the binary orbit, with the companion likely also possessing a small accretion disk \citep[see Fig.~5 in][]{2016ApJ...828...47P}. The shell phases are observed just before and after the inferior conjunction, i.e., when the companion passes in front of the Be star and the gas streams are projected onto the Be star photosphere. The presence of the gas streams and a circum-companion accretion disk may help explain the phase-dependent detectability of the companion in the FUV \citep{2016ApJ...828...47P}. 

The companion of HR~2142 was successfully detected interferometrically four times in the $H$ band and four times in the $K$ band by CHARA, as well as two times in the $K$ band by VLTI/GRAVITY (Table~\ref{tab:detections}). One of the CHARA measurements in the $H$ band (RJD$=59536.024$) was affected by bad weather conditions, and we had to fix $f_H$ at the average value from the other three good-quality measurements ($2.94$\%) for the companion position to converge. Similarly, $f_K$ had to be fixed for the two GRAVITY measurements, which initially resulted in almost equal fluxes for the two components, probably due to the sparser $(u,v)$ coverage and lower angular resolution compared to CHARA. Using the average $f_K = 3.49$\% derived from the three CHARA $K$-band measurements solved this issue. The higher companion flux ratio in the $K$ band is contrary to the expectations for a companion which is hotter than the primary Be star, and could be due to the presence of ionized circum-companion gas. 

\begin{figure*}
\epsscale{1.16}
\plottwo{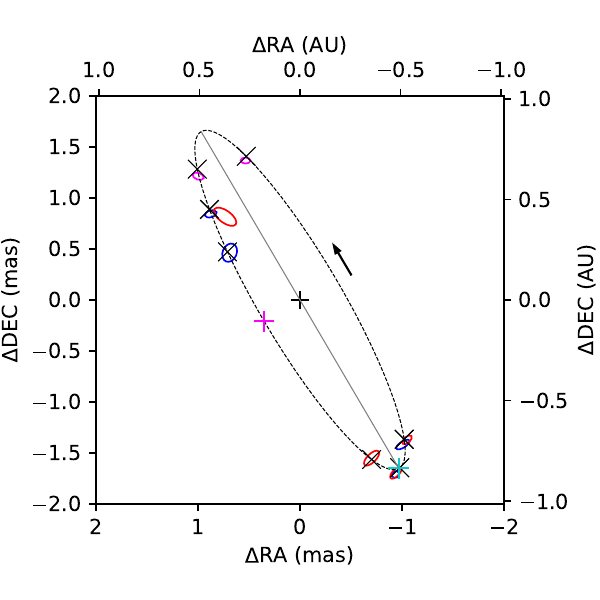}{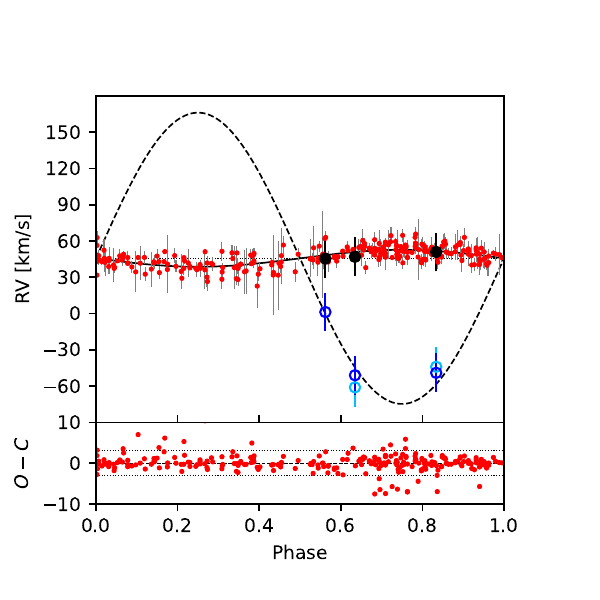}
\caption{Same as Fig.~\ref{fig:orb_28Cyg} but for HR~2142 with two additional GRAVITY detections (magenta) in the left panel. The RV curve in the right panel also includes GRAVITY-measured RVs for the Be star (black), and the sdO (light blue and dark blue determined from Br$\gamma$ and \ion{He}{1}\,$\lambda20587$, respectively). \label{fig:orb_HR2142}}
\end{figure*}

The orbital solution presented in Table~\ref{tab:orbital_params} and Fig.~\ref{fig:orb_HR2142} is based on ten interferometric detections (Table~\ref{tab:detections}) and 271 published RVs of the primary Be star derived from cross-correlation of FUV spectra as well as from the emission wings of H$\alpha$ \citep[Table~2 of][]{2016ApJ...828...47P}. All the RVs were assigned equal weights for the spectroscopic orbital solution of \citet{2016ApJ...828...47P}, so we did the same in our solution. 

The Gaia DR3 astrometric solution shows a slightly elevated RUWE value of 1.515, and the distance itself has an error of $\sim10$\%, which together with the high scatter of the available RV measurements enables only approximate computation of the dynamical masses: $M_{\mathrm{Be}} = 17.6\pm5.7$\,{\Msun} and $M_{\rm sdO} = 1.03\pm0.22$\,{\Msun}. The Be star spectral classification (B1Ve) is in agreement with the dynamical mass at the lower end of the error interval, while the best-fit mass would be expected for a B0V to B0.5V star according to \citet{2013ApJS..208....9P}. This is in line with the fact that the likely edge-on orientation of the Be star would make it appear cooler than implied by its mass due to gravity darkening. We note that including the RV error margins given by \citet{2016ApJ...828...47P} in the solution affects quite significantly the resulting $K_{\rm Be}$, which in this case converges to a lower value of $4.6\pm0.3$\,{\kms}. This then leads to a lower $q$ of $0.037\pm0.005$, while the total mass remains the same. 

Given the measured values of the near-IR flux ratios, $f_{\rm FUV}$ should be significantly higher than the reported lower limit of $0.9$\%, if the companion is indeed a hot sdO star. In order to investigate this in detail, we utilized model spectra from \citet{2003IAUS..210P.A20C}, UV spectra from the IUE, and near-IR photometry from \citet{2002yCat.2237....0D}. For the Be star and the interstellar reddening, we used $T_{\rm eff, Be} = 22.5$\,kK, $\log{g}_{\rm Be} = 3.35$, and $E(B-V)=0.14$\,mag taken from the spectral fitting of \citet{2018MNRAS.474.5287A}. We also take the Be star radius $R_{\rm Be}$ from the same work, but scale it to the larger Gaia distance, which results in $8.89$\,{\Rsun}, as opposed to $7.06$\,{\Rsun} given by the Hipparcos distance of $403$\,pc. We confirm that this combination of parameters results in an excellent fit to the averaged IUE spectra. For the sdO companion, we used $T_{\rm eff, sdO} = 43$\,kK and $\log{g}_{\rm sdO} = 5.0$ according to the FUV analysis \citep{2016ApJ...828...47P}. With these parameters, we obtain a radius of $R_{\rm sdO}=0.25$\,{\Rsun} assuming $f_{\rm FUV}=0.9$\%. On the other hand, in order to match the measured $H$-band flux ratio while keeping $T_{\rm eff, sdO}$ fixed, $R_{\rm sdO}$ has to be increased to $\sim0.60$\,\Rsun, which then results in an expected $f_{\rm FUV}$ of $\sim5.2$\%. These considerations therefore indicate that the sdO is probably indeed heavily obscured throughout its orbit, as suggested by \citet{2016ApJ...828...47P}.

The circumstellar disk of HR~2142 was partially resolved in the CHARA VIS2 data and marginally in the VLTI/GRAVITY data. Using only the CHARA data with the companion flux and position fixed at each epoch, and excluding the lower quality $H$-band CHARA dataset (RJD$=59536.024$), the 2D Gaussian parameters are ${\rm FWHM} = 0.36\pm0.02$\,mas, $i_{\rm disk}=54\pm4^{\circ}$, ${\rm PA}=55\pm6^{\circ}$ in the $H$ band and ${\rm FWHM} = 0.45\pm0.02$\,mas, $i_{\rm disk}=63\pm8^{\circ}$, ${\rm PA}=59\pm5^{\circ}$ in the $K$ band. Given that the disk should have a close to edge-on orientation (see above), these results might be underestimating $i_{\rm disk}$ due to the assumption of a geometrically thin disk. In a more realistic model, we introduced a UD representing the stellar photosphere with a diameter fixed at the radius of 0.163\,mas determined from SED fitting \citep{2018MNRAS.474.5287A}, which resulted in $i_{\rm disk}$ compatible with the orbital solution. The best-fit parameters for the disk in this case are, $i_{\rm disk}=71\pm10^{\circ}$, ${\rm PA}=56\pm6^{\circ}$ in the $H$ band, and $i_{\rm disk}=71\pm7^{\circ}$, ${\rm PA}=58\pm4^{\circ}$ in the $K$ band. Unfortunately, the flux contribution and the size of the disk are strongly correlated and cannot be determined independently. 

\begin{figure*}
\epsscale{1.16}
\plotone{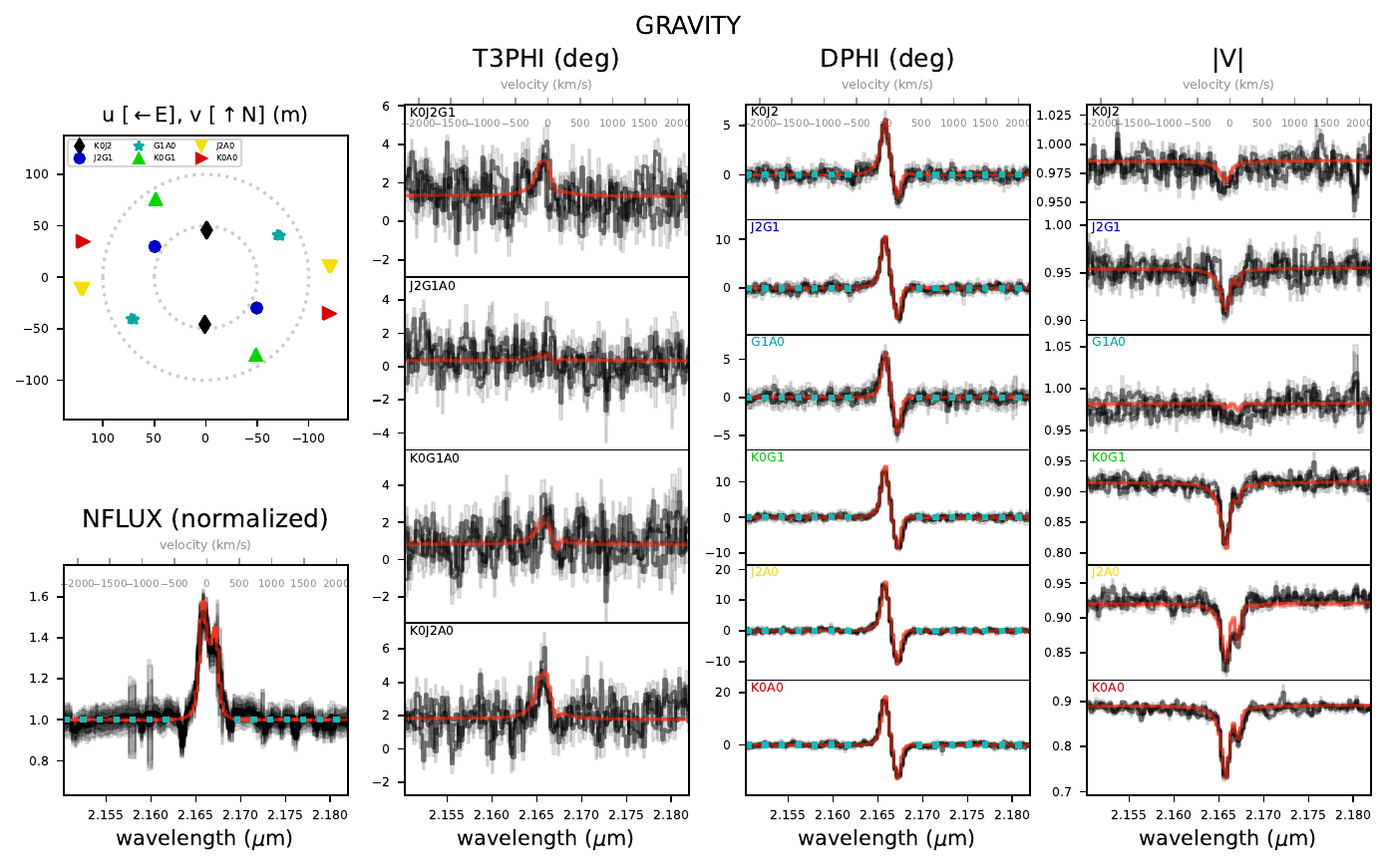}
\caption{PMOIRED model fit (red) to GRAVITY data (black with gray error bars) of HR~2142 obtained on ${\rm RJD} = 59932.760$ in the spectral region around Br$\gamma$. The upper left panel shows the $(u,v)$ coverage, the panel titled NFLUX shows the normalized spectrum, the T3PHI panel shows the closure phase for the four baseline triangles, and the DPHI and $|$V$|$ panels show the differential phases, and the absolute visibilities, respectively, for each baseline. Blue dots highlight the continuum regions used for normalization of NFLUX and DPHI. \label{fig:HR2142_Brg_DATA}}
\end{figure*}

Further insight into the HR\,2142 binary system can be gained with the two epochs of GRAVITY spectro-interferometry ($R\sim4000$) covering the entire $K$ band including the Br$\gamma$ and \ion{He}{1}\,$\lambda20587$ lines. The observables covering the Br$\gamma$ line show the typical signatures of a Be star with a circumstellar disk in Keplerian rotation, i.e., double-peaked emission line profile, S-shaped differential phases, and a V-shaped visibility profile \citep[e.g.,][]{2012A&A...538A.110M}, although there is an RV-shifted component in all observables introducing a clear asymmetry in the profiles. Fig.~\ref{fig:HR2142_Brg_DATA} shows the first epoch, where the asymmetric component is blue-shifted and coincides with the main blue emission peak of Br$\gamma$. 

For PMOIRED modeling of the observables covering the Br$\gamma$ line, we employed a geometric model consisting of two uniform disks representing the two photospheric stellar components, a 2D Gaussian representing continuum emission from the Be-star disk, a Keplerian rotating disk representing the line emission from the Be-star disk (see Sect.~\ref{sec:pmoired_keplerian}), and an additional unresolved emission line component (with a Lorentzian profile) which could originate from the companion surroundings or possibly from within the Be disk. Initially, only the two UDs representing the stellar photospheres were fixed, specifically at the values of 0.163\,mas for the Be star (determined from SED fitting), and 0.01\,mas for the sdO (representing a point source). The remaining parameters converged very close to the expected values for the Be disk, but also for the Be-star $v\sin{i}$ (which enters the model via the rotational velocity at the inner radius of the disk, see Sect.~\ref{sec:pmoired_keplerian}). The only discrepancy regards the disk inclination in the first epoch, where it converged to $\sim50${\degree}, while for the second epoch it agrees with the orbital inclination. In an updated PMOIRED model, we thus fixed the disk inclination at the orbital inclination ($77.7$\degree), which resulted in an equally good fit to the data in terms of the final residuals, indicating that the lower disk inclination resulting from the initial fit to the first epoch was not a significant result. In the final PMOIRED fit, we also fixed the Be-star $v\sin{i}$ at 330\,{\kms} according to \citet{2018MNRAS.474.5287A}, although this did not affect the results. 

\begin{figure*}
\epsscale{1.16}
\plotone{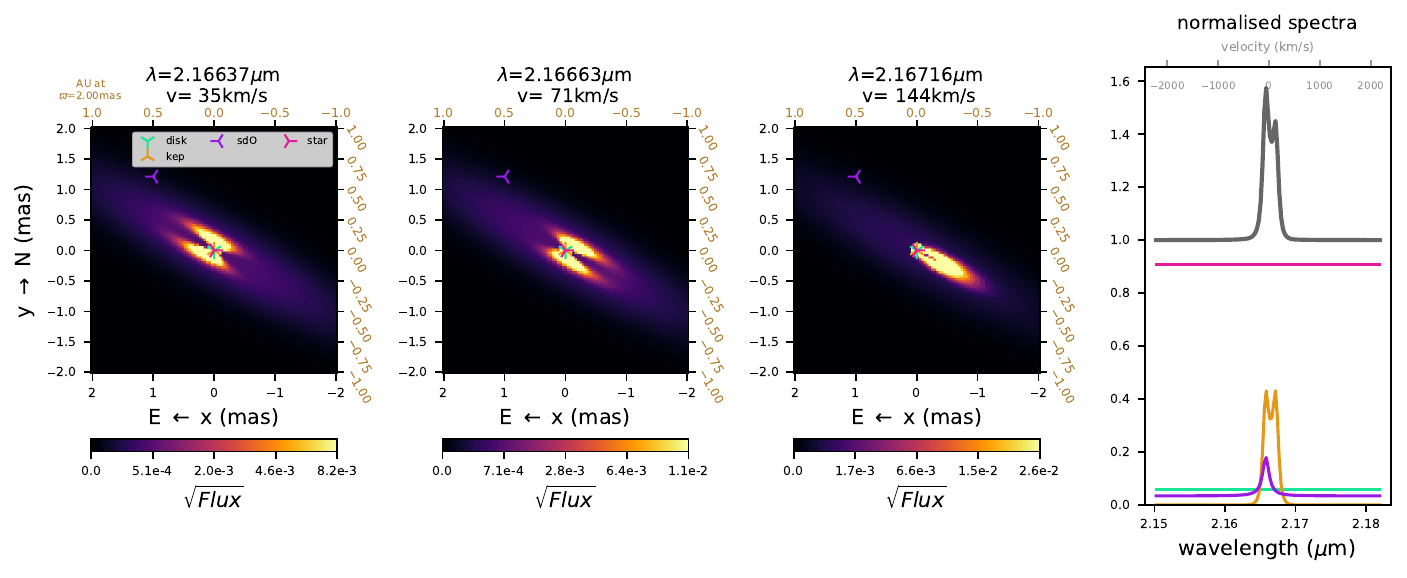}
\caption{Left: PMOIRED model images of HR~2142 for three different wavelengths (velocities) within the Br$\gamma$ line. The components include continuum flux from the primary Be star (labeled in the legend of the leftmost image as `star'), continuum flux (`disk') and emission line (`kep') from the Keplerian disk, and a continuum and line emission component from the sdO companion (`sdO'). Right: Total (black) and component (color-coded according to the legend in the leftmost model image) normalized spectra in the Br$\gamma$ line. \label{fig:HR2142_Brg_IMAGE}}
\end{figure*}

As for the location of the additional line component, it converged to within 0.1--0.2\,mas of the companion position when using the positions determined above (Table~\ref{tab:detections}) as the initial guesses, or did not converge at all. The GRAVITY Br$\gamma$ data thus indicate that the additional line emission originates from the vicinity of the companion and not from within the Be-star disk. Finally, the heliocentric rest wavelengths of the two line components can be compared with the RV curves from our orbital solution, revealing an excellent agreement (considering the wavelength calibration uncertainty of $\sim15$\,{\kms}) and thus providing further confidence to our orbital solution and to the origin of the line emission from the companion (right panel of Fig.~\ref{fig:orb_HR2142}). The PMOIRED fit to the data from the first epoch is shown in Fig.~\ref{fig:HR2142_Brg_DATA}, and the corresponding model images and normalised spectra for all model components are shown in Fig.~\ref{fig:HR2142_Brg_IMAGE}. 

As for the \ion{He}{1}\,$\lambda20587$ line, the GRAVITY spectra are unfortunately of low quality, but the spectro-interferometry shows a single spectral asymmetric feature in the two epochs. This feature could again only be reproduced by introducing an unresolved emission line component, whose location coincides with the location of the companion, and whose RV agrees with the one expected for the companion. Thus, in addition to Br$\gamma$, we are detecting emission from the vicinity of the sdO companion also in \ion{He}{1}\,$\lambda20587$, which indicates a hotter environment near the companion than in the Be-star disk, as would be expected. The two GRAVITY snapshots correspond to the orbital phase of $0.84$ and $0.64$, respectively, and all the measured RVs are plotted in the right panel of Fig.~\ref{fig:orb_HR2142}.


An additional GRAVITY snapshot was taken on ${\rm RJD}=59910.788$, but it could not be calibrated due to corrupted calibrator data, and was initially discarded. However, the normalized spectrum and the differential phases DPHI are still usable, as these do not require absolute calibration. This enabled us to constrain the Be star RV from Br$\gamma$ using the Keplerian disk model with most parameters fixed from the analysis above, as well as the sdO companion RV from \ion{He}{1}\,$\lambda20587$. The expected location of the companion (and the corresponding line component) for this epoch was derived from the orbital solution and used as an initial guess. The converged coordinates remained consistent with the expected position of the companion, albeit with larger error bars resulting from the limited dataset, and the derived RVs for the Be star and the companion are again in excellent agreement with the RV curves from our orbital solution (right panel of Fig.~\ref{fig:orb_HR2142}, this measurement corresponds to the orbital phase of $0.54$).

\textbf{HD 161306} (HIP\,86884) is a single-lined spectroscopic binary \citep{2014A&A...567A..57K} with the most recent spectral classification given as B3/5Vnne, where the slash indicates a similar probability for the two given spectral types \citep{1999MSS...C05....0H}. The spectroscopic orbit of \citet{2014A&A...567A..57K} is based on 43 RVs of the Be star measured from the emission wings of H$\alpha$. HD~161306 was proposed as another Be + sdOB binary due to anti-phased orbital motion of a narrow emission component of the \ion{He}{1}\,$\lambda6678$ line, likely originating from a heated-up part of the Be star disk facing the companion \citep{2014A&A...567A..57K}. The emission in H$\alpha$ has been slowly decreasing over the last 15 years, with the higher resolution spectra available in the BeSS database showing complex morphologies with varying emission peaks. Unfortunately, there are no FUV spectra available.

\begin{figure*}
\epsscale{1.16}
\plottwo{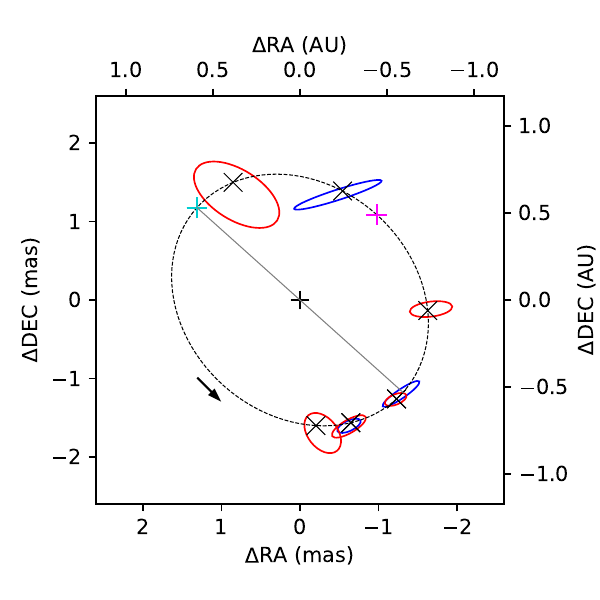}{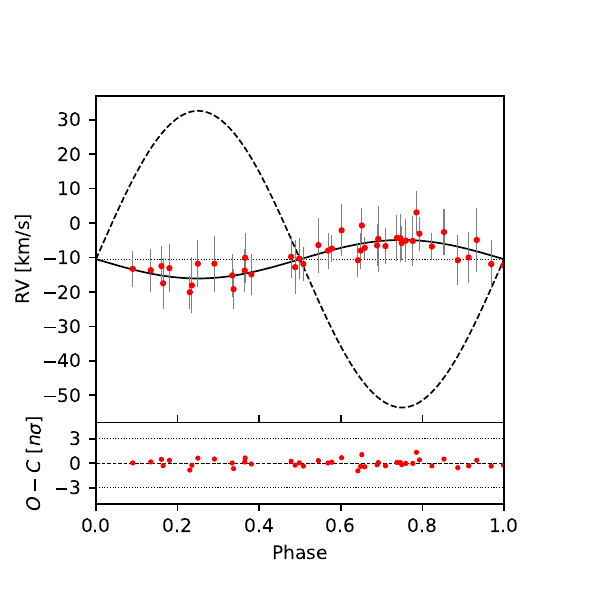}
\caption{Same as Fig.~\ref{fig:orb_28Cyg} but for HD~161306. \label{fig:orb_HD161306}}
\end{figure*}

New interferometric measurements which include a total of eight reliable detections of the companion (Table~\ref{tab:detections}) agree well with the previously derived spectroscopic orbit. Using the same 43 RV measurements, we obtain a well-constrained combined orbital solution (Table~\ref{tab:orbital_params} and Fig.~\ref{fig:orb_HD161306}). The average flux ratios are $f_H = 2.86\pm0.20$\% and $f_K = 2.03\pm0.07$\%. The difference in the flux contribution in the two bands is compatible with the companion being hotter than the primary, as well as with the increasing flux excess with wavelength from the circumstellar disk of the Be star. Finally, the Be star (and the companion) are unresolved in the VIS2 data.

The dynamical masses derived with the help of the Gaia distance are $M_{\rm Be} = 6.02\pm0.26$\,{\Msun} and $M_{\rm sdOB} = 0.784\pm0.074$\,{\Msun}. According to the tabulation of \citet{2013ApJS..208....9P}, the measured mass of the primary corresponds to a spectral type close to B2.5V, showing a reasonably good agreement with the B3 spectral classification mentioned above.

\textbf{7 Vul} (HR\,7409, HD\,183537, HIP\,95818) is a spectroscopic 69.3-day binary with a Be star primary and a low-mass secondary \citep{2011MNRAS.413.2760V, 2020A&A...639A..32H}. Spectroscopically, 7~Vul appears as a B5Vn star \citep{1968ApJS...17..371L} with very broad absorption lines and a weak emission spectrum with sharp shell absorption originating from the circumstellar disk seen close to edge-on orientation. The emission has been diminishing over the last $\sim15$ years, with H$\alpha$ transitioning from double-peaked emission reaching slightly above the continuum to a pure absorption profile \citep{2020A&A...639A..32H}. 7~Vul was suggested to be a post-mass transfer binary, based on a weak absorption component of the \ion{He}{1}\,6678 line that moves in antiphase with the orbital period \citep{2011MNRAS.413.2760V, 2020A&A...639A..32H}.

\begin{figure*}
\epsscale{1.16}
\plottwo{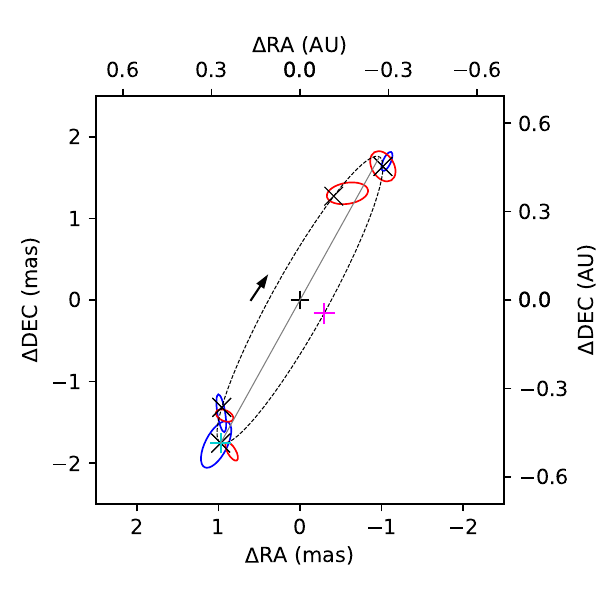}{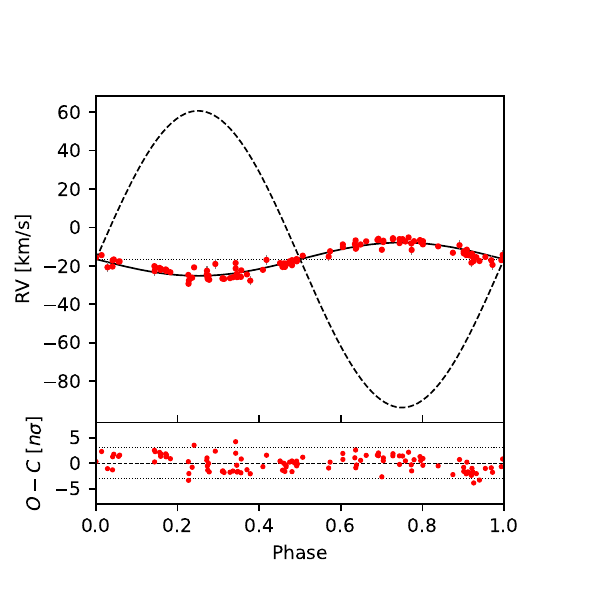}
\caption{Same as Fig.~\ref{fig:orb_28Cyg} but for 7~Vul. \label{fig:orb_7Vul}}
\end{figure*}

CHARA observations of 7~Vul led to a total of seven reliable interferometric detections of the faint companion at different orbital phases (Table~\ref{tab:detections}). For the combined astrometric and spectroscopic orbital solution, a total of 116 RV measurements of H$\alpha$ absorption were adopted from the most recent spectroscopic study \citep{2020A&A...639A..32H}. The solution shows a good agreement with the parameter estimates from previous studies (Table~\ref{tab:orbital_params} and Fig.~\ref{fig:orb_7Vul}). Four additional interferometric observations resulting in clear non-detections were found to correspond to orbital phases where the angular separation of the two components is expected to be $<0.5$\,mas. Thus, these non-detections are also compatible with the expected orbit, and were therefore included as additional constraints in the orbital solution. This, however, did not affect the best-fit orbital parameters in a significant way. The almost edge-on orbital inclination was found to be compatible with the shell spectrum appearance, and the orbital and disk planes must therefore be nearly parallel, as would be expected for a post-mass-transfer binary. On the other hand, the non-zero eccentricity reported in the previous studies \citep{2011MNRAS.413.2760V, 2020A&A...639A..32H} was not confirmed, with the combined circular orbit solution reproducing the data well. The Be star and its disk are unresolved in the visibility data.

The resulting dynamical masses computed with the help of the Gaia distance are $M_{\rm Be} = 4.25\pm0.23$\,{\Msun} and $M_{\rm sdOB} = 0.477\pm0.020$\,\Msun. The Be star mass corresponds almost exactly to a B6V star according to the classification of \citet{2013ApJS..208....9P}, which is close to the B5V type found in the literature, although for a Be star seen close to edge-on like 7~Vul, we would expect the apparent spectral type to be later than implied by the mass (cf. the cases of 60~Cyg and HR~2142 discussed above).

The companion flux ratios were found to be $f_{H}=1.44\pm0.08$\% and $f_{K}=1.74\pm0.14$\%, while we would expect a slightly lower flux ratio at longer wavelengths for a hot sdO companion. Even for an sdB companion with $T_\mathrm{eff}$ similar to that of the Be star \citep[cf. the case of $\kappa$~Dra;][]{2022ApJ...940...86K}, $f_K$ should be at most equal to $f_H$, but our measurements indicate that $f_K$ is higher by about 2$\sigma$. This discrepancy remains unresolved in the current study. 

\section{Marginal detections, non-detections, and new wide companions}
\label{sec:non-detections}

In this section, we discuss all the remaining targets from the different groups individually. Most of the stars described here show no signs of the presence of a companion, and the corresponding magnitude limits for each non-detection are listed in Table~\ref{tab:detlims}. In addition, the data for V780~Cas and 11~Cam reveal a possible companion, which however needs to be confirmed with further observations. Finally, for 59~Cyg, 48~Per, and V782~Cas, we detected previously unreported wide companions, which are however not the stripped sdOB stars that we were looking for. Unless noted otherwise, the stars were unresolved in the visibility data, which translates to a star and disk UD diameter of $\lesssim0.25$\,mas, or a Gaussian FWHM of $\lesssim0.15$\,mas.

\subsection{Be + sdOB binaries}

\begin{figure}
\epsscale{1.2}
\plotone{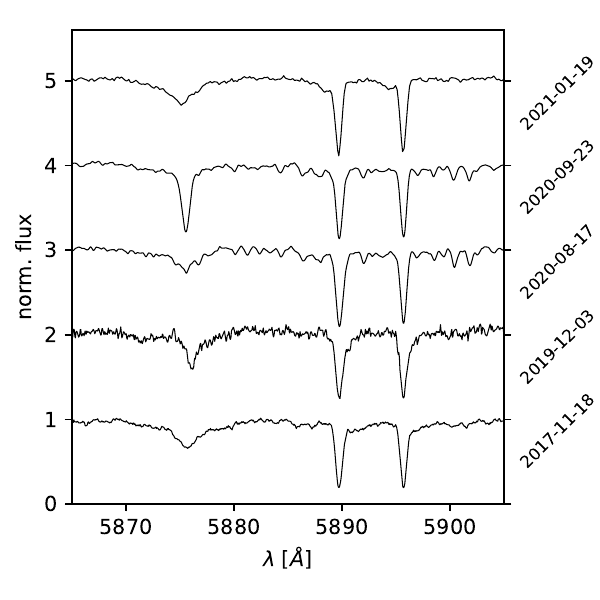}
\caption{Five BeSS echelle spectra for V780~Cas in the region around \ion{He}{1}\,$\lambda 5875$ and the Na D lines. All spectra except for the first epoch are shifted upwards in flux for clarity. The \ion{He}{1}\,$\lambda 5875$ line profile shows a transient phase with a deep shell absorption on 2021~Jan~19, and to a smaller extent on 2019~Dec~3, while the deep interstellar absorption in the Na D lines persists. \label{fig:V780Cas_spec}}
\end{figure}

\textbf{V780 Cas} (HD\,12302, HIP\,9538) was placed among Be + sdOB candidates based on our visual inspection of publically available spectra in the BeSS database. Similar to Be binaries such as for instance $\zeta$~Tau or $\kappa$~Dra, the strong H$\alpha$ emission line shows complex morphology including a deformed, flat-topped, and triple-peaked profile. Furthermore, in the five echelle spectra available from the BeSS database, one shows a strong shell absorption in several \ion{He}{1} lines (${\rm MJD}=59115.033$, Fig.~\ref{fig:V780Cas_spec}), which could make it similar to $\varphi$~Per or FY~CMa, i.e., Be stars with enhanced ionization in the outer parts of the disk facing the hot companion, where the shell absorption appears as a line-of-sight effect during inferior conjunction. If so, the occurrence of the shell phase implies close to edge-on orientation of the circumstellar disk plane. Alternatively, the transient shell absorption could occur due to a tilting or warping disk, or a variable disk flaring.

In our interferometric campaign, we detected a close companion in both $H$ band and $K$ band data at the first epoch (${\rm RJD}=59777.957$), with a separation of about 1.1\,mas, and $f_H \sim f_K \sim 1.2$\% (Table~\ref{tab:detections}). However, upon revisiting the target four more times during a period of more than two months following the initial observation, the detection could not be confirmed, as those data are fully compatible with a single star. Thus, it is possible that the detection at the first epoch was spurious, although it is unlikely that the signal comes from an asymmetry in the circumstellar disk of the Be star instead of a companion, as the disk is only marginally resolved in the visibility data $({\rm FWHM} \sim 0.2$\,mas in both bands). The other possibility is that due to the small expected size of the angular orbit together with the probably very high inclination, the companion is separated from the primary by $>0.5$\,mas only during a very limited orbital phase interval.

Given that we only have one detection (in both $H$ and $K$) in a total of five epochs, we conclude that further observations would be needed to confirm the presence of the companion. If the detection is real, $a''$ is at least $\sim1.1$\,mas and the epoch of the detection should be close to the orbital phase corresponding to quadrature, while the epoch of the spectrum showing shell absorption should be close to inferior conjunction. Unfortunately, given that the shell spectrum and the interferometric detection were obtained almost two years apart, we are not able to place any meaningful constraints on the possible orbital period.

\textbf{V1150 Tau} (HD\,29441, HIP\,21626) is a confirmed Be + sdO binary, based on cross-correlation analysis of HST/STIS spectra \citep[$f_{\rm FUV} = 2.7\pm0.3$\%;][]{2021AJ....161..248W}, although the orbital parameters are unknown. V1150~Tau was observed one time late in our campaign, resulting in a non-detection of the companion. With the target magnitude being at the very limit of the capabilities of MIRC-X and MYSTIC, the derived lower limits on the magnitude difference between the two components are not as tight as for the brighter targets in our sample (Table~\ref{tab:detlims}).

\textbf{QY Gem} (HD\,51354, HIP\,33493) is also a confirmed Be + sdO binary based on HST/STIS spectra \citep[$f_{\rm FUV} = 9.9\pm2.7$\%,][]{2021AJ....161..248W}. Similarly to V1150~Tau, there is no estimate of the orbital parameters. In our one CHARA observation in both $H$ and $K$ band, the companion remains undetected.

\textbf{59 Cyg} (HR\,8047, HD\,200120, HIP\,103632) was one of the first identified Be + sdO candidates based on observed features in optical spectra \citep{2000ASPC..214..581R, 2005PAICz..93...21M}, and it was later confirmed by cross-correlation analysis of 157 IUE spectra \citep[$f_{\rm FUV} = 4.0\pm0.5$\%,][]{2013ApJ...765....2P}. It is one of the few Be stars for which a double-lined spectroscopic orbital solution exists, with the reported orbit being slightly eccentric \citep[$e\sim0.14$;][]{2013ApJ...765....2P}. Inspection of WDS \citep{2023yCat....102026M} reveals that 59~Cyg also has a speckle companion first detected in 1981, but only with an `indeterminate' orbital solution available ($a''=208$\,mas, $P=161.5$\,yr, and $\Delta V \sim 2.8$\,mag). The Gaia DR3 catalog reveals an elevated value of RUWE at 2.724, which indicates that the astrometric solution might not be fully reliable. Indeed, the Gaia distance given by \citet{2021AJ....161..147B} has the largest uncertainty in our sample. A detailed orbital analysis, which might enable independent distance measurement based on the dynamical parallax, is reserved for an upcoming study (in prep.). In the NPOI survey by \citet{2021ApJS..257...69H}, only the wide speckle companion was detected (on two occasions).

Our one CHARA snapshot reveals a companion, which is however not the inner sdO one. It is separated by $\sim11$\,mas, and contributes $\sim2.4\%$ of the flux of the primary Be star in the $H$ and $K$ bands, while the estimated size of $a''$ for the sdO companion is only $\sim0.4$\,mas. In fact, this previously unknown companion had been already detected in several unpublished older CHARA snapshots, revealing orbital motion with a preliminary solution indicating $P\sim906$\,d, $i\sim144${\degree} and a significant eccentricity of $\sim0.5$ (Monnier \& Gardner, priv. comm.). The non-detection of this companion by NPOI \citep{2021ApJS..257...69H} is compatible with the contrast limits (magnitude difference of 3.0--3.5 in the visible for NPOI, while the measured $\Delta H \sim \Delta K \sim 4.05$\,mag.) Together with the wide speckle companion, 59~Cyg is thus (at least) a rather complex quadruple system, whose detailed analysis is reserved for a dedicated study (in prep.). The presence of the new companion is probably to blame for the elevated RUWE in Gaia DR3, as well as the non-zero eccentricity of the inner Be + sdO pair, which could have been acquired via dynamical interactions in the hierarchical system. Furthermore, the disk of 59~Cyg has experienced major disk tilts in the past \citep{1998A&A...330..243H,2023A&A...678A..47B}, which could have been also caused by the newly found companion.

Following a removal of the binary signal caused by the new companion from the interferometric data with CANDID, the inner sdO companion itself remains undetected in our one CHARA snapshot. This could be because the separation between the Be star and the sdO companion is indeed too small for the angular resolution of our data, and/or because the companion is fainter than the Be star by more than $\sim5$\,mag (Table~\ref{tab:detlims}). 

The disk of 59~Cyg is partially resolved in the VIS2 data, enabling a basic measurement of the disk size, PA, and inclination (after subtraction of the $\rho \sim 11$\,mas companion). The resulting values are Gaussian ${\rm FWHM} = 0.39\pm0.01$\,mas, $i_{\rm disk} = 54.2\pm0.8^{\circ}$, and ${\rm PA} = 28.7\pm0.5^{\circ}$ in the $H$ band, and a ${\rm FWHM} = 0.45\pm0.01$\,mas, $i_{\rm disk} = 52.4\pm0.8^{\circ}$, and ${\rm PA} = 31.2\pm0.4^{\circ}$ in the $K$ band. The derived $i_{\rm disk}$ is slightly lower than typically assumed for 59~Cyg \citep[60--80{\degree};][]{2005PAICz..93...21M}, and it is therefore possible that we are underestimating it by assuming a geometrically thin disk for the Gaussian fitting.

\begin{figure}
\epsscale{1.2}
\plotone{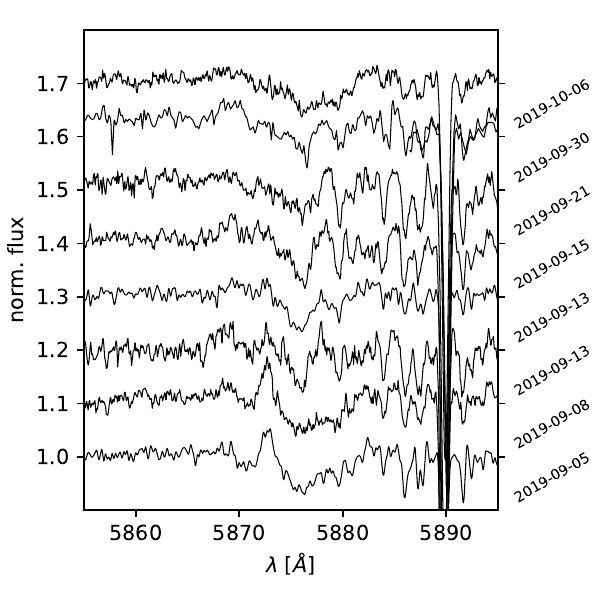}
\caption{Eight echelle spectra for V2162~Cyg taken within one month in the region around the \ion{He}{1}\,$\lambda 5875$ line. All spectra except for the first epoch are shifted upwards in flux for clarity. The \ion{He}{1}\,$\lambda 5875$ line profile shows a moving narrow emission peak on top of the broad absorption. The vertical dashed lines indicate our rough estimate of the stellar $v \sin{i}$ ($320\,$\kms).  \label{fig:V2162Cyg_spec}}
\end{figure}

\textbf{V2162 Cyg} (HD\,204722, HIP\,106079) is another candidate Be + sdOB binary on the basis of our inspection of publically available BeSS spectra. In addition to deformed and flat-topped H$\alpha$ line profiles, indicative of a possible disk truncation by an orbiting companion \citep{2018MNRAS.473.3039P,2019IAUS..346..105R}, there are narrow emission components in \ion{He}{1} lines that are well within the stellar $v \sin{i}$, which we interpret as a sign of external hard-UV irradiation heating up the outer, less rapidly rotating parts of the Be star disk (Fig.~\ref{fig:V2162Cyg_spec}). 

In our Be star sample, V2162~Cyg is the faintest target at $H=7.64$, which means the near-IR CHARA instruments were pushed to their brightness limits. We were able to obtain only a single snapshot in excellent observing conditions, with no companion detected, albeit at a limiting magnitude difference of only $\sim3$\,mag. 

\textbf{8 Lac A} (HR\,8603, HD\,214168, HIP\,111546) is the brighter component of a pair of stars separated by $\sim22.5$\,arcsec \citep{2023yCat....102026M}. 8~Lac~A itself is a binary star with components resolved by speckle interferometry for the first time in 1985 \citep{2023yCat....102026M}. The astrometric orbital solution for the 8~Lac~A binary listed in the WDS is `reliable' with a high eccentricity of $0.83\pm0.07$, a period of $33.1\pm3.6$\,yr, semimajor axis of $49.1\pm6.7$\,mas, inclination of $62.8\pm4.6^{\circ}$, and $\Delta V$ between the components of $\sim0.6$\,mag. 8\,Lac\,A was also identified as a Be + sdO binary candidate based on the analysis of 20 IUE spectra \citep[][incorrectly listed there as 8~Lac~B, which is not known as a Be star]{2018ApJ...853..156W}. However, the presence of the sdO companion was not confirmed in three more recent HST/STIS exposures \citep{2021AJ....161..248W}.

In our one CHARA snapshot, we clearly detect the 8~Lac~A speckle binary in both $H$ and $K$ with $\rho = 36.7$\,mas, PA$=143^{\circ}$, and $f=68$\%. After analytically removing the companion signal from the CP data with CANDID, a second search did not result in a detection of the possible inner sdO companion (Table~\ref{tab:detlims}). Thus, 8~Lac~A presently remains only a candidate Be + sdO binary.

\subsection{SB1 binaries}

\textbf{28 Tau} (Pleione, HR\,1180, HD\,23862, HIP\,17851) has been presented as a single-lined binary with a period of 218 days and an eccentricity of 0.6 or even more \citep{1996PASJ...48..317K, 2010A&A...516A..80N}, making it rather unusual among close Be star binaries, which typically have circular or nearly circular orbits. Furthermore, the disk and orbital planes are strongly misaligned in Pleione, which seems to result in periodic disk tearing events due to the tidal influence of the companion \citep{2022ApJ...928..145M}. These facts strongly disfavor the origin of Pleione in a mass transferring binary, as that should result in a (nearly) circular orbit and aligned planes of the orbit and the disk \citep[e.g.,][]{2010ApJ...724..546S}.

Evidence for a wider companion separated by $\gtrsim 0.2$\,arcsec was presented in speckle observations and later in adaptive optics imaging, although in all cases the quality of the detection did not enable constraining the flux ratio or orbital motion \citep{2023yCat....102026M}. However, the wide companion was also undetected on several occasions, so that it is probably close to the detection limit of the single-telescope methods and also possibly variable (with the primary Be star being variable as well).

The four $H$-band and the three $K$-band interferometric snapshots show a very small CP signal suggestive of a small-scale asymmetry, which the CANDID grid search interprets as a low-confidence detection of a point-source companion. However, we are unable to reconcile the data time-series with an orbiting companion due to the following reasons. First, the derived position of the companion disagrees between the $H$-band and $K$-band data, with the $K$-band data resulting in larger separations. This could be indicative of an asymmetry in the circumstellar disk of the Be star, as the $K$-band emitting region is larger than the $H$-band one. Second, the PA of the detected asymmetry at each epoch coincides with the PA of the major axis of the inclined Be disk, as implied by polarimetric data, modeling of the disk long-term evolution \citep{2022ApJ...928..145M}, as well as by the VIS2 data presented here (see below). Third, $f_H$ and $f_K$ differ by up to a factor of two in the individual snapshots, while it would be expected to remain roughly constant. Thus, although the orbital phase coverage of our data is not ideal - one snapshot from ${\rm RJD}=59567.704$, and three snapshots between ${\rm RJD}=59817$ and $59821$, i.e., a time difference of $\sim250$\,d, which differs by only 32 days from the published orbital period - we conclude that the asymmmetry originates from within the circumstellar disk and not the spectroscopic companion, and we use the data to derive a lower limit on the magnitude difference between the two components (Table~\ref{tab:detlims}). 

The circumstellar disk of the Be star is partially resolved in the CHARA data, and we use the best-quality dataset from ${\rm RJD}=59567.704$ (with the others showing small miscalibrations of the visibilities and overall noisier data) to fit a 2D Gaussian to determine the basic disk parameters. As expected, the disk is larger in the $K$ band with ${\rm FWHM} = 0.37\pm0.01$\,mas compared to $0.29\pm0.01$\,mas in the $H$ band. In the $H$ band, $i_{\rm disk}$ is $67.2\pm1.2^{\circ}$ and the PA is $163.3\pm0.8^{\circ}$, while in the $K$ band the result is $59.5\pm0.7^{\circ}$ and $157.6\pm0.6^{\circ}$, respectively. This is in excellent agreement with the published result based on geometrical fitting of old CHARA data \citep{2013ApJ...768..128T}.

\textbf{$\zeta$ Tau} (HR\,1910, HD\,37202, HIP\,26451) is a Be-shell star, which was extensively studied due its strong $V/R$ variability, which is a manifestation of a one-armed density wave in the disk with a nonperiodic timescale of 1400--1430 days \citep{2009A&A...504..929S,2009A&A...504..915C}. $\zeta$ Tau is an SB1 binary with a period of 133 days and a circular orbit \citep{2009A&A...506.1319R}, and it presents an SED turndown, indicative of disk truncation \citep{2019ApJ...885..147K}. $\zeta$ Tau has also been recently reported as a candidate for a source of $\gamma$~Cas-like X-rays \citep{2022MNRAS.516.3366N}, making it a potential Be + WD binary candidate. The companion was not detected with the NPOI interferometer \citep{2021ApJS..257...69H}.

The large circumstellar disk seen close to edge-on orientation is resolved in the CHARA $H$-band VIS2 data. In fact, $\zeta$~Tau is nearby enough so that even the stellar photosphere should be partially resolved. We fitted a two-component model composed of (1) a UD with a fixed diameter of 0.40\,mas, corresponding to the estiamate of the size of the photosphere \citep{2007ApJ...654..527G}, and (2) a 2D Gaussian representing the disk. This resulted in the following best-fit paramaters: ${\rm FWHM} = 1.77\pm0.03$\,mas, $i_{\rm disk} = 100.6\pm0.6^{\circ}$, and a ${\rm PA} = 122.1\pm0.5^{\circ}$, and a disk flux contribution to the total $H$-band flux of $54.1\pm1.5$\%. The geometrical parameters are in a good agreement with those previously derived from broad-band polarimetry and Br$\gamma$ AMBER spectro-interferometry \citep{2009A&A...504..915C}, as well as those derived from previous CHARA observations \citep{2010AJ....140.1838S,2013ApJ...768..128T}.

The spectroscopic companion was not detected in our one $H$ band snapshot, even though a CP signal indicating a small scale asymmetry is present in the data. However, similar to 28~Tau, this signal most probably comes from an asymmetry within the circumstellar disk, as indicated by the resulting $\rho\sim0.5$\,mas, i.e., very close to the stellar surface, and PA $\sim96^{\circ}$, i.e., close to the disk major axis. An asymmetry in the disk akin to the one presented here was also detected in older CHARA data \citep{2010AJ....140.1838S}. For comparison, the expected size of the binary orbit is $\sim8.5$\,mas, with the smallest separation at conjunction being $\sim1.5$\,mas if the disk and orbital inclinations are roughly equal. Thus, we conclude that in all likelihood, no signal from the spectroscopic companion is present in our data, and we derive the limiting magnitude difference between the components after subtracting the asymmetry signal from the data (Table~\ref{tab:detlims}).

\textbf{$\beta$ CMi} (HR\,2845, HD\,58715, HIP\,36188) was suggested as a possible binary based on an SED turndown \citep{2015A&A...584A..85K, 2017A&A...601A..74K}. Spectroscopic detection of the companion was reported by \citet{2017ApJ...836..112D} from periodic motions of the emission wings in H$\alpha$ ($P=170.4$\,d), although the result was later disputed by \citet{2019ApJ...875...13H}. Recently, \citet{2023Galax..11...83M} detected a clear periodicity in the H$\alpha$ $V/R$ variations with $P=182.5$\,d, which could correspond to the binary orbital period. $\beta$ CMi is the most nearby Be star in the sample with an estimated size of the orbit of $\sim19$\,mas, and the inclination is intermediate \citep{2015A&A...584A..85K}, so the companion should be detectable at all orbital phases if bright enough. Still, the suspected spectroscopic companion is undetected in the two CHARA snapshots, and we can only provide limiting magnitudes for the suspected companion (Table~\ref{tab:detlims}). We also inspected archival PIONIER data taken on three different nights, in which the CP is also consistent with zero, although the data SNR is lower than in the newer CHARA data. Likewise, no companion was detected in NPOI data \citep{2021ApJS..257...69H}.

The disk of $\beta$ CMi is resolved in the VIS2 data. For the estimation of basic disk parameters, we use the better quality $H$-band dataset from ${\rm RJD}=59185.989$. Fitting a 2D Gaussian fully reproduces the data with the following parameters: ${\rm FWHM}=0.51\pm0.02$\,mas, $i=41\pm1^{\circ}$, and ${\rm PA}=134.5\pm0.5^{\circ}$ (Fig.~\ref{fig:betCMi_mircx}). These are in reasonable agreement with the results from radiative transfer modeling of VLTI/AMBER data \citep{2015A&A...584A..85K}, as well as from geometrical joint modeling of the same VLTI/AMBER data together with older observations from CHARA/MIRC \citep{2012ApJ...744...19K}.

\textbf{4 Her} (HR\,5938, HD\,142926, HIP\,77986) is a Be-shell star that was reported as an SB1 binary with $P=46.18$\,d based on RV variations of the shell line absorption, and $V/R$ variations of \Ha\ were found to be phase-locked with the orbit \citep{1997A&A...328..551K, 2006A&A...459..137R}. We were not able to obtain a reliable detection of the spectroscopic companion in our two CHARA snapshots, which were taken on two subsequent nights. According to the linear ephemeris given by \citet{1997A&A...328..551K}, 4~Her was observed by CHARA at orbital phase $\sim$0.4, i.e., roughly between quadrature and conjunction. Still, given the estimated size of the orbit of 2.5\,mas (Table~\ref{tab:SB1_binaries}), the separation between the components should have been large enough to be resolved. An upper limit on the brightness of the companion was derived from the interferometric data (Table~\ref{tab:detlims}).

\textbf{88 Her} (z\,Her, HR\,6664, HD\,162732, HIP\,87280) is another Be-shell star with an SB1 period of 86.72\,days \citep{1988A&AS...75..311D,1974A&A....33..117H}. According to the ephemeris given by \citet{1988A&AS...75..311D}, 88~Her was observed by CHARA (one time) at phase $\sim0.15$, where a phase of 0.25 corresponds to the epoch of maximum RV. Thus, the CHARA dataset was taken in between quadrature and conjunction, and as the estimated size of the orbit $a$ is $\sim2$\,mas, the separation of the components should have been $\sim1$\,mas, which is well within the angular resolution of our data. The magnitude limits on the companion given in Table~\ref{tab:detlims} should therefore give a reliable upper limit on the true brightness of the companion. 

\textbf{V1294~Aql} (HD\,184279, HIP\,96196) was recently reported as an SB1 binary with a low-mass companion \citep{2022A&A...666A.136H}. This Be star was included in our sample at the end of the CHARA observing campaign, despite its distance being $\sim1.4$\,kpc, which means that the angular resolution of our data might be too low to resolve the two components if the size of the orbit is at the lower end of the typical interval for close Be binaries. Indeed, the companion remains undetected in our data, although the orbital phase closer to quadrature indicates that the angular separation should have been sufficient to be resolved. Recently, V1294~Aql was also reported to have a wider speckle companion separated by $\sim0.24$" with a magnitude difference of $\sim3.3$\,mag at $\lambda=824$\,nm \citep{2023yCat....102026M}.

\textbf{$\epsilon$ Cap} (HR\,8260, HD\,205637, HIP\,106723) is a Be-shell star and an SB1 binary with a period of 128.5 days, as reported by \citet{2006A&A...459..137R} based on RVs measured from the absorption cores of the H$\delta$ line, and recently confirmed by \citet{2023Galax..11...83M} from multiple Balmer lines. The latter study also presented evidence for orbital phase-locked $V/R$ variability in Balmer lines. According to the ephemeris, the one CHARA dataset was taken at phase 0.86, corresponding to an epoch between quadrature and conjunction, meaning that the binary separation should have been sufficient to enable resolving it. However, the companion was probably too faint to be detected. In the four archival GRAVITY snapshots, the companion likewise remains clearly undetected (Table~\ref{tab:detlims}). Lastly, we also inspected archival PIONIER data taken on three separate nights in 2014, but these did not bring any further insights other than that the CP is also consistent with zero, but with larger error bars than in the CHARA or GRAVITY data.

\subsection{$\gamma$~Cas-like binaries}

\textbf{$\gamma$ Cas} (HR\,264, HD\,5394, HIP\,4427) is an SB1 binary with a circular orbit and a period of 203.52\,d \citep{2012A&A...537A..59N}. It also has a strong SED turndown, but the overall SED structure suggests that the disk might in fact extend beyond the binary orbit, with the circumbinary part being visible only in the radio \citep{2017A&A...601A..74K}. According to the mass estimates based on the spectroscopic orbital solution and Hipparcos parallax, the angular size of the orbit is $\sim9.7$\,mas (Table~\ref{tab:gamCas_binaries}). Together with the intermediate inclination of the circumstellar disk plane \citep[e.g.,][]{2012A&A...545A..59S}, this implies that the companion should be well within the comfort zone of being detected by CHARA observations at all orbital phases if it is bright enough. No companion was detected in the survey by the NPOI \citep{2021ApJS..257...69H}. 

The companion remains undetected in our two extended CHARA/MIRC-X snapshots, the first of which was taken with a higher spectral resolution of $R\sim190$, which enables probing angular separations up to $\sim 200$\,mas. From the $R\sim190$ snapshot, we were also able to obtain a particularly tight limit on the minimum $\Delta H$ between the components of $\sim6.7$\,mag (Table~\ref{tab:detlims}). This limit enables ruling out an sdOB or an MS companion with the estimated spectroscopic mass of $\sim1$\,{\Msun} considering a realistic mass range for the primary Be star of 10--20\,{\Msun} (see Fig.~\ref{fig:masses}), although the typical literature estimates are $<15$\,{\Msun} \citep[][]{2012A&A...537A..59N}. Thus, we are left with a rather massive WD as the only viable companion of $\gamma$~Cas (see Sect.~\ref{sec:discussion}). Several additional archival CHARA/MIRC snapshots were analyzed by Labadie-Bartz et al. (in prep.), also resulting in no companion detection.

The Be star and its disk are resolved in the VIS2 data, which allows us to constrain the disk geometry. For this, we used a two-component model consisting of the stellar photosphere represented by a tilted UD (seen under the same inclination as the disk) with a diameter of $0.48$\,mas \citep{2012A&A...545A..59S}, and the circumstellar disk represented by a 2D Gaussian. Using our two CHARA $H$-band snapshots, the resulting parameters are ${\rm FWHM} = 1.10\pm0.02$\,mas, $i_{\rm disk} = 44.6\pm1.9${\degree}, and ${\rm PA} = 24.0\pm2.0${\degree}, which is in good agreement with the interferometric fitting results of \citet{2013ApJ...768..128T}.

\textbf{FR CMa} (HR\,2284, HD\,44458, HIP\,30214) is an SB1 binary candidate \citep{2022MNRAS.510.2286N}. We do not detect any sign of a companion in our CHARA snapshot (limits in Table~\ref{tab:detlims}).

\textbf{HR 2370} (HD\,45995, HIP\,31066) is an SB1 binary with a likely orbital period of $\sim100$\,d \citep{2022MNRAS.510.2286N}. In our one CHARA snapshot, the CP signal is fully consistent with zero and no companion is detected.

\textbf{V558 Lyr} (HR\,7403, HD\,183362, HIP\,95673) was reported as an SB1 binary ($P\sim83.3$\,d) based on RVs measured in several hydrogen and helium lines \citep{2022MNRAS.510.2286N}. In our CHARA snapshot, the spectroscopic companion remains undetected (Table~\ref{tab:detlims}).

\textbf{V782 Cas} (HD\,12882, HIP\,9997) was also reported to be an SB1 system with a primary Be star by \citet{2022MNRAS.510.2286N} with a period of $\sim122.0$\,d. Furthermore, a previously detected photometric variability with a period of $2.5132$\,d \citep{2017AJ....153..252L} was found in \ion{He}{1} line RVs, and was interpreted as originating from a close binary V782~Cas~B, making the SB1 binary with the Be star the A component of V782~Cas, which should therefore be a quadruple system.

With our three interferometric snapshots in $H$ and $K$ band, we were able to resolve the wide V782~Cas~AB binary, which is consistent with the quadruple scenario. The A and B components are separated by $\sim29.5$\,mas, with the fainter component contributing $\sim23$\% of the total flux in the $H$ band, and $\sim18$\% of the total flux in the $K$ band. This is in rough agreement with the approximate flux ratios derived by \citet{2022MNRAS.510.2286N} in the visible, in which the B component contributes about 43\% of the total light. The somewhat higher contrast in the near-IR compared to the visible can be explained by the higher near-IR contribution of the Be star disk in the brighter A component, and possibly also by the brighter component of the B binary being of an earlier spectral type than the Be star in the A binary. The same reasoning can be used to explain the difference between the contrast in the $H$ and $K$ bands. With our observations spanning only 78~days, we were not able to convincingly detect an orbital motion of the AB binary, as the derived positions are consistent within the respective error bars. 

We did not detect any signature from the invisible SB1 companion of the Be star after subtracting the strong signal from the wide and low-contrast AB binary. The detection limits for this SB1 companion are listed in Table~\ref{tab:detlims}.

\textbf{$\pi$ Aqr} (HR\,8539, HD\,212571, HIP\,110672) was found to emit $\gamma$~Cas-like X-rays by \citet{2017A&A...602L...5N}. It is the rare case of a Be star for which spectroscopic evidence was presented in favor of the presence of a close MS companion \citep[with an orbital period of $\sim84$\,d;][]{2002ApJ...573..812B}. However, the value of the velocity amplitude of the Be star was not confirmed in more recent studies, with the latter reporting an amplitude lower by about a factor of two \citep{2019A&A...632A..23N, 2023PASJ...75..177T}. This then favors a lower-mass companion, compatible with a WD or sdOB nature, and there is no need to invoke a higher-mass MS companion.

The RV curve for the companion was originally measured from a narrow moving emission peak in H$\alpha$ when the circumstellar disk was otherwise almost completely dissipated \citep{2002ApJ...573..812B}. If the emission peak indeed originates from the vicinity of the companion, its RVs should be reliable, as opposed to the Be-star RVs derived from the broad photospheric absorption, which turned out to be overestimated. A possible interpretation of the moving emission peak is the presence of residual circum-companion gas, possibly a disk, formed by accretion of the outer Be-star disk material. This is similar to the interpretation of \citet{2002ApJ...573..812B}, who suggested that the companion is a late B-type star `surrounded by a gaseous envelope' (motivated by the overestimated companion mass of $2$ to $3$\,{\Msun}). If the companion is instead a WD, the presence of an accretion disk around it would be in line with the suggestion that the $\gamma$~Cas-like X-rays originate from accretion onto WD companions \citep{2023ApJ...942L...6G}. The fact that the emission from this possible disk shows up specifically in H$\alpha$ could be explained by the WD being cooler than the sdO companions of $\varphi$~Per and 59~Cyg, where a circum-companion emission is observed in \ion{He}{2}\,$\lambda 4686$ \citep{1981PASP...93..297P,2000ASPC..214..581R}, as well as the companion of HD~55606, which has a circum-companion accretion disk emitting in \ion{He}{1} lines \citep[][see also the case of HR~2142 in Sect.~\ref{sec:orbits_new}]{2018ApJ...865...76C}. We note that \citet{2023ApJ...942L...6G} offered a different plausible interpretation for the moving emission peak, specifically that it was caused by a hot WD companion irradiating a residual outer part of the Be disk. 

A weak CP signal is observed in our three CHARA $H$-band snapshots spanning 18~days, but we are unable to reconcile it with an orbiting close companion or with an orbital phase-locked feature in the disk, as the changes in $\rho$ ($\sim0.6$\,mas) and PA ($\sim 62${\degree}) between the snapshots are too small given the orbital period of $84.1$\,d \citep[the CHARA data span orbital phases 0.11--0.32 according to the ephemeris given by][see Table~\ref{tab:detlims}]{2002ApJ...573..812B}. We also inspected two archival GRAVITY snapshots, as well as archival PIONIER data taken on four separate nights, but in all of these the measured CP is consistent with zero, and therefore there is no signature of a companion or any disk asymmetry.

Inspecting H$\alpha$ and other Balmer line profiles in the public BeSS spectra taken close to the epochs of the interferometric observations reveals that the emission was significantly smaller and more symmetric at the time of the PIONIER and GRAVITY observations (2014--2017) than at the time of the CHARA measurements (2020 June). Indeed, the strong $V>R$ asymmetry seems to have persisted at least throughout the latter half of 2020, although no spectra are available between 2020 February and July. Spectra taken in 2020 January also show a $V>R$ asymmetry, but weaker than from 2020 July onward.

We suggest that the observations can be interpreted as a signature of a global density wave, which propagates through the Be-star disk on a significantly longer time scale than the orbital period. This would explain the lack of any asymmetry in the interferometric data taken in 2014--2017, when the line emission was smaller and more symmetric, as well as the seemingly static asymmetry observed in the 20202 CHARA data, when there was a clear and persistent asymmetry in the emission lines as well.

Thus, for the derivation of the lower limits on the magnitude difference between the binary components (Table~\ref{tab:detlims}), we subtracted the CP signal, as we concluded that it is not coming from the companion. Similar to $\gamma$~Cas, the final magnitude limits (taken close to quadrature) are tight enough to rule out both sdO and MS companions that would have a mass close to the one implied by the spectroscopic solutions, i.e., $\sim1$\,{\Msun}, for a reasonable mass range of the primary Be star between 9 and 15\,{\Msun} \citep{2023PASJ...75..177T}. Therefore, it appears that similar to $\gamma$~Cas, a WD nature of the companion is the only remaining option (see Sect.~\ref{sec:discussion}). 

The Be disk is partially resolved, and fitting the visibilities simultaneously from all three epochs results in the following parameters: FWHM$=0.43\pm0.01$\,mas, $i=60.8\pm0.6${\degree}, and PA$=76.1\pm0.5${\degree}. The inclination is in rough agreement with that of $\sim45${\degree} derived from IR spectroscopy \citep{2023Galax..11...90C}.

\subsection{Candidate binaries}

\textbf{$\psi$ Per} (HR\,1087, HD\,22192) is a Be-shell star with a clear SED turndown. The radio part of the SED also suggests a possible circumbinarity of the disk, with a possible companion carving a gap through it \citep[similar to other sample stars, namely $\eta$~Tau, EW~Lac, and $\gamma$~Cas,][]{2017A&A...601A..74K}. Recently, \cite{2023Galax..11...83M} studied the weak $V/R$ variability in H$\alpha$ and suggested a possible period of $179.6$\,d. The results from the NPOI interferometric survey \citep{2021ApJS..257...69H} are compatible with $\psi$~Per being a single star at the corresponding contrast limits (magnitude difference of 3.0 to 3.5 mag in the visible).

In our CHARA dataset consisting of two $H$ band snapshots, the closure phase does not show clear deviations from 0{\degree}, indicating absence of any asymmetry. The circumstellar disk is partially resolved and the data are consistent with an inclined Gaussian with FWHM$=0.45\pm0.02$\,mas, $i_{\rm disk}=60.9\pm0.8${\degree}, and PA$=135.8\pm2.6${\degree}. The intermediate inclination is however at odds with the shell nature of $\psi$~Per, and $i_{\rm disk}$ might thus be underestimated due to the assumption of a geometrically thin disk. The study of \citet{2013ApJ...768..128T} resulted in a higher inclination of $75${\degree}, which is more consistent with the observed shell spectrum.

\textbf{$\eta$ Tau} (Alcyone, HR\,1165, HD\,23630) is a Pleiades Be star with an SED turndown, which also hints at extended, possibly circumbinary structure, if a disk truncation by a companion is indeed responsible for the turndown \citep{2019ApJ...885..147K}. However, no sign of a companion could be identified in our one CHARA $H$-band snapshot or in the previously published NPOI data \citep{2021ApJS..257...69H}. In the visibility data, the circumstellar disk is almost fully resolved and consistent with an inclined Gaussian with the following parameters: FWHM$=0.49\pm0.01$\,mas, $i_{\rm disk}=34.4\pm0.4${\degree}, and PA$=101.2\pm0.6${\degree}. 

\textbf{48 Per} (c\,Per, HR\,1273, HD\,25940, HIP\,19343) was previously suggested as a spectroscopic binary by \citet[][$P=16.9$\,d]{1971PASJ...23..159K} and \citet[][$P=16.6$\,d]{1989MNRAS.238.1085J}, but this status remains unconfirmed (although the latter study strangely lists 48~Per in their Table~1 as a `confirmed binary'). It also presents an SED turndown. No companion was reported based on NPOI interferometric observations \citep{2021ApJS..257...69H}. In our CHARA snapshot, we see no evidence for a close companion, but a small CP signal is reconciled with a previously undetected wide companion at $\rho \sim 152$\,mas, which is well outside the interferometric FoV (here $\sim 91$\,mas). The resulting $f_H$ is $1.74\pm0.07$\%, although this value should be treated with caution due to the large separation. The companion and its orbital motion were also detected in archival CHARA data, ruling out a chance alignment, but the orbital analysis for this wide component is reserved for an upcoming study (in prep.). The Be star disk is resolved in VIS2, and the best-fit 2D Gaussian parameters are FWHM$=0.36\pm0.01$\,mas, $i_{\rm disk}=28.5\pm0.5${\degree}, and PA$=107.7\pm1.1${\degree}. The low inclination was indeed expected from the observed single-peaked H$\alpha$ emission, and the parameters agree well with those given by \citet{2013ApJ...768..128T}.

\textbf{11 Cam} (HR\,1622, HD\,32343) is another bright northern Be star with a detected SED turndown. Despite the rather strong H$\alpha$ emission, the disk is only marginally resolved with a FWHM of $\sim0.25$\,mas and an unconstrained $i_{\rm disk}$ and PA. As for 48~Per, H$\alpha$ emission is single-peaked, so the inclination is probably also very low.

In our two interferometric snapshots in both $H$ and $K$ bands, we detect a signature of a possible companion, but the results derived from the two bands are not entirely compatible with each other (Table~\ref{tab:detections}). For the first epoch, the positions of the possible companion in the two bands are compatible within $\sim5\sigma$, with a separation of $\sim2.2$\,mas, PA of $\sim168^{\circ}$, and a flux ratio of $\sim2$\% of the primary. However, for the second epoch, the derived positions in the two bands differ significantly (although the companion lies in both cases in the general north direction and is separated by 2.3--2.7\,mas from the primary), while the flux ratio is still $\sim2$\% in the $H$ band, as opposed to $\sim1$\% in the $K$ band. Fixing the flux ratio in the $K$ band to 2\% does not reconcile the results, neither does an exploration of secondary minima resulting from the grid search. The most likely culprit for this discrepancy is a miscalibration issue due to varying atmospheric conditions at the second epoch, which can actually be clearly seen in the VIS2 data. Nevertheless, the companion detection seems to be real, although more observations are needed for a confirmation and for a first estimate of the orbit. As the companion moved by about $180^{\circ}$ in PA during 13 months, the orbital period could be $\sim3.5$, $\sim5$, $\sim9$, or $\sim26$~months. 

\textbf{105 Tau} (HR\,1660, HD\,32991) is another Be star with an SED turndown, but there is no sign of a companion in our two CHARA $H$ band snapshots. The Be disk is partially resolved, with a Gaussian fit to the better-quality dataset from ${\rm RJD}= 59182.822$ resulting in the following parameters: FWHM$=0.24\pm0.01$\,mas, $i_{\rm disk}=36.8\pm1.2${\degree}, and PA$=187.4\pm2.4${\degree}. The low inclination is in agreement with the rough estimate that can be made based on the appearance of emission lines.

\textbf{25 Ori} (HR\,1789, HD\,35439) has a complex H$\alpha$ profile indicating possible disk truncation by an orbiting companion. It was also the first Be star studied for its $V/R$ variability \citep{1936ApJ....84..180D}. Similar to the case of $\zeta$~Tau and $\pi$~Aqr, we detect an asymmetric signal in the CP, but it likely comes from a global density wave in the disk. This is because the implied separation between the two possible stellar components is very small at $<1.0$\,mas, and the two positions and flux ratios differ between the $H$ band and $K$ band measurements. The limiting magnitude differences were thus derived after subtracting the detected asymmetry signal (Table~\ref{tab:detlims}).

The disk is partially resolved in our CHARA snapshot, and the best-fit geometric parameters are FWHM$=0.33\pm0.01$\,mas, $i_{\rm disk}=56.9\pm0.9${\degree}, and PA$=26.5\pm0.6${\degree} in the $H$ band, and FWHM$=0.40\pm0.01$\,mas, $i_{\rm disk}=51.8\pm0.8${\degree}, and PA$=27.3\pm0.6${\degree} in the $K$ band. The intermediate inclination is indeed what would be expected from the morphology of the double-peaked emission lines.

\textbf{$\theta$ CrB} (HR\,5778, HD\,138749) is a nearby Be star with only a weak emission component in the wings of H$\alpha$, which is dominated by photospheric absorption. We observed $\theta$~CrB in both $H$ and $K$ on two occasions, with no asymmetrical signal detected in CP. The star with its weak disk is only marginally resolved with FWHM$=0.15\pm0.02$\,mas in the $H$ band, and FWHM=$0.17\pm0.02$\,mas in the $K$ band for the ${\rm RJD}= 59732.751$ dataset, while the VIS2 data from ${\rm RJD}= 59567.084$ are unusable due to calibration issues. $\theta$ CrB also has a wide speckle companion of a similar brightness, with a `preliminary' astrometric orbit available in the WDS with $P = 361.1\pm76.8$\,yr, $a" = 0.826\pm0.118"$, and $e=0.780\pm0.054$ \citep{2023yCat....102026M}.

\textbf{$o$ Her} (HR\,6779, HD\,166014) was observed by CHARA on a single occasion, revealing a completely flat CP signal at 0{\degree}. No companion was detected in NPOI data either \citep{2021ApJS..257...69H}. The disk is partially resolved and fitting an inclined Gaussian fully reproduces the visibility data with FWHM$=0.42\pm0.01$\,mas, $i_{\rm disk}=38.5\pm0.7${\degree}, and PA$=74.9\pm1.0${\degree}. The H$\alpha$ emission is weak and also narrow, which is in agreement with the rather low inclination indicated by the interferometry.

\textbf{12 Vul} (HR\,7565, HD\,187811) is the only classical Be star with tentative evidence for molecular emission, which could indicate the presence of a cool evolved companion \citep{2021A&A...647A.164C}. In our three $H$-band CHARA snapshots, spanning about 13 days, no sign of a companion was detected, with the CP being fully consistent with zero. 

\textbf{$o$ Aqr} (HR\,8402, HD\,209409) is a nearby Be-shell star with an SED turndown. In our one CHARA $H$-band snapshot, the visibilities are rather noisy, and fitting an inclined Gaussian results in unconstrained inclination and PA, but the FWHM is $\sim0.35$\,mas. No companion was detected in the CHARA data as well as in four archival GRAVITY snapshots (limits in Table~\ref{tab:detlims}).

\textbf{EW Lac} (HR\,8731, HD\,217050) is an SED-turndown Be-shell star which had a very weak H$\alpha$ emission at the time of our observation. The star with its disk remains unresolved in our one CHARA snapshot, and no signature of a companion was detected. 

\textbf{$\beta$ Psc} (HR\,8773, HD\,217891) is another Be star with a detected SED turndown, but no other sign of a companion, including a null result from NPOI interferometry \citep{2021ApJS..257...69H}. The strong emission in H$\alpha$ is narrow and single-peaked. The Be star with its disk is marginally resolved in the $H$-band VIS2 data, however, the resulting Gaussian FWHM of $\sim0.35$\,mas has to be taken with caution, as the data are rather noisy and subject to miscalibration at the short baselines. The CP data from our one CHARA snapshot and three archival GRAVITY snapshots do not indicate any asymmetry or a companion (magnitude limits are given in Table~\ref{tab:detlims}).

\section{Discussion \& Conclusions}
\label{sec:discussion}

We have presented a major part of the results from a two-year long interferometric program with the CHARA Array aimed at the detection of suspected faint binary companions and mapping the binary orbits among classical Be stars. The results include:
\begin{itemize}
\item updated 3D orbital solutions and (preliminary) dynamical masses for the \textit{Be + sdOB binaries} 28~Cyg, V2119~Cyg, and 60~Cyg,
\item the first 3D orbital solutions and (preliminary) dynamical masses for the \textit{Be + sdOB binaries} HR~2142, HD~161306, and 7~Vul,
\item marginal detections of faint and close companions to V780~Cas (\textit{Be + sdOB binary}) and 11~Cam (\textit{Candidate binary}),
\item the first (spectro-)interferometric detection of line emission from circum-companion gas in the case of the \textit{Be + sdOB binary} HR~2142,
\item stringent lower limits on the magnitude differences between the remaining 29 sample Be stars and their undetected (suspected) companions,
\item no companions detected in the \textit{$\gamma$~Cas-like binaries} subgroup, and MS companions ruled out for $\gamma$~Cas and $\pi$~Aqr, leaving WD companions as the most likely option.
\end{itemize}

3D orbital solutions and (preliminary) dynamical masses are now available for a total of eight Be + sdOB binaries (including the previously published results for $\varphi$~Per and $\kappa$~Dra). The sizes of the orbits range from $\sim1.8$ to $\sim7.4$\,mas. The amount and precision of the literature RVs are the main limiting factors in our solutions. This is the case especially for 28~Cyg, where we had to fix the $\gamma$-velocity in the solution, as we were unable to constrain both $\gamma$ and $K_{\rm sdO}$ due to the poor RV phase coverage. 

The newly derived dynamical masses of the primary Be stars are in good agreement with the masses of normal B-type dwarfs according to the spectral types \citep{2013ApJS..208....9P}, which is a significant preliminary result given the different formation histories of these two groups. However, a more stringent comparison would be desirable in the future, as the Be-star spectral types are affected by gravity darkening and the presence of the disk. The eight now-available dynamical mass estimates for the sdOB companions range from $\sim0.4$\,{\Msun} for $\kappa$~Dra to $\sim1.45$\,{\Msun} for 60~Cyg. In our sample, there is thus no strong candidate for a stripped supernova progenitor \citep[the considered limit being $>1.6$\,{\Msun} following][]{2019ApJ...878...49W}, which would lead to the formation of a Be X-ray binary with NS companion or a runaway Be star. Instead, the Be stars in these systems will likely be the first to evolve to the next stage, filling their Roche lobes, and initiating further stages of mass transfer between the components. 

The binary formation channel of classical Be stars is clearly not limited to early-type Be stars, but extends at least to mid-type Be stars, as exemplified by the case of the B6\,III star $\kappa$~Dra \citep{2022ApJ...940...86K} and the B5\,V star 7~Vul (this work). For $\kappa$~Dra, the companion is a cooler sdB star, and this could be the case also for 7~Vul, given that the masses for both components in 7~Vul are just slightly higher than in $\kappa$~Dra. As the FUV cross-correlation searches for stripped companions have been performed almost exclusively with very hot ($45$\,kK) model spectra, it would be worthwhile to extend them to include cooler templates, especially for mid- or late- type Be stars. On the other hand, the contamination of the sdB signal by the Be star with a similar temperature might pose a significant obstacle to this method.

\begin{figure}
\epsscale{1.2}
\plotone{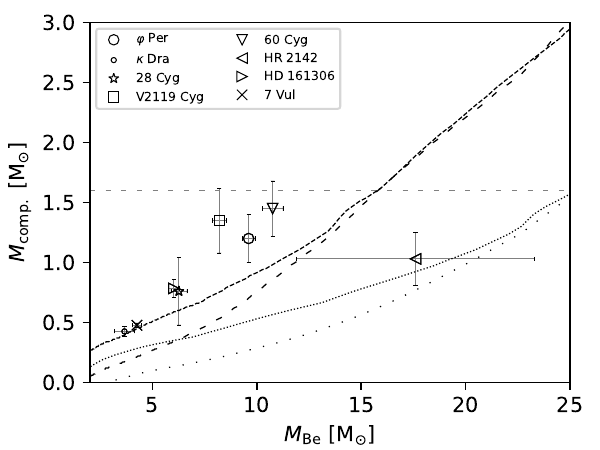}
\caption{Dynamical masses for the components of Be + sdOB binaries. The horizontal dashed grey line shows the limit of $1.6$\,{\Msun}, above which the stripped stars could explode in core-collapse supernovae. The dashed and dotted black curves indicate an estimate of the CHARA detection limits for $\Delta H = 5.0$ and $\Delta H = 6.4$, respectively, where for a given Be star mass, companions with masses above the corresponding curve should have been detected (see text). The loosely dashed and dotted lines are for MS companions, and the denser ones are for sdOB companions. \label{fig:masses}}
\end{figure}

We investigated a possible correlation between the masses of the primary Be stars and the sdOB companions. The eight cases (including 28~Cyg with its mass ratio fixed at 0.12) are plotted in Fig.~\ref{fig:masses}, indeed showing a correlation for the six well-constrained masses, while the one clear outlier from the trend is HR~2142, which is a special case in other aspects as well (see below). Fig.~\ref{fig:masses} also contains estimates for the $H$-band interferometric detectability of sdOB and MS companions to Be stars in terms of their masses, which were determined using absolute $H$-band magnitudes from stellar evolutionary models. For the magnitudes of Be stars and the possible MS companions, we used a MESA MIST\footnote{\url{https://waps.cfa.harvard.edu/MIST/}} model \citep{2016ApJS..222....8D,2016ApJ...823..102C} based on an $5$\,Myr isochrone, while for the sdOB companions, the magnitudes were obtained from the models of \citet{2018A&A...615A..78G} at the middle of the He core-burning lifetime. Then, the minimum mass of a detectable MS or sdOB companion was found for a grid of Be-star masses and for a given magnitude difference $\Delta H$ between the components. The two values of $\Delta H$ were chosen to represent the best-performance CHARA limit of $6.4$\,mag, and the more typically achieved limit of $5.0$\,mag.

\begin{figure}
\epsscale{1.2}
\plotone{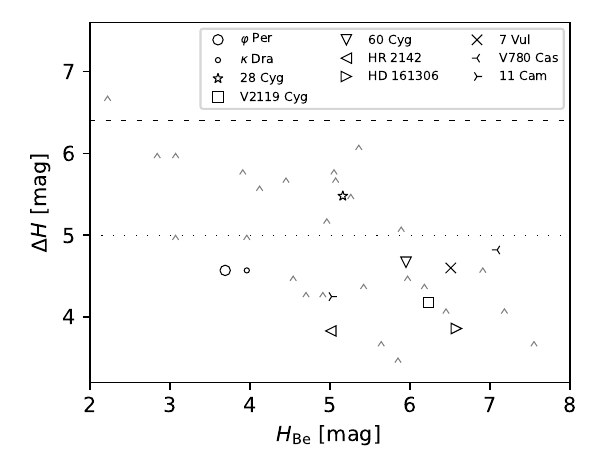}
\caption{$H$-band magnitude differences for detected companions (specified in the legend) or their lower limits for non-detections averaged for $\rho<25$\,mas (caret symbols) vs. the apparent $H$-band magnitude of the Be stars. The dashed and dotted horizontal lines show the best-performance limit of the CHARA Array of $\Delta H = 6.4$\,mag, and the typical limit for average conditions of $\Delta H = 5.0$\,mag, respectively. \label{fig:mag_limits}}
\end{figure}

In our subsample of 12 \textit{Be + sdOB binaries}, we failed to detect the companions for five stars, but this should not be considered as a conclusive result, as the observational challenges posed by these targets prevented the CHARA Array from attaining its best performance. The challenges included (1) being below the angular resolution limit (59~Cyg), (2) being fainter than the limit for average atmospheric conditions (V1150~Tau, QY~Gem, V2162~Cyg), (3) the presence of a bright wider companion contaminating the data (59~Cyg, 8~Lac~A), and (4) only a single snapshot being available (all five). It is therefore worthwhile to continue with the CHARA Array interferometric monitoring of nearby Be + sdOB candidates specifically, although stars fainter than $H \sim K \sim 7$\,mag require excellent weather conditions. 

Be star companions of the sdOB type are common but seem to be far from ubiquitous. The very low companion detection rate among the \textit{SB1 binaries}, \textit{$\gamma$~Cas-like binaries}, and \textit{Candidate binaries} subgroups indicates that the companions might be of a different type, although it should be mentioned that the distribution of spectral types is very different within the three subgroups, and the sample stars might thus not be fully representative of them. Among the \textit{SB1 binaries}, the only detected companion is a cooler sdB companion to the mid-type Be star $\kappa$~Dra, and faint sdB stars or possibly WDs could be the companions for the other mid- to late-type Be stars in this subgroup as well (28~Tau, $\beta$~CMi, 4~Her, and 88~Her). As for the early-type stars in this subgroup, we should again mention that $\zeta$~Tau is also a candidate $\gamma$~Cas-like X-ray source \citep{2022MNRAS.516.3366N}, while V1294~Aql might be too distant ($\sim 1.4$\,kpc) for the binary to be resolved. 

It is significant that we did not detect a single companion among the six \textit{$\gamma$~Cas-like binaries}, five of which are confirmed SB1 binaries (six when including $\zeta$~Tau), and one is a strong SB1 binary candidate \citep{2022MNRAS.510.2286N}. Be stars with $\gamma$~Cas-like X-rays are possible Be + WD binary systems, although it remains a debated issue whether the characteristic X-rays have in fact anything to do with accretion on a companion \citep{2016AdSpR..58..782S}. If accretion onto a WD is not responsible for the X-rays, at least some of these stars should have detectable sdOB or MS companions \citep{2023ApJ...942L...6G}. For $\gamma$~Cas and $\pi$~Aqr, we obtained tight limits which enable ruling out MS companions, given the reasonably constrained companion masses ($\sim1${\Msun}) from the SB1 spectroscopic orbits and inclination estimates based on emission line morphologies. This can also be seen in Fig.~\ref{fig:masses}, where for instance for a massive Be star with a mass of $15${\Msun} and for $\Delta H =6.4$, we should be able to detect an sdOB companion down to $M_{\rm sdO} \sim 0.75$, and an MS companion down to $M_{\rm MS} \sim 0.55$. Thus, at least for $\gamma$~Cas and $\pi$~Aqr, the WD nature of the companion is likely the only viable option, as the expected masses appear too low for a NS companion, which would be the only other remaining possibility.

Regarding the \textit{Candidate binaries} subgroup and specifically those identified from an SED turndown, it could be that these are genuinely single and the turndown is caused by a different effect \citep[see][]{2022A&A...664A.185C}, although the similarity between the SED shapes of confirmed close binaries such as $\gamma$~Cas to those of the seemingly single Be stars like $\psi$~Per, $\eta$~Tau, and EW~Lac remains striking \citep{2017A&A...601A..74K,2019ApJ...885..147K}. Lastly, the Be stars in these three subgroups have not yet been found to present any evidence for the outer disk regions being heated by a hot companion.

Generally, the derived magnitude limits for non-detections (Table~\ref{tab:detlims}) are not tight enough to rule out low-mass MS companions, although higher-mass MS companions should have in many cases been detected. In Fig.~\ref{fig:mag_limits}, we show $\Delta H$ for the detected sdOB companions, and the lower limits on $\Delta H$ for the non-detections (excluding 59~Cyg and V1294~Aql, where the separation between the binary components might be too low). As can be seen, the CHARA best-performance limit was (nearly) achieved only for the brighter Be stars in our sample. Inspecting the detection limits for the expected masses of these Be stars in Fig.~\ref{fig:masses} can then give us an estimate on the minimum mass of an MS (or sdOB) companion that should have been detected. For instance, for a Be star mass of $10$\,{\Msun}, we should have detected MS companions with masses down to $0.3$ and $0.7$\,{\Msun}, respectively, for $\Delta H = 6.4$\,mag and $\Delta H = 5.0$\,mag. The fact that there are still no confirmed close MS companions remains a significant result for classical Be stars as a class \citep[cf.][]{2020A&A...641A..42B}.

For HR~2142, we presented the first interferometric detection of line emission components coming from the vicinity of the sdO companion, indicating a circum-companion structure such as a disk accreting material from the outer parts of the Be-star disk. The presence of circum-companion gas would offer an explanation for the apparently attenuated FUV flux from the hot companion, whose detectability also appears to be dependent on the orbital phase. Unfortunately, the low spectral-resolution CHARA data presented here do not have the capability of detecting such structures, which was possible only in the high spectral-resolution GRAVITY data covering the Br$\gamma$ and \ion{He}{1}\,$\lambda20587$. The GRAVITY data also enabled the measurement of RVs associated with the Be star as well as the circum-companion gas structure, revealing an excellent agreement with the RVs predicted by the orbital solution. We note that HR~2142 seems to be the only Be star with obvious obscuration of the hot companion (as the case of 28~Cyg appears more consistent with a varying quality of the high-cadence FUV spectra), and its configuration thus deserves to be studied in further detail.


In conclusion, the formation of Be stars in mass-transferring binaries is clearly common and possibly dominant. The prospects are good for finding more examples of post-mass transfer Be stars, namely by looking for the typical features in high-resolution visible spectra, extending the FUV searches to cooler companions, and continuing with interferometric monitoring of suitable candidates. In fact, an ongoing program of a similar scope is currently in progress with VLTI/GRAVITY, which is expected to double the number of Be + sdOB binaries with 3D orbital solutions and dynamical masses. While the binary fraction among Be stars remains unconstrained \citep[all of the sample stars with non-detections could have invisible (pre-)WD companions, cf.\ the case of Regulus,][]{2020ApJ...902...25G}, the first robust calibration of (asteroseismologically-guided) evolutionary models should soon be possible.

\section*{Acknowledgments}
Based on observations collected at the European Southern Observatory under ESO programmes 093.C-0503(A), 093.D-0571(A), 093.D-0571(B), 096.D-0803(A), 097.D-0533(A), 099.D-2009(C), 099.D-2009(D), 110.23Z7.001.
This research has made use of the Washington Double Star Catalog maintained at the U.S. Naval Observatory.
This work is based upon observations obtained with the Georgia State University Center for High Angular Resolution Astronomy Array at Mount Wilson Observatory.  The CHARA Array is supported by the National Science Foundation under Grant No. AST-1636624 and AST-2034336.  Institutional support has been provided from the GSU College of Arts and Sciences and the GSU Office of the Vice President for Research and Economic Development. 
This work has made use of the BeSS database, operated at LESIA, Observatoire de Meudon, France: \url{http://basebe.obspm.fr}. 
This research has made use of the Jean-Marie Mariotti Center Aspro, SearchCal, and MOIO AMHRA service at \url{https://amhra.jmmc.fr/services}. This research has made use of the Jean-Marie Mariotti Center OiDB service available at \url{http://oidb.jmmc.fr}. 
This work has made use of data from the European Space Agency (ESA) mission
{\it Gaia} (\url{https://www.cosmos.esa.int/gaia}), processed by the {\it Gaia}
Data Processing and Analysis Consortium (DPAC,
\url{https://www.cosmos.esa.int/web/gaia/dpac/consortium}). Funding for the DPAC
has been provided by national institutions, in particular the institutions
participating in the {\it Gaia} Multilateral Agreement.
We warmly thank Anthony Meilland for sharing his reduced AMBER observations of $\alpha$~Col and for answering our questions about the Keplerian disk model.
SK acknowledges funding for MIRC-X received from the European Research Council (ERC) under the European Union's Horizon 2020 research and innovation programme (Starting Grant No. 639889 and Consolidated Grant No. 101003096). JDM acknowledges funding for the development of MIRC-X (NASA-XRP NNX16AD43G, NSF-AST 1909165) and MYSTIC (NSF-ATI 1506540, NSF-AST 1909165).


\facilities{CHARA,VLTI,IUE,HIPPARCOS,Gaia}


\appendix
\restartappendixnumbering

\section{Detailed results from the analysis of interferometric data}

Table~\ref{tab:detections} gives all epochs with companion detections. All of the CHARA data that led to detections were obtained with the default $R=50$. In Table~\ref{tab:detlims}, we list all epochs without companion detection, with the lower limits on the magnitude difference averaged for different ranges of the angular separation $\rho$. Values in the column for $\rho<100$ are given for datasets taken with higher spectral resolution of $\sim100$ (12 epochs) and $\sim190$ (one epoch). Finally, in Table~\ref{tab:interf_calibrators}, we list the $H$-band angular diameters for all of the calibrators used for the observations presented here.

\startlongtable
\begin{deluxetable}{lLCCCCCCCc}
\tablecaption{Interferometric detections of sdOB companions. \label{tab:detections}}
\tablewidth{0pt}
\tablehead{
\colhead{Star} & \colhead{HJD} & \colhead{$\rho$} & \colhead{PA} & \colhead{$\sigma$-$a$} & \colhead{$\sigma$-$b$} & \colhead{$\sigma$-PA} & \colhead{$f$} & \colhead{Band} & \colhead{Calibrator}\\ 
\nocolhead{Star} & \colhead{-2,400,000} & \colhead{[mas]} & \colhead{[{\degree}]} & \colhead{[mas]} & \colhead{[mas]} & \colhead{[{\degree}]} & \colhead{[\% primary]}  & \nocolhead{n$\sigma$} & \nocolhead{Band}
}
\startdata
\hline
V780~Cas & 59777.957 & 1.122 & 113.491 & 0.056 & 0.027 & 176.8 & 1.18\pm0.14 & H & HD\,8992, SAO\,11910 \\
         & 59777.957 & 1.056 & 112.948 & 0.124 & 0.111 & 138.4 & 1.26\pm0.25 & K & HD\,8992, SAO\,11910 \\
11~Cam & 59480.905 & 2.227 & 167.347 & 0.020 & 0.014 & 34.5 & 1.80\pm0.10 & H & HD\,30166 \\
       & 59480.905 & 2.229 & 169.909 & 0.024 & 0.014 & 177.1 & 2.24\pm0.06 & K & HD\,30166 \\
       & 59817.025 & 2.266 & 4.411 & 0.017 & 0.013 & 157.8 & 2.09\pm0.05 & H & HD\,237280 \\
       & 59817.025 & 2.685 & 320.544 & 0.033 & 0.024 & 85.5 & 1.11\pm0.04 & K & HD\,237280 \\
HR~2142 & 59481.015 & 1.950 & 208.710 & 0.013 & 0.011 & 110.3 & 2.92\pm0.05 & H & HD\,45855 \\ 
        & 59536.024 & 0.831 & 56.086 & 0.023 & 0.031 & 67.6 & 2.94\tablenotemark{a} & H & HD\,45855 \\ 
        & 59567.776 & 1.740 & 215.452 & 0.025 & 0.010 & 122.5 & 2.76\pm0.07 & H & HD\,38072, HD\,46737 \\ 
        & 59855.974 & 1.216 & 46.016 & 0.020 & 0.011 & 105.8 & 3.31\pm0.09 & H & HD\,37236, HD\,46737\\
        & 59481.015 & 1.935 & 209.274 & 0.024 & 0.011 & 137.9 & 3.18\pm0.09 & K & HD\,45855 \\
        & 59567.776 & 1.724 & 217.256 & 0.018 & 0.011 & 129.8 & 4.44\pm0.09 & K  & HD\,38072, HD\,46737\\
        & 59855.973 & 1.108 & 42.099 & 0.043 & 0.018 & 51.4 & 3.96\pm0.27 & K & HD\,37236, HD\,46737\\
        & 59878.947 & 1.705 & 204.351 & 0.030 & 0.014 & 133.3 & 3.24\pm0.08 & K & HD\,37236, HD\,50115\\
        & 59932.760\tablenotemark{b} & 1.571 & 39.446 & 0.019 & 0.011 & 71.1 & 3.49\tablenotemark{a} & K & HD\,41644\\ 
        & 59997.530\tablenotemark{b} & 1.468 & 21.348 & 0.016 & 0.010 & 87.8 & 3.49\tablenotemark{a} & K & HD\,41644\\
HD~161306 & 59725.830 & 1.721 & 201.265 & 0.054 & 0.022 & 115.2 & 2.94\pm0.25 & H & HD\,145225 \\
          & 59732.854 & 1.758 & 227.091 & 0.094 & 0.019 & 124.3 & 2.36\pm0.51 & H & HD\,155662 \\
          & 59763.708 & 1.425 & 340.211 & 0.195 & 0.022 & 107.8 & 2.98\pm0.40 & H & HD\,155662 \\
          & 59725.830 & 1.728 & 201.143 & 0.082 & 0.028 & 120.1 & 1.98\pm0.10 & K & HD\,145225 \\
          & 59732.854 & 1.759 & 223.903 & 0.049 & 0.022 & 112.6 & 2.10\pm0.16 & K & HD\,155662 \\
          & 59745.825 & 1.674 & 266.030 & 0.091 & 0.032 & 97.1 & 2.14\pm0.25 & K & HD\,169252\\
          & 59777.716 & 1.566 & 31.054 & 0.204 & 0.106 & 57.7 & 2.41\pm0.70 & K & HD\,156971, HD\,166161 \\
          & 59820.677 & 1.720 & 189.675 & 0.097 & 0.063 & 38.5 & 2.27\pm0.17 & K & HD\,156971\\
7~Vul & 59725.385 & 2.048 & 149.876 & 0.104 & 0.045 & 153.0 & 1.42\pm0.12 & H & HD\,230476 \\
      & 59732.431 & 1.690 & 145.200 & 0.078 & 0.016 & 9.0 & 1.35\pm0.26 & H & HD\,351160 \\
      & 59745.369 & <0.5 & - & - & - & - & - & H & HD\,186962\\
      & 59763.349 & 2.005 & 327.894 & 0.042 & 0.016 & 155.8 & 1.47\pm0.11 & H & HD\,230337 \\
      & 59777.308 & <0.5 & - & - & - & - & - & H & HD\,230337 \\
      & 59725.386 & 2.037 & 155.736 & 0.042 & 0.017 & 30.0 & 3.17\pm0.67 & K & HD\,230476 \\
      & 59732.432 & 1.691 & 146.892 & 0.038 & 0.022 & 66.3 & 1.73\pm0.18 & K & HD\,351160 \\
      & 59745.369 & <0.7 & - & - & - & - & - & K & HD\,186962 \\
      & 59763.349 & 1.928 & 328.215 & 0.067 & 0.045 & 31.0 & 1.56\pm0.33 & K & HD\,230337 \\
      & 59777.308 & <0.7 & - & - & - & - & - & K & HD\,230337 \\
      & 59820.230 & 1.431 & 336.032 & 0.085 & 0.043 & 99.1 & 1.80\pm0.37 & K & HD\,230337 \\
28~Cyg & 59364.956 & 4.152 & 199.530 & 0.038 & 0.030 & 142.7 & 1.13\pm0.09 & H & HD\,188256\\ 
       & 59398.850 & 6.270 & 165.203 & 0.063 & 0.044 & 142.6 & 0.85\pm0.10 & H & HD\,188365  \\ 
       & 59399.842 & 6.401 & 164.649 & 0.042 & 0.037 & 162.8 & 0.74\pm0.04 & H & HD\,188365, HD\,196881  \\ 
       & 59480.634 & 5.739 & 120.031 & 0.038 & 0.033 & 90.5  & 1.01\pm0.06 & H & HD\,188256, HD\,204171 \\ 
       & 59725.930 & 4.237 & 195.516 & 0.034 & 0.028 & 154.9 & 0.64\pm0.05 & H & HD\,188256 \\ 
       & 59732.903 & 4.756 & 188.311 & 0.030 & 0.027 & 151.8 & 0.78\pm0.04 & H & HD\,188365 \\ 
       & 59745.915 & 5.468 & 175.353 & 0.079 & 0.037 & 0.9   & 0.71\pm0.06 & H & HD\,195647  \\ 
       & 59878.646 & 3.565 & 66.517  & 0.058 & 0.032 & 147.2 & 1.05\pm0.17 & H & HD\,188365  \\ 
       & 59725.930 & 4.349 & 195.929 & 0.035 & 0.027 & 122.0 & 0.70\pm0.04 & K & HD\,188256 \\ 
       & 59732.903 & 4.745 & 187.623 & 0.033 & 0.031 & 114.1 & 0.74\pm0.04 & K & HD\,188365 \\ 
       & 59745.915 & 5.618 & 174.757 & 0.051 & 0.031 & 105.7 & 0.78\pm0.06 & K & HD\,195647 \\ 
       & 59820.779 & 6.938 & 131.436 & 0.068 & 0.044 & 16.6 & 0.68\pm0.06 & K & HD\,188256 \\ 
V2119~Cyg & 59397.917 & 1.207 & 190.307 & 0.031 & 0.009 & 164.1 & 2.21\pm0.13 & H & HD\,188256, HD\,196881  \\
          & 59399.943 & 1.197 & 210.013 & 0.021 & 0.013 & 10.7 & 2.20\pm0.14 & H & HD\,188256, HD\,196215 \\
          & 59439.846 & 1.568 & 85.991 & 0.017 & 0.014 & 64.9 & 2.01\pm0.11 & H & HD\,195647 \\
          & 59479.753 & 1.783 & 300.491 & 0.031 & 0.016 & 15.2 & 2.23\pm0.09 & H & HD\,196215, HD\,188256 \\
          & 59745.932 & 1.196 & 16.264 & 0.030 & 0.012 & 153.9 & 1.83\pm0.18 & H & HD\,195647 \\
          & 59479.754 & 1.823 & 301.663 & 0.038 & 0.017 & 47.4 & 2.54\pm0.15 & K & HD\,196215, HD\,188256 \\        
          & 59745.932 & 1.097 & 20.709 & 0.130 & 0.032 & 122.4 & 2.08\pm0.12 & K & HD\,195647 \\
60~Cyg & 59364.996 & 2.387 & 191.538 & 0.028 & 0.016 & 151.2 & 2.49\pm0.32& H & HD\,196881 \\
       & 59397.972 & 3.413 & 197.346 & 0.023 & 0.022 & 151.2 & 1.30\pm0.05 & H & HD\,196881, HD204876 \\
       & 59399.926 & 3.332 & 198.096 & 0.019 & 0.018 & 139.3 & 1.40\pm0.04 & H & HD\,196881, HD\,188256 \\
       & 59439.898 & 0.815 & 0.689 & 0.033 & 0.018 & 154.8 & 1.27\pm0.06 & H & HD\,204876 \\
       & 59479.728 & 1.227 & 29.013 & 0.019 & 0.012 & 170.4 & 1.59\pm0.10 & H & HD\,188256, HD\,204876 \\
       & 59725.977 & 0.442 & 284.872 & 0.155 & 0.071 & 175.1 & 2.3\pm2.3 & H & HD\,204171 \\
       & 59745.973 & 2.178 & 14.428 & 0.105 & 0.018 & 132.3 & 2.09\pm0.51 & H & HD\,204171 \\
       & 59479.731 & 1.246 & 26.045 & 0.027 & 0.020 & 30.6 & 1.43\pm0.06 & K &  HD\,188256, HD\,204876\\
       & 59745.973 & 2.228 & 14.102 & 0.031 & 0.016 & 114.1 & 1.44\pm0.06 & K & HD\,204171 \\
\enddata
\tablecomments{$\rho$ is the angular separation between the components, PA is the position angle (from North to East), $\sigma$-$a$ and $\sigma$-$b$ are the major and minor axes of the error ellipse, respectively, $\sigma$-PA is the position angle of the error ellipse (from North to East), and Band gives the near-IR observing band.}
\tablenotetext{a}{Fixed at the average value determined from the other better-quality detections in the same band (see text).}
\tablenotetext{b}{VLTI/GRAVITY data.}
\end{deluxetable}

\startlongtable
\begin{deluxetable}{llCCCCCCc}
\tablecaption{Limiting magnitude differences for non-detections\label{tab:detlims}}
\tablewidth{0pt}
\tablehead{
\colhead{Target} & \colhead{HJD} & \colhead{Phase\tablenotemark{a}} & \multicolumn{3}{c}{$\Delta H_\mathrm{lim}$ [mag]} & \multicolumn{2}{c}{$\Delta K_\mathrm{lim}$ [mag]} & \colhead{Calibrator}\\
\nocolhead{Target} & \colhead{- 2,400,000} & \nocolhead{Orb. phase} & \colhead{$\rho<25$\,mas} & \colhead{$\rho<50$\,mas} & \colhead{$\rho<100$\,mas} & \colhead{$\rho<30$\,mas} & \colhead{$\rho<60$\,mas} & \nocolhead{Calibrator}
}
\startdata
V780~Cas    & 59815.013 & \nodata & 3.9\pm0.2 & 3.9\pm0.1 & \nodata & 4.0\pm0.2 & 4.0\pm0.2 & HD\,1166 \\ 
            & 59820.856 & \nodata & 4.9\pm0.2 & 4.9\pm0.1 & \nodata & 4.6\pm0.2 & 4.5\pm0.1 & HD\,13453 \\ 
            & 59855.796 & \nodata & 4.4\pm0.2 & 4.4\pm0.1 & \nodata & 3.5\pm0.2 & 3.5\pm0.1 & HD\,15665 \\ 
            & 59878.786 & \nodata & 3.8\pm0.2 & 3.7\pm0.1 & \nodata & 3.5\pm0.2 & 3.5\pm0.1 & HD\,8992, HD\,236982 \\ 
V1150~Tau   & 59855.887 & \nodata & 3.7\pm0.3 & 3.8\pm0.2 & \nodata & 3.3\pm0.3 & 3.3\pm0.2 & HD\,26937, HD\,32525 \\ 
QY~Gem      & 59856.024 & \nodata & 4.1\pm0.2 & 4.1\pm0.1 & \nodata & 3.7\pm0.2 & 3.7\pm0.1 & HD\,59202\\ 
59~Cyg      & 59480.746 & 0.70    & 4.9\pm0.1 & 4.7\pm0.2 & \nodata & 5.1\pm0.2 & 5.0\pm0.1 & HD\,204482, HD\,209236 \\ 
V2162~Cyg   & 59878.712 & \nodata & 2.9\pm0.1 & 2.9\pm0.1 & \nodata & 2.8\pm0.2 & 2.7\pm0.2 & HD\,196881 \\ 
8~Lac~A     & 59480.841 & \nodata & 3.5\pm0.1 & 3.3\pm0.2 & \nodata & 3.0\pm0.2 & 2.9\pm0.1 & HD\,209236 \\ 
\hline
28~Tau       & 59567.704 & 0.27 & 5.7\pm0.1 & 5.7\pm0.1 & \nodata & 5.4\pm0.2 & 5.4\pm0.1 & HD\,20680 \\ 
             & 59816.988 & 0.41 & 4.7\pm0.3 & 4.8\pm0.2 & \nodata & \nodata & \nodata & HD\,20680 \\ 
             & 59818.994 & 0.42 & 5.2\pm0.2 & 5.2\pm0.2 & \nodata & 5.3\pm0.2 & 5.3\pm0.1 & HD\,29835 \\ 
             & 59820.975 & 0.43 & 5.1\pm0.2 & 5.1\pm0.1 & \nodata & 5.4\pm0.2 & 5.4\pm0.1 & HD\,22107 \\ 
$\zeta$~Tau  & 59185.870 & 0.19 & 5.0\pm0.2 & 5.2\pm0.2 & 5.2\pm0.1 & \nodata & \nodata & HD\,40981, HD\,33172 \\ 
$\beta$~CMi  & 59185.989 & 0.41 & 6.0\pm0.2 & 6.2\pm0.2 & 6.1\pm0.1 & \nodata & \nodata & HD\,53358, HD\,67873 \\ 
             & 59566.869 & 0.65 & 4.5\pm0.1 & 4.5\pm0.1 & \nodata & 5.2\pm0.2 & 5.1\pm0.1 & HD\,65525, HD\,64685 \\ 
4~Her        & 59398.749 & 0.31 & 5.1\pm0.1 & 5.0\pm0.1 & \nodata & \nodata & \nodata & HD\,139440\\ 
             & 59399.714 & 0.33 & 5.1\pm0.2 & 5.0\pm0.1 & \nodata & \nodata & \nodata & HD\,137870, HD\,158414\\ 
88~Her       & 59399.772 & 0.90 & 4.6\pm0.1 & 4.5\pm0.1 & \nodata & \nodata & \nodata & HD\,158414, HD\,188365 \\ 
V1294~Aql    & 59763.799 & 0.86 & 4.0\pm0.2 & 4.0\pm0.1 & \nodata & 4.1\pm0.2 & 4.1\pm0.1 & HD\,178282 \\ 
$\epsilon$~Cap & 57960.790\tablenotemark{b} & 0.69 & \nodata & \nodata & \nodata & 3.7\pm0.1 & 4.2\pm0.1 & HD\,205903 \\ 
             & 57962.835\tablenotemark{b} & 0.70 & \nodata & \nodata & \nodata & 4.5\pm0.1 & 5.0\pm0.1 & HD\,205903\\ 
             & 57990.732\tablenotemark{b} & 0.92 & \nodata & \nodata & \nodata & 4.3\pm0.1 & 4.8\pm0.1 & HD\,205903\\ 
             & 57994.758\tablenotemark{b} & 0.95 & \nodata & \nodata & \nodata & 4.7\pm0.1 & 5.1\pm0.1 & HD\,205903\\ 
             & 59185.620 & 0.22 & 4.3\pm0.3 & 4.5\pm0.2 & 4.5\pm0.1 & \nodata & \nodata & HD\,201013, HD\,209925 \\ 
\hline
$\gamma$~Cas  & 59400.947 & 0.21 & 6.7\pm0.1 & 6.8\pm0.1 & 6.7\pm0.1 & \nodata & \nodata & HD\,3250 \\
              & 59567.603 & 0.03 & 5.6\pm0.2 & 5.6\pm0.1 & \nodata & 5.7\pm0.1 & 5.7\pm0.1 & HD\,8906 \\ 
FR~CMa        & 59878.970 & \nodata & 3.7\pm1.0 & 3.8\pm0.5 & \nodata & 4.9\pm0.3 & 5.0\pm0.2 & HD\,50115 \\ 
HR~2370       & 59879.019 & 0.14 & 4.5\pm0.2 & 4.4\pm0.2 & \nodata & 5.0\pm0.3 & 5.0\pm0.2 & HD\,52767 \\ 
V558~Lyr      & 59745.784 & 0.78 & 4.4\pm0.3 & 4.3\pm0.2 & \nodata & 4.3\pm0.2 & 4.3\pm0.1 & SAO\,67647 \\ 
V782~Cas~A    & 59778.008 & 0.44 & 4.1\pm0.2 & 4.0\pm0.2 & \nodata & 4.5\pm0.2 & 4.5\pm0.1 & SAO\,11910 \\ 
              & 59820.903 & 0.79 & 3.2\pm0.2 & 3.2\pm0.1 & \nodata & 4.4\pm0.2 & 4.4\pm0.1 & HD\,13543, HD\,14519 \\ 
              & 59855.768 & 0.07 & 3.9\pm0.3 & 3.9\pm0.2 & \nodata & 3.6\pm0.3 & 3.8\pm0.2 & HD\,9296, HD\,15665 \\  
$\pi$~Aqr     & 57959.811\tablenotemark{b} & 0.59 & \nodata & \nodata & \nodata & 4.8\pm0.1 & 5.0\pm0.1 & HD\,212778 \\ 
              & 57962.800\tablenotemark{b} & 0.63 & \nodata & \nodata & \nodata & 4.3\pm0.1 & 4.7\pm0.1 & HD\,212778 \\ 
              & 59012.970 & 0.11 & 5.4\pm0.2 & 5.3\pm0.1 & \nodata & \nodata & \nodata & HD\,209237, HD\,216701 \\ 
              & 59023.967 & 0.24 & 6.1\pm0.2 & 6.1\pm0.1 & \nodata & \nodata & \nodata & HD\,209237, HD\,216701 \\ 
              & 59030.988 & 0.32 & 5.0\pm0.2 & 4.9\pm0.2 & \nodata & \nodata & \nodata & HD\,209237 \\ 
\hline
$\psi$~Per    & 59184.768 & \nodata & 5.6\pm0.2 & 5.7\pm0.1 & 5.6\pm0.1 & \nodata & \nodata & HD\,19111 \\  
              & 59439.989 & \nodata & 5.5\pm0.2 & 5.5\pm0.1 & \nodata & \nodata & \nodata & HD\,18768, HD\,25030 \\ 
$\eta$~Tau    & 59182.772 & \nodata & 6.0\pm0.2 & 6.2\pm0.2 & 6.2\pm0.1 & \nodata & \nodata & HD\,20457, HD\,27370  \\ 
48~Per        & 59185.784 & \nodata & 5.8\pm0.1 & 5.8\pm0.1 & 5.7\pm0.2 & \nodata & \nodata & HD\,23178 \\ 
105~Tau       & 59182.822 & \nodata & 5.2\pm0.2 & 5.3\pm0.1 & 5.3\pm0.1 & \nodata & \nodata & HD\,27370, HD\,245711 \\ 
              & 59821.029 & \nodata & 4.8\pm0.2 & 4.7\pm0.1 & \nodata & 5.4\pm0.2 & 5.4\pm0.1 & HD\,22107 \\ 
25~Ori        & 59480.981 & \nodata & 4.4\pm0.3 & 4.5\pm0.2 & \nodata & 4.5\pm0.2 & 4.6\pm0.2 & HD\,38225, HD\,45855  \\ 
$\theta$~CrB  & 59567.084 & \nodata & 4.5\pm0.2 & 4.5\pm0.2 & \nodata & 5.6\pm0.2 & 5.5\pm0.1 & HD\,134129 \\ 
              & 59732.751 & \nodata & 5.7\pm0.2 & 5.7\pm0.1 & \nodata & 6.0\pm0.3 & 6.0\pm0.2 & HD\,145225 \\ 
$o$~Her       & 59354.845 & \nodata & 5.0\pm0.2 & 5.0\pm0.1 & 4.9\pm0.1 & \nodata & \nodata & HD\,168914, HD\,172132 \\ 
12~Vul        & 59354.963 & \nodata & 4.2\pm0.2 & 4.2\pm0.1 & 4.1\pm0.2 & \nodata & \nodata & HD\,183418, HD\,192712 \\ 
              & 59364.408 & \nodata & 5.8\pm0.2 & 5.7\pm0.1 & \nodata & \nodata & \nodata & HD\,183418 \\ 
              & 59366.918 & \nodata & 5.2\pm0.3 & 5.2\pm0.2 & \nodata & \nodata & \nodata & HD\,189235, HD\,344262 \\ 
$o$~Aqr       & 57959.776\tablenotemark{b} & \nodata & \nodata & \nodata & \nodata & 4.5\pm0.1 & 4.9\pm0.1 & HD\,209778 \\
              & 57983.769\tablenotemark{b} & \nodata & \nodata & \nodata & \nodata & 4.1\pm0.1 & 4.6\pm0.1 & HD\,209778  \\
              & 57993.773\tablenotemark{b} & \nodata & \nodata & \nodata & \nodata & 5.1\pm0.1 & 5.6\pm0.1 & HD\,209778 \\
              & 57998.745\tablenotemark{b} & \nodata & \nodata & \nodata & \nodata & 4.2\pm0.1 & 4.6\pm0.1 & HD\,209778 \\
              & 59184.627 & \nodata & 4.3\pm0.3 & 4.5\pm0.2 & 4.4\pm0.1 & \nodata & \nodata & HD\,204391, HD\,213276 \\ 
EW~Lac        & 59182.626 & \nodata & 5.5\pm0.2 & 5.6\pm0.1 & 5.4\pm0.2 & \nodata & \nodata & HD\,213176, HD\,220702 \\ 
$\beta$~Psc   & 57961.899\tablenotemark{b} & \nodata & \nodata & \nodata & \nodata & 4.4\pm0.1 & 4.7\pm0.1 & HD\,218441 \\
              & 57962.885\tablenotemark{b} & \nodata & \nodata & \nodata & \nodata & 4.6\pm0.1 & 5.1\pm0.1 & HD\,218441 \\
              & 57982.844\tablenotemark{b} & \nodata & \nodata & \nodata & \nodata & 3.9\pm0.1 & 4.4\pm0.1 & HD\,218441 \\
              & 59184.680 & \nodata & 4.5\pm0.3 & 4.7\pm0.2 & 4.6\pm0.1 & \nodata & \nodata & HD\,213276, HD\,221235 \\ 
\enddata
\tablenotetext{a}{Orbital phase equal to 0.0 corresponds to the inferior conjunction, i.e., when the companion is in front of the primary Be star, compatible with the RV plots such as Fig.~\ref{fig:orb_28Cyg}. Maximum RV offset thus occurs at phases equal to 0.25 and 0.75.}
\tablenotetext{b}{VLTI/GRAVITY data.}
\end{deluxetable}

\startlongtable
\begin{deluxetable}{lChh}
\tablecaption{Calibrators\label{tab:interf_calibrators}}
\tablewidth{0pt}
\tablehead{
\colhead{Target} & \colhead{UD diam. ($H$-band)} & \nocolhead{$m_H$} & \nocolhead{Spectral type} \\
\nocolhead{Target} & \colhead{[mas]} & \nocolhead{$m_H$} & \nocolhead{Spectral type}
}
\startdata
HD\,1166 & 0.426\pm0.010 & ... & ...      \\
HD\,3250 & 0.477\pm0.011 & ... & ...      \\
HD\,8906 & 0.420\pm0.012 & ... & ...      \\
HD\,8992 & 0.353\pm0.010 & ... & ...      \\
HD\,9296 & 0.302\pm0.007 & ... & ...      \\
HD\,13453 & 0.395\pm0.014 & ... & ...      \\
HD\,15665 & 0.399\pm0.009 & ... & ...      \\
HD\,18768 & 0.383\pm0.009 & ... & ...      \\
HD\,20680 & 0.530\pm0.013 & ... & ...      \\
HD\,22107 & 0.422\pm0.010 & ... & ...      \\
HD\,25030 & 0.500\pm0.013 & ... & ...      \\
HD\,26937 & 0.280\pm0.007 & ... & ...      \\
HD\,29835 & 0.555\pm0.013 & ... & ...      \\
HD\,30166 & 0.446\pm0.010 & ... & ...      \\
HD\,32525 & 0.274\pm0.007 & ... & ...      \\
HD\,33172 & 0.591\pm0.015 & ... & ...      \\
HD\,37236 & 0.374\pm0.011  & ... & ...      \\
HD\,38072 & 0.505\pm0.017 & ... & ...      \\
HD\,38225 & 0.434\pm0.013 & ... & ...      \\
HD\,40981 & 0.527\pm0.015 & ... & ...      \\
HD\,41644 & 0.582\pm0.020 & ... & ...      \\
HD\,45855 & 0.441\pm0.012 & ... & ...      \\
HD\,46737 & 0.490\pm0.013 & ... & ...      \\
HD\,50115 & 0.426\pm0.010 & ... & ...      \\
HD\,52767 & 0.430\pm0.011 & ... & ...      \\
HD\,53358 & 0.623\pm0.021  & ... & ...      \\
HD\,59202 & 0.320\pm0.008 & ... & ...      \\
HD\,65525 & 0.442\pm0.014 & ... & ...      \\
HD\,64685 & 0.402\pm0.011 & ... & ...      \\
HD\,67873 & 0.495\pm0.045 & ... & ...      \\
HD\,137870 & 0.272\pm0.007 & ... & ...      \\
HD\,139440 & 0.340\pm0.008 & ... & ...      \\
HD\,145225 & 0.362\pm0.008 & ... & ...      \\
HD\,155662 & 0.431\pm0.012 & ... & ...      \\
HD\,156971 & 0.283\pm0.008 & ... & ...      \\
HD\,158414 & 0.295\pm0.008 & ... & ...      \\
HD\,168914 & 0.483\pm0.030 & ... & ...      \\
HD\,166161 & 0.382\pm0.010 & ... & ...      \\
HD\,169252 & 0.389\pm0.010 & ... & ...      \\
HD\,172132 & 0.439\pm0.012 & ... & ...      \\
HD\,178282 & 0.321\pm0.007 & ... & ...      \\
HD\,183418 & 0.393\pm0.009 & ... & ...      \\
HD\,186962 & 0.417\pm0.013 & ... & ...      \\
HD\,188256 & 0.322\pm0.008 & ... & ...      \\
HD\,188365 & 0.327\pm0.008 & 5.78 & G5      \\
HD\,189235 & 0.172\pm0.005 & ... & ...      \\
HD\,192712 & 0.460\pm0.013 & ... & ...      \\
HD\,195647 & 0.471\pm0.011 & 5.17 & K0\,III \\
HD\,196215 & 0.415\pm0.009 & 5.45 & K0       \\
HD\,196881 & 0.306\pm0.007 & 5.98 & G5  \\
HD\,201013 & 0.600\pm0.020 & ... & ...      \\
HD\,204171 & 0.405\pm0.011 & ... & ...      \\
HD\,204482 & 0.450\pm0.011 & ... & ...      \\
HD\,204876 & 0.356\pm0.008 & 5.73 & G5     \\
HD\,205903 & 0.522\pm0.012 & & \\
HD\,209236 & 0.441\pm0.011 & ... & ...      \\
HD\,209778 & 0.582\pm0.014 & & \\
HD\,209925 & 0.514\pm0.017 & ... & ...      \\
HD\,212778 & 0.341\pm0.009 & & \\
HD\,218441 & 0.633\pm0.021 & & \\
HD\,230337 & 0.347\pm0.010 & ... & ...      \\
HD\,230476 & 0.360\pm0.009 & ... & ...      \\
HD\,236982 & 0.376\pm0.009 & ... & ...      \\
HD\,237280 & 0.431\pm0.010 & ... & ...      \\
HD\,344262 & 0.384\pm0.009 & ... & ...      \\
HD\,351160 & 0.389\pm0.010 & ... & ...      \\
SAO\,11910 & 0.353\pm0.009 & ... & ...      \\
SAO\,67647 & 0.325\pm0.007 & ... & ...      \\
\enddata
\end{deluxetable}

\section{Spectro-interferometric rotating disk model}
\label{sec:pmoired_keplerian}

\subsection{Description of the model}

PMOIRED model composition assumes that the geometry and the spectra of each component are independent: in other words, that the 3D flux can be written as $F(x,y,\lambda) = I(x,y)\times S(\lambda)$ where $x$, $y$ are the angular coordinate on sky, and $\lambda$ is the wavelength. $I$ is the achromatic image and $S$ the spectrum of the component. An emission line from a flat rotating disk cannot be represented accurately using a linear combination of a reasonable number of components which satisfy this decomposition. A dedicated rotating disk model has been therefore implemented, following the prescription of \cite{2011A&A...529A..87D}, and using the same parametrization described in Sect.\ 5.3 of \cite{2012A&A...538A.110M}. The disk has the following parameters:

\begin{itemize}

\item the disk's inclination and major axis position angle
\item the rest wavelength for each spectral line we want to model
\item the equivalent width of each line
\item the angular FWHM in the continuum and in the line(s) (we assume the disk has a Gaussian intensity profile) 
\item the inner diameter of the disk (this value must be set to the diameter of the inner star, which is modeled by another PMOIRED component, for instance a UD)
\item the total flux of the disk in the continuum
\item the Keplerian rotational velocity at the inner diameter (i.e., the surface equatorial velocity, since we assume the disk starts at the equator of the star)
\item the power law exponent of the rotation law $\beta$, fixed at 0.5 for a Keplerian disk

\end{itemize}

Note that because of the way PMOIRED combines components, the star is fully visible through the disk. Even if this will produce inaccuracies in the modelling of the visibilities, this would only be apparent at a high angular resolution we usually do not reach, as the inner star is much smaller than the disk.

\subsection{Validation}

We have performed two validation tests of the Keplerian model implementation. First, we obtained (A.~Meilland, private communication) the AMBER observations of $\alpha$~Col in the Brackett~$\gamma$ line analysed in \cite{2012A&A...538A.110M}. We find slightly different model parameters (although compatible within 1$\sigma$), which is likely due to the difference in the model optimization, as \cite{2012A&A...538A.110M} use a grid of models, whereas PMOIRED uses a gradient descent algorithm.

Secondly, we used the Analysis and Modelling at High Angular Resolution service\footnote{https://amhra.oca.eu/AMHRA/index.htm} (AMHRA) from the JMMC. This utility allows to create synthetic data cubes (images as a function of wavelength) based on the model from \cite{2011A&A...529A..87D}. With PMOIRED, we generate noiseless synthetic GRAVITY data from these images, and fit the data with PMOIRED's implementation of the rotating disk. We found a perfect match, both in the image and the complex visibility spaces. 

\section{Data plots}

\begin{figure*}
\epsscale{1.1}
\plotone{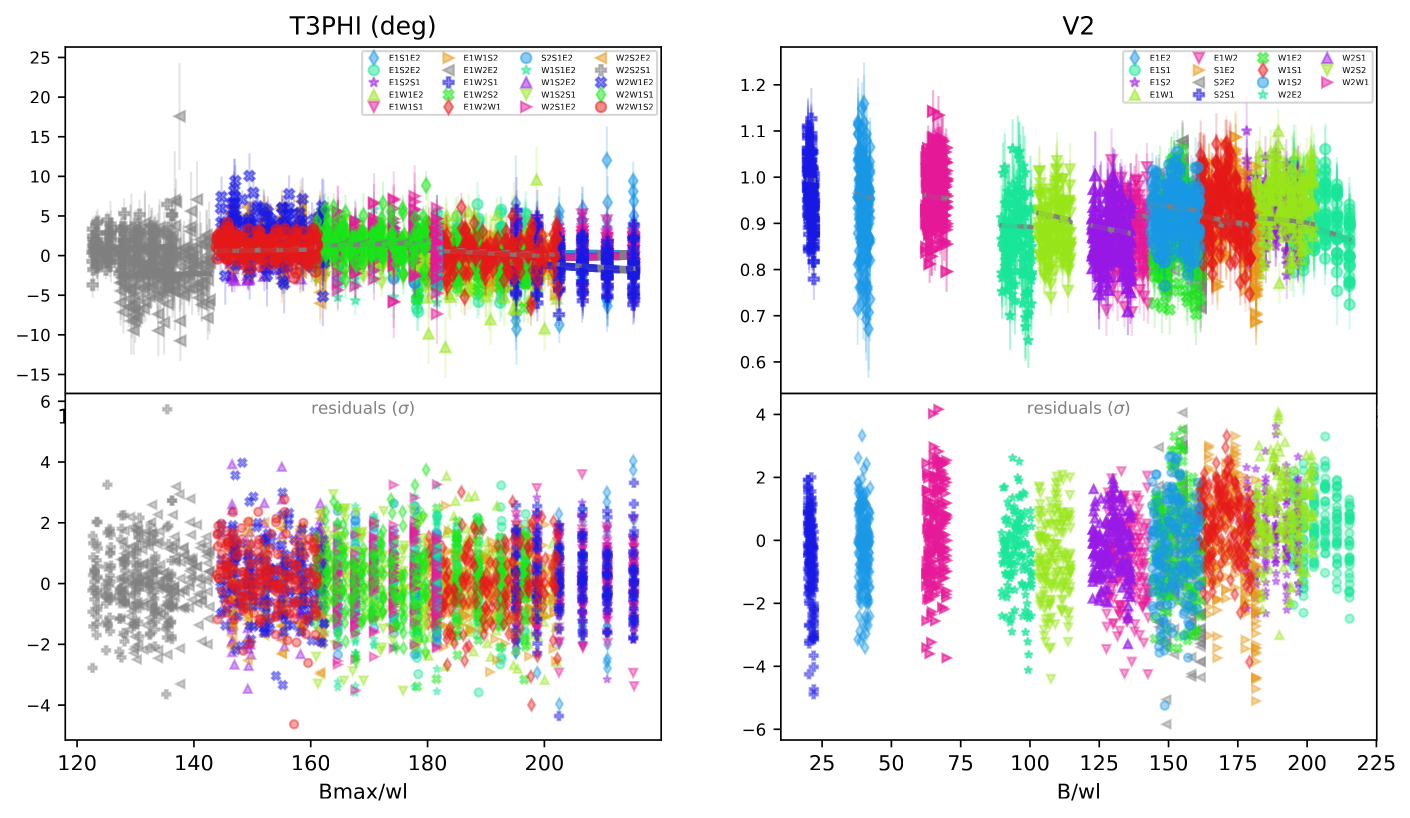}
\caption{Closure phases in units of degrees (T3PHI, left panel) and squared visibilities (V2, right panel) plotted versus the projected baseline (B) in units of the observed wavelength (wl) for V2119~Cyg obtained with CHARA/MIRC-X ($H$ band) on ${\rm RJD} = 59399.943$. The upper panels show the data (points with error bars colored according to the baselines specified in the legends) and the best-fit geometrical model (solid lines), while the lower panels show the residuals in units of $\sigma$. The geometrical model is composed of a Gaussian representing the Be star and its disk (FWHM of $0.13\pm0.02$\,mas), and a point source corresponding to the companion, which contributes $\sim2.2$\% of the flux of the Be star and its disk, and is separated by $\sim 1.20$\,mas at a PA $\sim 210${\degree}. The resulting reduced $\chi^2$ is $1.60$. \label{fig:V2119Cyg_mircx}}
\end{figure*}

\begin{figure*}
\epsscale{1.1}
\plotone{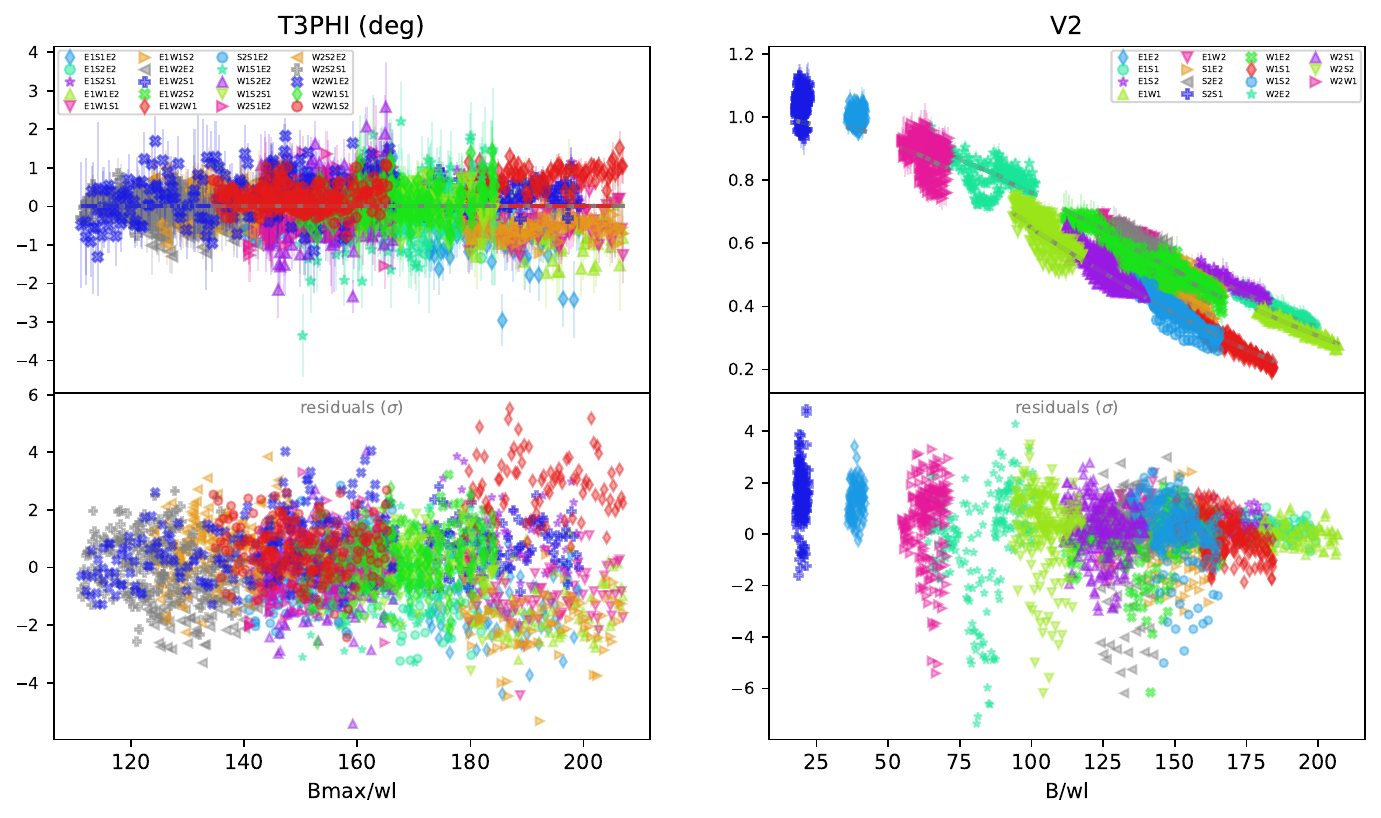}
\caption{Same as Fig.~\ref{fig:V2119Cyg_mircx}, but for $\beta$~CMi obtained on ${\rm RJD} = 59185.989$. The geometrical model is composed of an inclined Gaussian representing the Be star and its disk with the following parameters: ${\rm FWHM}=0.51\pm0.02$\,mas, $i=41\pm1^{\circ}$, and ${\rm PA}=134.5\pm0.5^{\circ}$. The resulting reduced $\chi^2$ is $1.74$. \label{fig:betCMi_mircx}}
\end{figure*}

In this Section, we provide additional plots showing the interferometric data and how the adopted models reproduce them (Figs.~\ref{fig:V2119Cyg_mircx} and \ref{fig:betCMi_mircx}).

\bibliography{Be_binaries}{}
\bibliographystyle{aasjournal}

\end{document}